\documentclass{elsarticle}
  
\usepackage[margin=1in]{geometry}

\usepackage{bbm}
\usepackage{graphicx}
\usepackage{pstool}
\usepackage{wrapfig}
\usepackage{caption}
\usepackage{subcaption}
\usepackage[section]{placeins} 

\usepackage{fancyhdr} 
\usepackage{lastpage} 
\usepackage{extramarks} 

\usepackage{xcolor} 
\usepackage{enumerate} 
\usepackage{paralist} 
\usepackage{amsmath, amsthm, amssymb, mathtools}
\usepackage{mathabx, pifont, stmaryrd} 
\usepackage[explicit]{titlesec} 
\usepackage{etoolbox} 
\usepackage{bibentry} 
\makeatletter\let\saved@bibitem\@bibitem\makeatother 
\usepackage[colorlinks, bookmarksopen, bookmarksnumbered,
            citecolor=red,urlcolor=red]{hyperref} 
\makeatletter\let\@bibitem\saved@bibitem\makeatother 
\usepackage{rotating}

\usepackage{algorithm}
\usepackage{algorithmic}

\newtheorem{remark}{Remark}

\theoremstyle{definition}

\usepackage{bm} 
\usepackage{amsfonts} 

\DeclareMathOperator*{\argmin}{arg\,min}

\newcommand{\func}[3]{\ensuremath{#1 : #2 \rightarrow #3}}

\newcommand{\norm}[1]{\ensuremath{\left\| #1 \right\|}}

\newcommand{\suchthat}{\mathrel{}\middle|\mathrel{}}

\newcommand{\pder}[2]{\ensuremath{\frac{\partial #1}{\partial #2}}}

\newcommand{\Ccal}{\ensuremath{\mathcal{C}}}
\newcommand{\Dcal}{\ensuremath{\mathcal{D}}}

\newcommand{\Fcal}{\ensuremath{\mathcal{F}}}
\newcommand{\Gcal}{\ensuremath{\mathcal{G}}}
\newcommand{\Hcal}{\ensuremath{\mathcal{H}}}
\newcommand{\Ical}{\ensuremath{\mathcal{I}}}
\newcommand{\Jcal}{\ensuremath{\mathcal{J}}}
\newcommand{\Kcal}{\ensuremath{\mathcal{K}}}

\newcommand{\Mcal}{\ensuremath{\mathcal{M}}}

\newcommand{\Ocal}{\ensuremath{\mathcal{O}}}
\newcommand{\Pcal}{\ensuremath{\mathcal{P}}}
\newcommand{\Qcal}{\ensuremath{\mathcal{Q}}}
\newcommand{\Rcal}{\ensuremath{\mathcal{R}}}
\newcommand{\Scal}{\ensuremath{\mathcal{S}}}

\newcommand{\Vcal}{\ensuremath{\mathcal{V}}}
\newcommand{\Wcal}{\ensuremath{\mathcal{W}}}
\newcommand{\Xcal}{\ensuremath{\mathcal{X}}}
\newcommand{\Ycal}{\ensuremath{\mathcal{Y}}}

\newcommand{\Rbb}{\ensuremath{\mathbb{R} }}

\newcommand\Rbm{{\ensuremath{\bm{R}}}}

\newcommand\Xbm{{\ensuremath{\bm{X}}}}

\newcommand\cbm{{\ensuremath{\bm{c}}}}

\newcommand\rbm{{\ensuremath{\bm{r}}}}

\newcommand\ubm{{\ensuremath{\bm{u}}}}

\newcommand\xbm{{\ensuremath{\bm{x}}}}
\newcommand\ybm{{\ensuremath{\bm{y}}}}

\newcommand\phibold{{\ensuremath{\boldsymbol{\phi}}}}

\newcommand\xibold{{\ensuremath{\boldsymbol{\xi}}}}

\newcommand\zerobold{\ensuremath{\mathbf{0}}}

\usepackage{tikz}
\usepackage{pgfplots}
\usepackage{pgfplotstable, filecontents, booktabs}
\pgfplotsset{compat=1.9}

\usepackage{tikz-3dplot} 
\usetikzlibrary{pgfplots.groupplots}
\usepgfplotslibrary{fillbetween}
\usetikzlibrary{calc,fit,matrix,arrows,automata,positioning,shapes}
\usetikzlibrary{arrows.meta}

\pgfplotsset{select coords between index/.style 2 args={
    x filter/.code={
        \ifnum\coordindex<#1\fi
        \ifnum\coordindex>#2\fi
    }
}}

\tikzset{
 invisible/.style={opacity=0},
 visible on/.style={alt={#1{}{invisible}}},
 alt/.code args={<#1>#2#3}{%
   \alt<#1>{\pgfkeysalso{#2}}{\pgfkeysalso{#3}}
 },
}

\pgfplotsset{
   colormap={parula}{rgb255=(62,38,168) rgb255=(62,39,172) rgb255=(63,40,175) rgb255=(63,41,178) rgb255=(64,42,180) rgb255=(64,43,183) rgb255=(65,44,186) rgb255=(65,45,189) rgb255=(66,46,191) rgb255=(66,47,194) rgb255=(67,48,197) rgb255=(67,49,200) rgb255=(67,50,202) rgb255=(68,51,205) rgb255=(68,52,208) rgb255=(69,53,210) rgb255=(69,55,213) rgb255=(69,56,215) rgb255=(70,57,217) rgb255=(70,58,220) rgb255=(70,59,222) rgb255=(70,61,224) rgb255=(71,62,225) rgb255=(71,63,227) rgb255=(71,65,229) rgb255=(71,66,230) rgb255=(71,68,232) rgb255=(71,69,233) rgb255=(71,70,235) rgb255=(72,72,236) rgb255=(72,73,237) rgb255=(72,75,238) rgb255=(72,76,240) rgb255=(72,78,241) rgb255=(72,79,242) rgb255=(72,80,243) rgb255=(72,82,244) rgb255=(72,83,245) rgb255=(72,84,246) rgb255=(71,86,247) rgb255=(71,87,247) rgb255=(71,89,248) rgb255=(71,90,249) rgb255=(71,91,250) rgb255=(71,93,250) rgb255=(70,94,251) rgb255=(70,96,251) rgb255=(70,97,252) rgb255=(69,98,252) rgb255=(69,100,253) rgb255=(68,101,253) rgb255=(67,103,253) rgb255=(67,104,254) rgb255=(66,106,254) rgb255=(65,107,254) rgb255=(64,109,254) rgb255=(63,110,255) rgb255=(62,112,255) rgb255=(60,113,255) rgb255=(59,115,255) rgb255=(57,116,255) rgb255=(56,118,254) rgb255=(54,119,254) rgb255=(53,121,253) rgb255=(51,122,253) rgb255=(50,124,252) rgb255=(49,125,252) rgb255=(48,127,251) rgb255=(47,128,250) rgb255=(47,130,250) rgb255=(46,131,249) rgb255=(46,132,248) rgb255=(46,134,248) rgb255=(46,135,247) rgb255=(45,136,246) rgb255=(45,138,245) rgb255=(45,139,244) rgb255=(45,140,243) rgb255=(45,142,242) rgb255=(44,143,241) rgb255=(44,144,240) rgb255=(43,145,239) rgb255=(42,147,238) rgb255=(41,148,237) rgb255=(40,149,236) rgb255=(39,151,235) rgb255=(39,152,234) rgb255=(38,153,233) rgb255=(38,154,232) rgb255=(37,155,232) rgb255=(37,156,231) rgb255=(36,158,230) rgb255=(36,159,229) rgb255=(35,160,229) rgb255=(35,161,228) rgb255=(34,162,228) rgb255=(33,163,227) rgb255=(32,165,227) rgb255=(31,166,226) rgb255=(30,167,225) rgb255=(29,168,225) rgb255=(29,169,224) rgb255=(28,170,223) rgb255=(27,171,222) rgb255=(26,172,221) rgb255=(25,173,220) rgb255=(23,174,218) rgb255=(22,175,217) rgb255=(20,176,216) rgb255=(18,177,214) rgb255=(16,178,213) rgb255=(14,179,212) rgb255=(11,179,210) rgb255=(8,180,209) rgb255=(6,181,207) rgb255=(4,182,206) rgb255=(2,183,204) rgb255=(1,183,202) rgb255=(0,184,201) rgb255=(0,185,199) rgb255=(0,186,198) rgb255=(1,186,196) rgb255=(2,187,194) rgb255=(4,187,193) rgb255=(6,188,191) rgb255=(9,189,189) rgb255=(13,189,188) rgb255=(16,190,186) rgb255=(20,190,184) rgb255=(23,191,182) rgb255=(26,192,181) rgb255=(29,192,179) rgb255=(32,193,177) rgb255=(35,193,175) rgb255=(37,194,174) rgb255=(39,194,172) rgb255=(41,195,170) rgb255=(43,195,168) rgb255=(44,196,166) rgb255=(46,196,165) rgb255=(47,197,163) rgb255=(49,197,161) rgb255=(50,198,159) rgb255=(51,199,157) rgb255=(53,199,155) rgb255=(54,200,153) rgb255=(56,200,150) rgb255=(57,201,148) rgb255=(59,201,146) rgb255=(61,202,144) rgb255=(64,202,141) rgb255=(66,202,139) rgb255=(69,203,137) rgb255=(72,203,134) rgb255=(75,203,132) rgb255=(78,204,129) rgb255=(81,204,127) rgb255=(84,204,124) rgb255=(87,204,122) rgb255=(90,204,119) rgb255=(94,205,116) rgb255=(97,205,114) rgb255=(100,205,111) rgb255=(103,205,108) rgb255=(107,205,105) rgb255=(110,205,102) rgb255=(114,205,100) rgb255=(118,204,97) rgb255=(121,204,94) rgb255=(125,204,91) rgb255=(129,204,89) rgb255=(132,204,86) rgb255=(136,203,83) rgb255=(139,203,81) rgb255=(143,203,78) rgb255=(147,202,75) rgb255=(150,202,72) rgb255=(154,201,70) rgb255=(157,201,67) rgb255=(161,200,64) rgb255=(164,200,62) rgb255=(167,199,59) rgb255=(171,199,57) rgb255=(174,198,55) rgb255=(178,198,53) rgb255=(181,197,51) rgb255=(184,196,49) rgb255=(187,196,47) rgb255=(190,195,45) rgb255=(194,195,44) rgb255=(197,194,42) rgb255=(200,193,41) rgb255=(203,193,40) rgb255=(206,192,39) rgb255=(208,191,39) rgb255=(211,191,39) rgb255=(214,190,39) rgb255=(217,190,40) rgb255=(219,189,40) rgb255=(222,188,41) rgb255=(225,188,42) rgb255=(227,188,43) rgb255=(230,187,45) rgb255=(232,187,46) rgb255=(234,186,48) rgb255=(236,186,50) rgb255=(239,186,53) rgb255=(241,186,55) rgb255=(243,186,57) rgb255=(245,186,59) rgb255=(247,186,61) rgb255=(249,186,62) rgb255=(251,187,62) rgb255=(252,188,62) rgb255=(254,189,61) rgb255=(254,190,60) rgb255=(254,192,59) rgb255=(254,193,58) rgb255=(254,194,57) rgb255=(254,196,56) rgb255=(254,197,55) rgb255=(254,199,53) rgb255=(254,200,52) rgb255=(254,202,51) rgb255=(253,203,50) rgb255=(253,205,49) rgb255=(253,206,49) rgb255=(252,208,48) rgb255=(251,210,47) rgb255=(251,211,46) rgb255=(250,213,46) rgb255=(249,214,45) rgb255=(249,216,44) rgb255=(248,217,43) rgb255=(247,219,42) rgb255=(247,221,42) rgb255=(246,222,41) rgb255=(246,224,40) rgb255=(245,225,40) rgb255=(245,227,39) rgb255=(245,229,38) rgb255=(245,230,38) rgb255=(245,232,37) rgb255=(245,233,36) rgb255=(245,235,35) rgb255=(245,236,34) rgb255=(245,238,33) rgb255=(246,239,32) rgb255=(246,241,31) rgb255=(246,242,30) rgb255=(247,244,28) rgb255=(247,245,27) rgb255=(248,247,26) rgb255=(248,248,24) rgb255=(249,249,22) rgb255=(249,251,21) },
}

\newcommand{\colorbarMatlabParula}[5]{
\begin{tikzpicture}
\begin{axis}[
   hide axis, scale only axis,
   height=0pt, width=0pt,
   colormap={parula}{rgb255=(62,38,168) rgb255=(62,39,172) rgb255=(63,40,175) rgb255=(63,41,178) rgb255=(64,42,180) rgb255=(64,43,183) rgb255=(65,44,186) rgb255=(65,45,189) rgb255=(66,46,191) rgb255=(66,47,194) rgb255=(67,48,197) rgb255=(67,49,200) rgb255=(67,50,202) rgb255=(68,51,205) rgb255=(68,52,208) rgb255=(69,53,210) rgb255=(69,55,213) rgb255=(69,56,215) rgb255=(70,57,217) rgb255=(70,58,220) rgb255=(70,59,222) rgb255=(70,61,224) rgb255=(71,62,225) rgb255=(71,63,227) rgb255=(71,65,229) rgb255=(71,66,230) rgb255=(71,68,232) rgb255=(71,69,233) rgb255=(71,70,235) rgb255=(72,72,236) rgb255=(72,73,237) rgb255=(72,75,238) rgb255=(72,76,240) rgb255=(72,78,241) rgb255=(72,79,242) rgb255=(72,80,243) rgb255=(72,82,244) rgb255=(72,83,245) rgb255=(72,84,246) rgb255=(71,86,247) rgb255=(71,87,247) rgb255=(71,89,248) rgb255=(71,90,249) rgb255=(71,91,250) rgb255=(71,93,250) rgb255=(70,94,251) rgb255=(70,96,251) rgb255=(70,97,252) rgb255=(69,98,252) rgb255=(69,100,253) rgb255=(68,101,253) rgb255=(67,103,253) rgb255=(67,104,254) rgb255=(66,106,254) rgb255=(65,107,254) rgb255=(64,109,254) rgb255=(63,110,255) rgb255=(62,112,255) rgb255=(60,113,255) rgb255=(59,115,255) rgb255=(57,116,255) rgb255=(56,118,254) rgb255=(54,119,254) rgb255=(53,121,253) rgb255=(51,122,253) rgb255=(50,124,252) rgb255=(49,125,252) rgb255=(48,127,251) rgb255=(47,128,250) rgb255=(47,130,250) rgb255=(46,131,249) rgb255=(46,132,248) rgb255=(46,134,248) rgb255=(46,135,247) rgb255=(45,136,246) rgb255=(45,138,245) rgb255=(45,139,244) rgb255=(45,140,243) rgb255=(45,142,242) rgb255=(44,143,241) rgb255=(44,144,240) rgb255=(43,145,239) rgb255=(42,147,238) rgb255=(41,148,237) rgb255=(40,149,236) rgb255=(39,151,235) rgb255=(39,152,234) rgb255=(38,153,233) rgb255=(38,154,232) rgb255=(37,155,232) rgb255=(37,156,231) rgb255=(36,158,230) rgb255=(36,159,229) rgb255=(35,160,229) rgb255=(35,161,228) rgb255=(34,162,228) rgb255=(33,163,227) rgb255=(32,165,227) rgb255=(31,166,226) rgb255=(30,167,225) rgb255=(29,168,225) rgb255=(29,169,224) rgb255=(28,170,223) rgb255=(27,171,222) rgb255=(26,172,221) rgb255=(25,173,220) rgb255=(23,174,218) rgb255=(22,175,217) rgb255=(20,176,216) rgb255=(18,177,214) rgb255=(16,178,213) rgb255=(14,179,212) rgb255=(11,179,210) rgb255=(8,180,209) rgb255=(6,181,207) rgb255=(4,182,206) rgb255=(2,183,204) rgb255=(1,183,202) rgb255=(0,184,201) rgb255=(0,185,199) rgb255=(0,186,198) rgb255=(1,186,196) rgb255=(2,187,194) rgb255=(4,187,193) rgb255=(6,188,191) rgb255=(9,189,189) rgb255=(13,189,188) rgb255=(16,190,186) rgb255=(20,190,184) rgb255=(23,191,182) rgb255=(26,192,181) rgb255=(29,192,179) rgb255=(32,193,177) rgb255=(35,193,175) rgb255=(37,194,174) rgb255=(39,194,172) rgb255=(41,195,170) rgb255=(43,195,168) rgb255=(44,196,166) rgb255=(46,196,165) rgb255=(47,197,163) rgb255=(49,197,161) rgb255=(50,198,159) rgb255=(51,199,157) rgb255=(53,199,155) rgb255=(54,200,153) rgb255=(56,200,150) rgb255=(57,201,148) rgb255=(59,201,146) rgb255=(61,202,144) rgb255=(64,202,141) rgb255=(66,202,139) rgb255=(69,203,137) rgb255=(72,203,134) rgb255=(75,203,132) rgb255=(78,204,129) rgb255=(81,204,127) rgb255=(84,204,124) rgb255=(87,204,122) rgb255=(90,204,119) rgb255=(94,205,116) rgb255=(97,205,114) rgb255=(100,205,111) rgb255=(103,205,108) rgb255=(107,205,105) rgb255=(110,205,102) rgb255=(114,205,100) rgb255=(118,204,97) rgb255=(121,204,94) rgb255=(125,204,91) rgb255=(129,204,89) rgb255=(132,204,86) rgb255=(136,203,83) rgb255=(139,203,81) rgb255=(143,203,78) rgb255=(147,202,75) rgb255=(150,202,72) rgb255=(154,201,70) rgb255=(157,201,67) rgb255=(161,200,64) rgb255=(164,200,62) rgb255=(167,199,59) rgb255=(171,199,57) rgb255=(174,198,55) rgb255=(178,198,53) rgb255=(181,197,51) rgb255=(184,196,49) rgb255=(187,196,47) rgb255=(190,195,45) rgb255=(194,195,44) rgb255=(197,194,42) rgb255=(200,193,41) rgb255=(203,193,40) rgb255=(206,192,39) rgb255=(208,191,39) rgb255=(211,191,39) rgb255=(214,190,39) rgb255=(217,190,40) rgb255=(219,189,40) rgb255=(222,188,41) rgb255=(225,188,42) rgb255=(227,188,43) rgb255=(230,187,45) rgb255=(232,187,46) rgb255=(234,186,48) rgb255=(236,186,50) rgb255=(239,186,53) rgb255=(241,186,55) rgb255=(243,186,57) rgb255=(245,186,59) rgb255=(247,186,61) rgb255=(249,186,62) rgb255=(251,187,62) rgb255=(252,188,62) rgb255=(254,189,61) rgb255=(254,190,60) rgb255=(254,192,59) rgb255=(254,193,58) rgb255=(254,194,57) rgb255=(254,196,56) rgb255=(254,197,55) rgb255=(254,199,53) rgb255=(254,200,52) rgb255=(254,202,51) rgb255=(253,203,50) rgb255=(253,205,49) rgb255=(253,206,49) rgb255=(252,208,48) rgb255=(251,210,47) rgb255=(251,211,46) rgb255=(250,213,46) rgb255=(249,214,45) rgb255=(249,216,44) rgb255=(248,217,43) rgb255=(247,219,42) rgb255=(247,221,42) rgb255=(246,222,41) rgb255=(246,224,40) rgb255=(245,225,40) rgb255=(245,227,39) rgb255=(245,229,38) rgb255=(245,230,38) rgb255=(245,232,37) rgb255=(245,233,36) rgb255=(245,235,35) rgb255=(245,236,34) rgb255=(245,238,33) rgb255=(246,239,32) rgb255=(246,241,31) rgb255=(246,242,30) rgb255=(247,244,28) rgb255=(247,245,27) rgb255=(248,247,26) rgb255=(248,248,24) rgb255=(249,249,22) rgb255=(249,251,21) },
   colorbar horizontal,
   point meta min=#1, point meta max=#5,
   scaled ticks=false,
   colorbar style={width=10cm, xtick={#1,#2,#3,#4,#5}}
]
\addplot [draw=none] coordinates {(0,0)};
\end{axis}
\end{tikzpicture}
}

\usepackage{setspace}

\newbool{fastcompile}
\setbool{fastcompile}{false}

\begin{document}
\title{An implicit shock tracking method for simulation of shock-dominated flows over complex domains using mesh-based parametrizations}

\author[rvt1]{Alexander M. P\'{e}rez Reyes\fnref{fn1}}
\ead{aperez23@nd.edu}

\author[rvt1]{Matthew J. Zahr\fnref{fn2}\corref{cor1}}
\ead{mzahr@nd.edu}

\address[rvt1]{Department of Aerospace and Mechanical Engineering, University
               of Notre Dame, Notre Dame, IN 46556, United States}
\cortext[cor1]{Corresponding author}

\fntext[fn1]{Graduate Student, Department of Aerospace and Mechanical
             Engineering, University of Notre Dame}
\fntext[fn2]{Assistant Professor, Department of Aerospace and Mechanical
             Engineering, University of Notre Dame}


\begin{abstract}
	A mesh-based parametrization is a parametrization of a geometric object that
	is defined solely from a mesh of the object, e.g., without an analytical
	expression or computer-aided design (CAD) representation of the object.
	In this work, we propose a mesh-based parametrization of an arbitrary
	$d'$-dimensional object embedded in a $d$-dimensional space using tools
	from high-order finite elements. Using mesh-based parametrizations,
	we construct a boundary-preserving parametrization of the nodal coordinates
	of a computational mesh that ensures all nodes remain on all their original
	boundaries. These boundary-preseving parametrizations allow the nodes of the
	mesh to move only in ways that will not change the computational domain. They
	also ensure nodes will not move between boundaries, which would cause issues
	assigning boundary conditions for partial differential equation simulations
	and lead to inaccurate geometry representations for non-smooth boundary transitions.
	Finally, we integrate boundary-preserving, mesh-based parametrizations into
	high-order implicit shock tracking, an optimization-based discontinuous
	Galerkin method that moves nodes to align mesh faces with non-smooth flow
	features to represent them perfectly with inter-element jumps, leaving the
	intra-element polynomial basis to represent smooth regions of the flow with
	high-order accuracy. Mesh-based parametrizations enable implicit shock tracking
	simulations of shock-dominated flows over geometries without simple analytical
	parametrizations. Several demonstrations of mesh-based parametrizations are
	provided to: (1) give concrete examples of the formulation, (2) show that accurate
	parametrizations can be obtained despite the surrogate surfaces only being
	$C^0$, (3) show they integrate seemlessly with implicit shock tracking and can
	be used to parametrize surfaces without explicit expressions (e.g., RAE2822
	airfoil), and (4) effectively parametrize complex geometries and prevent
	nodes from moving off their original boundaries.
\end{abstract}

\maketitle
\section{Introduction}
\label{sec:intro}
High-order discontinuous Galerkin (DG) methods
\cite{cockburn_rungekutta_2001, hesthaven_nodal_2008}
offer unique advantages for fluid flow simulations such as high accuracy
per degree of freedom, low dissipation, and the ability to handle complex
geometries. However, high-speed flows remain challenging because spurious
oscillations arise near shock waves and contact discontinuities that impact
the accuracy of the flow approximation and lead to solver failure. These
issues are exacerbated as the Mach and Reynolds number of the flow increase
or the complexity and bluntness of the vehicle increase. Robust and accurate
methods to stabilize DG discretizations for realistic shock-dominated flows are
needed to make them competitive for real-world aerospace applications
\cite{wang_high-order_2013}.

Shock-capturing methods, including limiting \cite{van1979towards},
reconstruction \cite{harten_uniformly_1987, liu_weighted_1994, jiang_efficient_1996}, and
artificial viscosity \cite{persson_sub-cell_2006,barter_shock_2010}, are the
most popular stabilization methods. They smear non-smooth features sufficiently so they are
well-represented on the grid and, as such, they can handle complex shock structures.
However, they are first-order accurate, at best, near shocks and contacts, which
must be offset with extremely fine meshes in these regions. On the other hand,
shock tracking methods move the grid to align mesh faces with non-smooth features
\cite{moretti2002thirty, salas2009shock}, which do not require stabilization because
the non-smooth features are represented perfectly with inter-elements jumps in the
basis. As such, shock tracking methods are much more accurate than shock capturing
methods; however, they are usually limited to simple problems with basic shock structures
\cite{shubin1981steady,shubin1982steady, rosendale1994floating}
and require dimension-dependent algorithms
\cite{harten1983self, glimm_conservative_2003, bell1982fully}.

A new class of high-order methods, \textit{implicit shock tracking}, has recently
emerged, including the Moving Discontinuous Galerkin Method with Interface Condition
Enforcement (MDG-ICE)  \cite{corrigan2019moving,kercher2020least,kercher2020moving}
and High-Order implicit Shock Tracking (HOIST)
method \cite{zahr2018shktrk, zahr_implicit_2020, 2022_huang_shktrk}.
Like traditional shock tracking, these methods seek to align mesh faces with solution
features; however, the grid is implicitly defined as the solution of an optimization
problem. As such, they transform a shock-agnostic mesh into one aligned with all
non-smooth flow features, and simultaneously compute a high-order accurate flow
solution on the aligned grid. This key innovation has enabled the benefits of shock tracking
to be realized for flows with complex shock structure (e.g., shock intersections, curved
shocks, shock formation) \cite{corrigan2019moving, corrigan2019convergence, corrigan2019unsteady, zahr_implicit_2020, 2022_huang_shktrk, naudet2024space}.

An integral component of implicit shock tracking methods is an approach to ensure the
boundaries of the domain are unchanged as nodes move to track the shocks. This
condition will be satisfied if the nodes on each domain boundary are fixed to
their original position; however, this stringent requirement will prevent the grid
from aligning with shocks that impinge on a domain boundary (e.g., transonic, inviscid
flows) or lead to highly distorted grids for problems with shocks near boundaries (e.g.,
hypersonic flows or shock-boundary layer interaction). A more realistic approach
uses explicit parametrizations---a mapping from generalized coordinates in the intrinsic
dimension of a geometric object to the object itself in the embedding space---of the
object on which each node lies (e.g., a boundary or intersection of boundaries) to
slide nodes along their original boundaries \cite{zahr_implicit_2020, 2022_huang_shktrk}.
That is, for each node $i$ in the computational mesh, we introduce a generalized
coordinate $\ybm_i$ and a mapping $\phibold_i$ from the parameter domain to the
physical nodal coordinate $\xbm_i=\phibold_i(\ybm_i)$ that is constructed to ensure
$\xbm_i$ remains on its original boundaries for any $\ybm_i$. Then, instead of directly
optimizing for $\xbm_i$---which would need to be constrained to lie on its original
boundaries---we optimize for $\ybm_i$ and reconstruct the nodal coordinate
as $\phibold_i(\ybm_i)$, which will, by construction, lie on the same boundaries
as node $i$ before mesh motion. These boundary-preserving parametrizations have
been effectively integrated into implicit shock tracking methods; however,
they have only proven useful for relatively simple domains, i.e., domains
with planar boundaries or boundaries that can be explicitly parametrized
with analytical functions
\cite{zahr_implicit_2020, 2022_huang_shktrk, corrigan2019convergence}.
This work introduces a general procedure to construct boundary-preserving
parametrizations directly from a conforming, high-order mesh of the domain,
which allows implicit shock tracking to be applied to a new class of problems
with complex geometries.

In this work, we parametrize $d'$-dimensional objects embedded in a
$d$-dimensional space by constructing a surrogate object from a high-order
mesh and use finite element tools to parametrize the surrogate.
We refer to such parametrizations as \textit{mesh-based parametrizations}
because they are defined exclusively from a high-order mesh of an object.
Because mesh-based parametrizations are defined elementwise, efficient
implementation relies on a fast, reliable approach to find the elements
in a high-order mesh in which an arbitrary point in the parameter space
lies. We introduce an efficient search algorithm
that associates a collection of points with elements of the mesh and
uses KD trees to efficient find the points (and therefore elements)
closest to an arbitrary evaluation point. Next, mesh-based parametrizations
are used to parametrize the nodal coordinates of a computational mesh
in such a way that nodes slide along all their original boundaries.
This ensures the shape of the boundaries will not change and a
computational face will never straddle two boundaries, which would
lead to difficulties assigning boundary conditions and poor geometry
representation if the transition between boundaries is not smooth.
Finally, these boundary-preserving mesh-based parametrizations are
integrated into an implicit shock tracking method so these methods
can be applied to a wider class of shock-dominated flow problems,
e.g., flows over geometries that do not admit simple explicit
parametrizations. The novelty of this work lies in:
(1) the formalization of mesh-based parametrizations,
(2) an efficient search algorithm to locate elements containing an
    arbitrary point in a high-order mesh,
(3) the use of mesh-based parametrizations to construct boundary-preserving
    parametrizations of the nodal coordinates of a computational mesh, and
(4) the integration of boundary-preserving mesh-based parametrization into
    implicit shock tracking and demonstration of the new capabilities.

Another common approach to parametrize objects, particularly in the mesh
generation community, uses computer-aided design (CAD) representation of the object
\cite{xie_2012_generation, toulorge_optimizing_2016, geuzaine_generation_2015}.
Despite being the most general approach to parametrize objects, it
comes with a number of limitations in the context of implicit shock
tracking. First, CAD geometries are often not sufficiently clean to parametrize
objects such that mesh nodes can seamlessly transition between different NURBS
patches, for example. Second, there are many CAD file formats, most of which support
a huge number of geometric objects, which makes working with them
cumbersome. Finally, derivatives of the parametrization, required by implicit shock
tracking solvers, are usually not available through CAD systems. Explicit
parametrizations can be avoided by combining an implicit representation of the
surface, i.e., a level set function, with a variational penalty function to force
nodes onto surfaces \cite{knupp_adaptive_2021,2023_barrera,mittal2024mixedordermeshesrpadaptivitysurface}.
While these methods effectively place faces (and therefore nodes) of a mesh onto
the implicitly defined surface, they cannot handle non-smooth features or transitions,
which exist in all relevant geometries, particularly aerospace vehicles
(e.g., Section~\ref{sec:numexp:euler}-\ref{sec:numexp:cone}).

The remainder of the paper is organized as follows. Section~\ref{sec:mbp}
introduces the concept of mesh-based parametrizations of arbitrary geometric
objects, uses them to construct a boundary-preserving parametrization of the
nodes of a computational mesh, and introduces an
efficient search algorithm to locate the elements of a high-order mesh
in which an arbitrary point lies, a central ingredient for the implementation of
mesh-based parametrizations. Section~\ref{sec:ist} integrates mesh-based parametrizations
into the High-Order Implicit Shock Tracking method. Section~\ref{sec:numexp:surf}
provides two concrete examples of the mesh-based parametrization ingredients
and shows the error between the surrogate and true surfaces rapidly decreases
under mesh refinement. Sections~\ref{sec:numexp:advec}-\ref{sec:numexp:euler}
show mesh-based parametrizations integrate cleanly with implicit shock
tracking, where solver convergence and highly accurate, grid-aligned
solutions are obtained. Section~\ref{sec:numexp:cone} demonstrates mesh-based
parametrizations for a complex geometry (sliced cone flap) with multiple surfaces
and non-smooth transitions between them.

\section{Mesh-based parametrization}
\label{sec:mbp}
In this section we introduce the concept of a mesh-based parametrization,
a parametric representation of geometric objects defined solely from a high-order
(geometry) mesh of the objects, assuming a complete mathematical or CAD description
is unavailable (Section~\ref{sec:mbp:obj}). These parametrizations
are used to construct a mesh-based parametrization of an entire computational
mesh (Section~\ref{sec:mbp:msh}) that guarantees nodes remain on their original
surfaces regardless of perturbations to them \textit{in the parameter domain}.
We close this section with an efficient strategy to identify the element(s)
of a high-order (curved) mesh in which an arbitrary point lies
(Section~\ref{sec:mbp:pnteval}), which is integral to efficient
implementation of mesh-based parametrization.

\subsection{Mesh-based parametrization of embedded geometric object}
\label{sec:mbp:obj}
Consider a $d'$-dimensional geometric object embedded in $d$ dimensions,
$\Scal \subset \Rbb^d$ ($d' < d$). We are mainly interested in curves ($d' = 1$)
and surfaces ($d' = d-1$); however, the development in this section is general.
Let $\hat\Scal_{h',q'}\in 2^{\Rbb^d}$ be a high-order mesh of $\Scal$ with
mesh size parameter (longest edge) $h' \in \Rbb_{\ge 0}$ and polynomial
degree $q'$, where $2^S$ is the power set (set of all subsets) of $S$.
That is, $\hat\Scal_{h',q'}$ is a discretization of $\Scal$ into
non-overlapping, potentially curved, $d'$-dimensional elements
(embedded in $\Rbb^{d}$). Because the mesh $\hat\Scal_{h',q'}$ will
not, in general, perfectly represent the original object $\Scal$,
we define the surrogate object $\Scal_{h',q'} \subset \Rbb^d$ as
\begin{equation}
 \Scal_{h',q'} \coloneqq \bigcup_{K \in \hat\Scal_{h',q'}} K.
\end{equation}
We assume the mesh $\hat\Scal_{h',q'}$ is built from conforming Lagrangian finite elements
to ensure the surrogate $\Scal_{h',q'}$ is globally $C^0$. The remainder of this section
will construct a parametrization of the surrogate object $\Scal_{h',q'}$, i.e., a mapping
$\Mcal_{h',q'} : \Rcal_{h',q'} \rightarrow \Scal_{h',q'}$, where $\Rcal_{h',q'}\subset\Rbb^{d'}$ is
the parameter domain. This parametrization guarantees that for any $r \in \Rcal_{h',q'}$,
we have $\Mcal_{h',q'}(r) \in \Scal_{h',q'}$.

To construct a mesh-based parametrization, we first define each element of
the mesh $K \in \hat\Scal_{h',q'}$ as the image of a polynomial mapping,
$\Qcal_{h',q'}^K \in \left[\Pcal_{q'}(\Omega_\square)\right]^d$, applied to an
idealized parent element $\Omega_\square \subset \Rbb^{d'}$,
i.e.,
\begin{equation}
	K = \Qcal_{h',q'}^K(\Omega_\square),
\end{equation}
where $\Pcal_{q'}(\Omega_\square)$ is an appropriate polynomial space of degree
$q'$ over $\Omega_\square$. The mapping is defined as
\begin{equation} \label{eqn:Qmap}
    \Qcal_{h',q'}^K : \xi \mapsto \xbm_{i}^{K} \phi_{i}(\xi),
\end{equation}
where $\{\phi_i\}_{i=1}^{n_{q'}}$ is a nodal basis of $\Pcal_{q'}(\Omega_\square)$
associated with the nodes $\{\xibold_1, \dots, \xibold_{n_{q'}}\}\subset\Omega_\square$,
$n_{q'} = \dim \Pcal_{q'}(\Omega)$, and
$\{\xbm_i^K,\dots,\xbm_{n_{q'}}^K\}\subset K$ are the high-order nodes of the mesh
$\hat\Scal_{h',q'}$ associated with element $K\in\hat\Scal_{h',q'}$.

Next, let $\Pi_{h',q'} : \Scal_{h',q'} \rightarrow \Rbb^{d'}$ be any injective mapping from
the surrogate object to $\Rbb^{d'}$. We use this mapping to define the parameter
space $\Rcal_{h',q'} \subset \Rbb^{d'}$ as
\begin{equation} \label{eqn:paramsp}
	\Rcal_{h',q'} \coloneqq \Pi_{h',q'}(\Scal_{h',q'}).
\end{equation}
Furthermore, we define a mesh $\hat\Rcal_{h',q'} \in 2^{\Rbb^{d'}}$ of the parameter
space by applying $\Pi_{h',q'}$ to each element of the original mesh $\hat{S}_{h',q'}$, i.e.,
\begin{equation} \label{eqn:paramsp_msh}
	\hat\Rcal_{h',q'} \coloneqq
	\left\{\Pi_{h',q'}(K) \suchthat K \in \hat\Scal_{h',q'}\right\}.
\end{equation}
Each element $\check{K}\in\hat\Rcal_{h',q'}$, where $\check{K} = \Pi_{h',q'}(K)$
for some $K\in\hat\Scal_{h',q'}$, can be written as the image of a polynomial
mapping, $\Gcal_{h',q'}^{\check{K}} \in \left[\Pcal_{q'}(\Omega_\square)\right]^{d'}$,
applied to the parent element $\Omega_\square$, i.e.,
\begin{equation}
	\check{K} = \Gcal_{h',q'}^{\check{K}}(\Omega_\square).
\end{equation}
The mapping, $\Gcal_{h',q'}^{\check{K}}$, is defined as
\begin{equation} \label{eqn:Gmap}
	\Gcal_{h',q'}^{\check{K}} : \xi \mapsto \Pi_{h',q'}(\xbm_i^K)\phi_i(\xi)
\end{equation}
and we denote its inverse as $\Hcal_{h',q'}^{\check{K}} : \Rbb^{d'}\rightarrow \Rbb^{d'}$,
i.e., $\Hcal_{h',q'}^{\check{K}}(\Gcal_{h',q'}^{\check{K}}(\xi)) = \xi$ and
$\Gcal_{h',q'}^{\check{K}}(\Hcal_{h',q'}^{\check{K}}(r)) = r$.

\begin{figure}
    \centering
    \resizebox{0.65\textwidth}{!}{
        \input{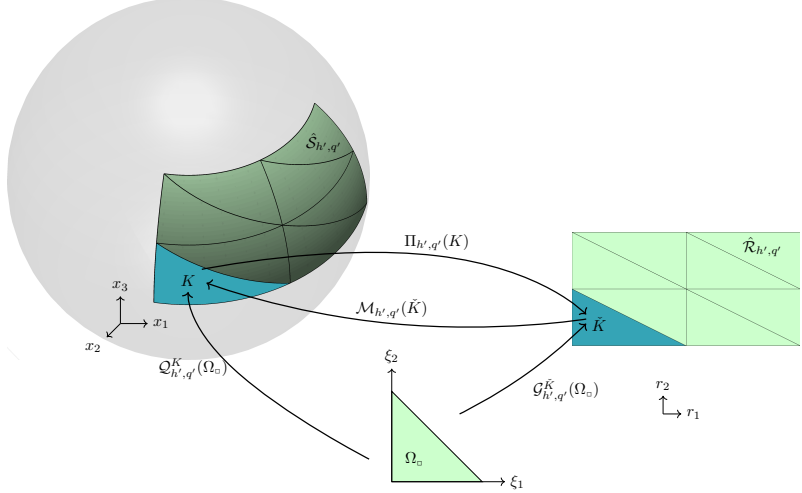}
    }
    \caption{
	    Schematic of the surface mesh parametrization $\Mcal_{h',q'}$ built as the
	    composition of $\Qcal_{h',q'}^K$ and $\Gcal_{h',q'}^{\check{K}}$ mappings
	    (Section~\ref{sec:mbp:obj}) for a surface patch of a sphere in $d=3$
	    dimensions ($d' = 2$).
    }
    \label{fig:surf_map_demo}
\end{figure}

Finally, the mesh-based parametrization of the surrogate object is defined as
the composition of the mappings in (\ref{eqn:Qmap}) and (\ref{eqn:Gmap})
\begin{equation} \label{eqn:mbp_surf}
	\left.\Mcal_{h',q'}\right|_{\check{K}} : r \mapsto
	\Qcal_{h',q'}^K\left(\Hcal_{h',q'}^{\check{K}}(r)\right) = 
	\xbm_i^K \phi_i\left(\Hcal_{h',q'}^{\check{K}}(r)\right).
\end{equation}
By construction, $\Mcal_{h',q'}(r) \in \Scal_{h',q'}$ for any $r\in\Rcal_{h',q'}$. The
construction of the mesh-based parametrization $\Mcal_{h',q'}$ is illustrated in
Figure~\ref{fig:surf_map_demo} for a sphere surface patch ($d=3$, $d'=2$).

The partial derivatives of the three mappings introduced in this section will
be needed to construct the first derivative of the implicit shock tracking objective
and constraint in Section~\ref{sec:ist}, which, in turn, will be needed to solve the
constrained optimization problem using gradient-based methods. The derivative of
$\Qcal_{h',q'}^K$ in (\ref{eqn:Qmap}), denoted
$Q_{h',q'}^K : \Omega_\square \rightarrow \Rbb^{d\times d'}$, is defined as
\begin{equation}
	Q_{h',q'}^K : \xi \mapsto \pder{\Qcal_{h',q'}^K}{\xi}(\xi) =
	\xbm_i^K\pder{\phi_i}{\xi}(\xi),
\end{equation}
where $\pder{\phi_i}{\xi} : \Rbb^{d'} \rightarrow \Rbb^{1\times d'}$ is the
derivative of the nodal basis function $\phi_i$ for $i=1,\dots,n_{q'}$.
The derivative of $\Gcal_{h',q'}^{\check{K}}$ in (\ref{eqn:Gmap}), denoted
$G_{h',q'}^{\check{K}} : \Omega_\square \rightarrow \Rbb^{d'\times d'}$, is defined as
\begin{equation}
	G_{h',q'}^{\check{K}} : \xi \mapsto \pder{\Gcal_{h',q'}^{\check{K}}}{\xi}(\xi) =
	\Pi_{h',q'}(\xbm_i^K)\pder{\phi_i}{\xi}(\xi).
\end{equation}
Finally, the derivative of $\Mcal_{h',q'}$ in (\ref{eqn:mbp_surf}), denoted
$M_{h',q'} : \Rcal_{h',q'} \rightarrow \Rbb^{d \times d'}$, is defined as
\begin{equation} \label{eqn:mbp_surf_deriv}
	M_{h',q'} : r \mapsto \pder{\Mcal_{h',q'}}{r}(r), \qquad
	\left.\pder{\Mcal_{h',q'}}{r}\right|_{\check{K}} : r \mapsto
	Q_{h',q'}^K(\Hcal_{h',q'}^{\check{K}}(r)) \left[G_{h',q'}^{\check{K}}(\Hcal_{h',q'}^{\check{K}}(r))\right]^{-1}.
\end{equation}

\begin{remark}
	The mapping $\Pi_{h',q'}$ is non-unique. In this work, we use a simple approach
	to define $\Pi_{h',q'}$ by orthogonally projecting points in $\Scal_{h',q'}$ onto
	a predefined plane ($d' = d-1$) or line ($d' = 1$); see~\ref{sec:surfproj}
	for a complete description.
\end{remark}

\begin{remark} \label{rem:mbp_deriv}
	From the elemental definition of the parametrization $\Mcal_{h',q'}$
	(\ref{eqn:mbp_surf}) and the assumption that the surrogate object
	$\hat\Scal_{h',q'}$ is globally $C^0$ (not $C^1$),
	the derivative of $\Mcal_{h',q'}$ is discontinuous across element
	boundaries. In this work, we investigate two approaches to obtain a single
	value of $M_{h',q'}(r)$ for any $r\in\bigcap_{\check{K}\in\Xcal_r}\check{K}$,
	where $\Xcal_r\subset\hat\Rcal_{h',q'}$ is the collection of all elements
	containing the point $r$.
	The simplest and most efficient option applies (\ref{eqn:mbp_surf_deriv}) for
	a single, arbitrary $\hat{K}\in\Xcal_r$. A more accurate (relative to the
	derivative of a parametrization of the true, not surrogate, surface) but more
	expensive approach averages the derivative over all elements in $\Xcal_r$, i.e.,
	\begin{equation} \label{eqn:mbp_surf_deriv1}
		\pder{\Mcal_{h',q'}}{r} : r \mapsto
		\frac{1}{|\Xcal_r|}\sum_{\check{K}\in\Xcal_r}
        Q_{h',q'}^K(\Hcal_{h',q'}^{\check{K}}(r)) \left[G_{h',q'}^{\check{K}}(\Hcal_{h',q'}^{\check{K}}(r))\right]^{-1}.
	\end{equation}
	While the definition in (\ref{eqn:mbp_surf_deriv1}) is more accurate than
	(\ref{eqn:mbp_surf_deriv}), the implementation in (\ref{eqn:mbp_surf_deriv})
	is easier and more efficient because it only requires finding a single element
        $\check{K}\in\hat\Rcal_{h',q'}$ containing $r$. Both methods converge to the
	derivative of a parametrization of the true surface as the mesh is refined
	$h'\rightarrow 0$, especially for higher polynomial degrees $q'$. In this
	work, we use the more accurate implementation in (\ref{eqn:mbp_surf_deriv1}),
	aided by our efficient search algorithm in Section~\ref{sec:mbp:pnteval}.
\end{remark}

\subsection{Boundary-preserving, mesh-based parametrization of computational mesh}
\label{sec:mbp:msh}
Implicit shock tracking dynamically moves the nodes of a computational mesh to align
element faces with non-smooth flow features. While internal nodes can move freely,
boundary nodes must always remain on their original boundaries. Allowing nodes to
move on or off boundaries would lead to faces that straddle multiple boundaries.
This would make it impossible to assign a single boundary condition to such faces.
Furthermore, complex geometries are often non-smooth across boundary surfaces so
allowing a face to straddle multiple boundaries would lead to oscillations in the
high-order element. In this section we use the mesh-based parametrizations
(Section~\ref{sec:mbp:obj}) to parametrize all nodes of a computational mesh
in such a way that they are guaranteed to remain on their original boundaries.
Such a parametrization was introduced in \cite{zahr_implicit_2020, 2022_huang_shktrk, corrigan2019convergence} for domains with planar boundaries and for domains with curved
boundaries assuming an explicit parametrization of the curved boundaries was available.
Our approach only requires a mesh of the boundaries and their intersections.

Consider a computational domain $\Omega \subset \Rbb^d$ with smooth
boundaries $\partial\Omega_i \subset \Rbb^d$ for $i = 1,\dots,N_\mathrm{b}$.
Let $\hat\Omega_{h,q} \in 2^{\Rbb^d}$ be a high-order mesh of $\Omega$
with mesh size parameter $h$ and polynomial degree $q$. This will be
the mesh used for the implicit shock tracking simulation
in Sections~\ref{sec:ist}-\ref{sec:numexp}. We assume the
mesh $\hat\Omega_{h,q}$ is built from conforming Lagrangian
finite elements to ensure the surrogate domain
\begin{equation} \label{eqn:surdom}
	\Omega_{h,q} \coloneqq \bigcup_{K \in \hat\Omega_{h,q}} K
\end{equation}
is globally $C^0$, and let $\{\Xbm_I\}_{I=1}^{N_x} \subset \Rbb^d$ be the nodes
(global numbering) associated with the Lagrange elements. Next, define
$B_i \in 2^{\{1,\dots,N_\mathrm{b}\}}$ as the collection of boundaries on
which node $i$ lies for $i=1,\dots,N_x$, i.e.,
$I \notin B_i\implies\Xbm_i\notin\partial\Omega_I$.
Let $\Ical_s$ denote the collection of
all nodes that lie on exactly $s$ boundaries, i.e.,
\begin{equation}
	\Ical_s \coloneqq \left\{i \in \{1,\dots,N_x\} \suchthat |B_i| = s\right\},
\end{equation}
for $s = 0, 1, \dots, N_\mathrm{b}$, where $|S|$ is the cardinality of the set $S$.
Each $i\in\{1,\dots,N_x\}$ lies in exactly one of the sets
$\Ical_0,\dots,\Ical_{N_\mathrm{b}}$.

Next, we construct a mesh-based parametrization of the computational mesh
$\hat\Omega_{h,q}$ as a parametrization of its nodal coordinates.
Let $\hat\Scal_{h',q'}^S$ be a high-order mesh of the intersection of
the boundaries in $S \in 2^{\{1,\dots,N_\mathrm{b}\}}$, i.e.,
$\bigcap_{s\in S} \partial\Omega_s$, with mesh size parameter $h'$ and
polynomial degree $q'$, and let $\Scal_{h',q'}^S = \bigcup_{K\in\hat\Scal_{h',q'}^S} K$
be the surrogate intersection. We call $\hat\Scal_{h',q'}^S$ a \textit{geometry}
mesh in the remainder; see Remark~\ref{rem:geommsh} for approaches to construct
$\hat\Scal_{h',q'}$. Then, let $\Mcal_{h',q'}^S$ (\ref{eqn:mbp_surf}) be a mesh-based
parametrization of the intersection of the boundaries in $S$ based on the geometry
mesh $\hat\Scal_{h',q'}^S$ with parameter domain $\Rcal_{h',q'}^S$
and mesh $\hat\Rcal_{h',q'}^S$ (\ref{eqn:paramsp_msh}),
and physical-to-parameter mapping $\Pi_{h',q'}^S$ (\ref{eqn:paramsp}).
From these definitions, the domain of the computational mesh parametrization
$\Ycal_{h',q'}\subset\Rbb^{N_y}$ is
\begin{equation} \label{eqn:paramdom-msh}
	\Ycal_{h',q'} \coloneqq \Rcal_{h',q'}^{B_1} \times \cdots \Rcal_{h',q'}^{B_{N_x}},
\end{equation}
where $\dim\Rcal_{h',q'}^{B_i} = \max\{d - s,0\}$ for $i \in \Ical_s$
(Remark~\ref{rem:intdim}), which makes the embedding dimension
$N_y = \sum_{s=0}^{d-1} (d-s) |\Ical_s|$.
Owing to the form of the parameter domain (\ref{eqn:paramdom-msh}), any
$\ybm \in \Ycal_{h',q'}$ has a nodal structure $\ybm = (\ybm_1,\dots,\ybm_{N_x})$
where $\ybm_i \in \Rcal_{h',q'}^{B_i}\subset\Rbb^{\max\{d-s,0\}}$
for $i \in \Ical_s$. Finally, we define the mesh-based parametrization of the computational
mesh $\hat\Omega_{h,q}$ as $\phibold : \Ycal_{h',q'} \rightarrow \Rbb^{d N_x}$,
$\phibold : \ybm \mapsto (\phibold_1(\ybm_1), \dots, \phibold_{N_x}(\ybm_{N_x}))$.
The nodal parametrization $\phibold_i : \Rcal_{h',q'}^{B_i} \rightarrow \Rbb^d$
is defined as
\begin{equation} \label{eqn:mbp_msh}
	\phibold_i : \ybm_i \mapsto
	\begin{cases}
		\ybm_i & i \in \Ical_0 \\
		\Mcal_{h',q'}^{B_i}(\ybm_i) & i \in \Ical_1 \cup \cdots \cup \Ical_{d-1} \\
		\Xbm_i & i \in \Ical_d \cup \cdots \cup \Ical_{N_\mathrm{b}}
	\end{cases}
\end{equation}
for $i = 1,\dots,N_x$. We call $\ybm\in\Ycal_{h',q'}$ the unconstrained degrees of freedom
because, for any $\ybm_i\in\Rcal_{h',q'}^{B_i}$, $\xbm_i = \phibold_i(\ybm_i)\in\Rbb^d$
is guaranteed, by construction, to lie on the same (surrogate) boundaries as
$\Xbm_i\in\Rbb^d$, i.e., $\xbm_i\in\Scal_{h',q'}^{B_i}$. The Jacobian of the
mesh parametrization,
$\pder{\phibold}{\ybm} : \Rbb^{N_y} \rightarrow \Rbb^{dN_x \times N_y}$,
is obtained from direct differentiation of (\ref{eqn:mbp_msh}) to yield
\begin{equation} \label{eqn:mbp_msh_deriv}
	\pder{\phibold}{\ybm} : \ybm \mapsto
	\begin{bmatrix}
		\pder{\phibold_1}{\ybm_1}(\ybm) & & \\
		& \ddots & \\
		& & \pder{\phibold_{N_x}}{\ybm_{N_x}}(\ybm)
	\end{bmatrix}, \qquad
	\pder{\phibold_i}{\ybm_i} : \ybm_i \mapsto
	\begin{cases}
		I_{d\times d} & i \in \Ical_0 \\
		M_{h',q'}^{B_i}(\ybm_i) & i \in \Ical_1 \cup \cdots \cup \Ical_{d-1} \\
		\zerobold_{d\times 0} & i \in \Ical_d \cup \cdots \cup \Ical_{N_\mathrm{b}},
	\end{cases}
\end{equation}
where $\pder{\phibold_i}{\ybm_i}(\ybm_i) \in \Rbb^{d\times \max\{d-s,0\}}$ for $i\in\Ical_s$
and $M_{h',q'}^S = \pder{\Mcal_{h',q'}^S}{r}$ is the derivative of the mesh-based
parametrization $\Mcal_{h',q'}^S$ (\ref{eqn:mbp_surf_deriv}).

\begin{remark}
	Any non-smooth features on $\partial\Omega$, the boundary of the computational
	domain $\Omega$, are assumed to lie at the intersection of distinct boundaries,
	which means each boundary itself is smooth. This is standard practice in both
	CAD modeling and mesh generation and therefore does not increase the burden of
	constructing mesh-based parametrizations.
\end{remark}

\begin{remark} \label{rem:geommsh}
	The geometry mesh, $\hat\Scal_{h',q'}$, can be extracted from the computational
	mesh $\hat\Omega_{h,q}$, in which case, $h' = h$ and $q' = q$. It can
	also easily be built with different resolution ($h' \neq h$, $q' \neq q$)
	than the computational mesh and can be defined without a volume mesh. A
	straightforward strategy to build a geometry mesh is to create a conforming
	mesh of all boundaries with sufficient resolution ($h'$ and $q'$) for the
	\textit{geometry} of the vehicle without regard to the flow conditions or
	features expected. This is usually a byproduct of the computational mesh
	$\hat\Omega_{h,q}$ generation process so it is not an additional burden
	in the workflow. The geometry mesh of all boundaries and their intersections
	can be extracted directly from this surface mesh.
\end{remark}

\begin{remark} \label{rem:intdim}
	The subscript $s$ in $\Ical_s$ denotes the number of constraints imposed on
	the nodes in the set, which means node $i \in \Ical_s$ will lie on a geometric
	object of dimension $\max\{d - s,0\}$. For example, in $d = 3$, any node in
	$\Ical_0$ is unconstrained in three-dimensional volume, any node in $\Ical_1$
	lies on exactly one surface (intrinsic dimension $d' = 2$), any node in $\Ical_2$
	lies at the intersection of two surfaces (a curve with intrinsic dimension
	$d' = 1$), and any node in $\Ical_3\cup\cdots\cup\Ical_{N_\mathrm{b}}$ lies
	at the intersection of three or more surfaces (a point with intrinsic dimension
	$d' = 0$). This approach could overconstrain node motion because it is possible
	that a completely smooth surfaces is arbitrarily split into two boundaries. In
	this setting, nodes would be required to remain on the curve at the intersection
	of these two boundaries. The proposed approach could be modified to allow nodes
	to move between boundaries if the transition between them is smooth and they use
	the same boundary condition. However, there is rarely a practical benefit of this
	approach due to its additional cost and the increased software complexity.
\end{remark}

\begin{remark}
	The mapping from unconstrained degrees of freedom ($\ybm$) to nodal coordinates
	($\xbm$) is well-defined; however, for most values of $\xbm\in\Rbb^{dN_x}$, it
	is not possible to define $\ybm \in \Rbb^{N_y}$ such that $\xbm = \phibold(\ybm)$
	(i.e., invert the mapping $\phibold$). We define a pseduo-inverse of $\phibold$
	as $\phibold^\dagger : \Rbb^{d N_x} \rightarrow \Rbb^{N_y}$, where
	$\phibold^\dagger : \xbm \mapsto (\phibold_1^\dagger(\xbm_1),\dots,\phibold_{N_x}^\dagger(\xbm_{N_x}))$ and $\phibold_i^\dagger : \Rbb^d \mapsto \Rbb^{\max\{d-s,0\}}$ with $i\in\Ical_s$ for $i = 1, \dots, N_x$ is defined as
\begin{equation}
        \phibold_i^\dagger(\xbm) =
        \begin{cases}
                \xbm_i & i \in \Ical_0 \\
		\Pi_{h',q'}^{B_i}(\xbm_i) & i \in \Ical_1 \cup \cdots \cup \Ical_{d-1} \\
		\Xbm_i & i \in \Ical_d\cup\cdots\cup\Ical_{N_\mathrm{b}}.
        \end{cases}
\end{equation}
\end{remark}

\subsection{Efficient element search in high-order mesh}
\label{sec:mbp:pnteval}
Evaluation of $\Mcal_{h',q'}(r)$ (\ref{eqn:mbp_surf}) or $M_{h',q'}(r)$ for any
$r \in \Rcal_{h',q'}$ requires a search through the mesh $\hat\Rcal_{h',q'}$ to find the
element(s) $\check{K}\in\hat\Rcal_{h',q'}$ in which $r$ lies. The condition $r\in\check{K}$
is difficult to check if $\check{K}$ is a curved element; however, $r \in \check{K}$
if and only if $\xi \in \Omega_\square$ where $r = \Gcal_{h',q'}^{\check{K}}(\xi)$.
The later condition $\xi \in \Omega_\square$ is easy to check because $\Omega_\square$
is a straight-sided parent element, but requires inversion of the element mapping,
which can be expensive. Furthermore, we must search through each element
$\check{K}\in\hat\Rcal_{h',q'}$ until one (or more) containing $r\in\Rcal_{h',q'}$ is
found. This strategy can be accelerated by eliminating elements that do not need
to be checked based on the location of
$r\in\Rcal_{h',q'}$ \cite{krause1996fast,noorani2015informal}; however,
this can be tricky for curved elements. We propose a variant of the search acceleration
techniques in \cite{noorani2015informal} specialized for high-order (curved) meshes.
The algorithm is defined for the mesh $\hat\Rcal_{h',q'}$ of the domain $\Rcal_{h',q'}$
with element mapping $\Gcal_{h',q'}^{\check{K}}$ because the notation is established;
however, it could be applied to any high-order mesh of a given domain.

Given a point $r \in \Rcal_{h',q'}$, we wish to find all elements
$\check{K}\in\hat\Rcal_{h',q'}$ such that $r \in \check{K}$. First,
we distribute points in each element of the mesh $\hat\Rcal_{h',q'}$.
Let $\Dcal_{\check{K}} \subset \check{K}$ with $|\Dcal_{\check{K}}| = n$
denote this set of points (Remark~\ref{rem:pntdist}) and let
$\Dcal\subset\Rcal_{h',q'}$ be the union of all points, i.e.,
\begin{equation}
	\Dcal = \bigcup_{\check{K}\in\hat\Rcal_{h',q'}} \Dcal_{\check{K}}.
\end{equation}
Next, we define the $k$th \textit{nearest point} of $r \in \Rcal_{h',q'}$ as
$y_r^k \in \Dcal$, where
\begin{equation}
	y_r^k \coloneqq \argmin_{y \in \mathcal{D} \setminus \{y_r^1,\dots,y_r^{k-1}\}} \| r - y \|
\end{equation}
for $k > 0$ (Remark~\ref{rem:kdtree}). We associate the nearest points with a
collection of elements (each point may lie in multiple elements),
$\Jcal_r^k \subset \hat\Rcal_{h',q'}$, where
\begin{equation}
	\Jcal_r^k \coloneqq
	\left\{
		\check{K} \in \hat\Rcal_{h',q'} \suchthat
		y_r^k \in \Dcal_{\check{K}}
	\right\} \setminus \bigcup_{k'=1}^{k-1} \Jcal_r^{k'}.
\end{equation}
The elements of the sets $\Jcal_r^1,\dots\Jcal_r^{k-1}$ are not
included in $\Jcal_r^k$ to avoid searching for the point $r$ in
the same element multiple times. Finally, we build the collection
of elements containing $r$ as $\Kcal_r \coloneqq \Kcal_r^1 \cup \cdots \cup \Kcal_r^k$,
where $\Kcal_r^k \subset \hat\Rcal_{h',q'}$ is defined as
\begin{equation}
	\Kcal_r^k \coloneqq
	\left\{
		\check{K} \in \Jcal_r^k \suchthat
		\xi \in \Omega_\square \text{ where }
		r = \Gcal_{h',q'}^{\check{K}}(\xi)
	\right\}.
\end{equation}
That is, the set $\Kcal_r^k$ is built from inverting the element mapping of each
element in $\Jcal_r^k$ at $r$ and determining if that new point lies in $\Omega_\square$.
As written, this method will find all elements in $\Jcal_r^1\cup\cdots\cup\Jcal_r^k$
containing $r$. In practice, $k$ is used as a safeguard to prevent the search from
including all elements of the mesh (we take $k = 5$ in this work).
If we only seek \textit{one} element containing $r$, the sets
$\Kcal_r^1,\dots,\Kcal_r^k$ are built sequentially and the search
is terminated after the first element containing $r$ is found (usually
during the construction of $\Kcal_r^1$). The proposed acceleration
search algorithm is demonstrated in Figure~\ref{fig:point:query}.

\begin{remark} \label{rem:pntdist}
	The point distributions, $\Dcal_{\check{K}}$, can be constructed by
	first defining a distribution in the parent element
	$\Dcal_\square\subset\Omega_\square$ and using the element transformation
	to map them to the physical element $\check{K}\in\hat\Rcal_{h',q'}$, i.e.,
	\begin{equation}
		\Dcal_{\check{K}} =
		\left\{\Gcal_{h',q'}^{\check{K}}(\xi) \suchthat \xi \in \Dcal_\square\right\}.
	\end{equation}
\end{remark}
\begin{remark} \label{rem:kdtree}
	The nearest neighbors $y_r^j$ for $j=1,\dots,k$ can be efficiently found
	using binary trees.
	In this work we use the implementation in the $\mathtt{NearestNeighbors.jl}$
	Julia package \cite{2024_Carlsson_NN}.
\end{remark}
\begin{remark} \label{rem:algeff}
	This algorithm is efficient because the point $r$ almost always lies in one of the
	elements of $\Jcal_r^1$. In practice, this only fails when $n$ is too small
	or when the mesh is highly skewed. In those cases, the search must proceed
	to $\Jcal_r^2$. We have not encountered a situation, in practice, where the
	search must continue to $\Jcal_r^3$ to find the first element containing $r$.
\end{remark}

\begin{figure}
    \centering
    \begin{tikzpicture}
\begin{groupplot} [
group style={group size = 1 by 3, vertical sep = 2mm}]
\nextgroupplot[axis equal image, axis lines=none, width=0.65\textwidth]
\addplot [blue, mark=*, mark size=1, mark options={solid, thin}, only marks]
coordinates {
(-1.00000000e+00,  0.00000000e+00)
(-6.25000000e-01,  0.00000000e+00)
(-2.50000000e-01,  0.00000000e+00)
( 1.25000000e-01,  0.00000000e+00)
(-1.00000000e+00,  2.50000000e-01)
(-6.25000000e-01,  2.50000000e-01)
(-2.50000000e-01,  2.50000000e-01)
(-1.00000000e+00,  5.00000000e-01)
(-6.25000000e-01,  5.00000000e-01)
(-1.00000000e+00,  7.50000000e-01)};\label{point:t1}

\addplot [magenta, mark=square*, mark size=1, mark options={solid, thin}, only marks]
coordinates {
( 1.25000000e-01,  2.50000000e-01)
(-2.50000000e-01,  5.00000000e-01)
(-6.25000000e-01,  7.50000000e-01)
(-1.00000000e+00,  1.00000000e+00)};\label{point:t12}

\addplot [gray, mark=star, mark size=2, mark options={solid}, only marks]
coordinates {
( 5.00000000e-01,  0.00000000e+00)};\label{point:t1234}

\addplot [green!60!black, mark=diamond*, mark size=1.5, mark options={solid}, only marks]
coordinates {
( 4.12500000e-01,  2.50000000e-01)
( 3.25000000e-01,  5.00000000e-01)
( 2.37500000e-01,  7.50000000e-01)
( 1.50000000e-01,  1.00000000e+00)};\label{point:t23}

\addplot [red, mark=triangle*, mark size=1.5, mark options={solid, thin}, only marks]
coordinates {
( 3.75000000e-02,  5.00000000e-01)
(-5.00000000e-02,  7.50000000e-01)
(-1.37500000e-01,  1.00000000e+00)
(-3.37500000e-01,  7.50000000e-01)
(-4.25000000e-01,  1.00000000e+00)
(-7.50000000e-01,  1.00000000e+00)};\label{point:t2}

\addplot [orange, mark=pentagon*, mark size=1.5, mark options={solid}, only marks]
coordinates {
( 1.50000000e+00,  1.00000000e+00)
( 7.50000000e-01,  2.50000000e-01)
( 1.00000000e+00,  5.00000000e-01)
( 1.25000000e+00,  7.50000000e-01)};\label{point:t34}

\addplot [cyan, mark=+, mark size=2, mark options={solid}, only marks]
coordinates {
( 6.62500000e-01,  5.00000000e-01)
( 9.12500000e-01,  7.50000000e-01)
( 1.16250000e+00,  1.00000000e+00)
( 5.75000000e-01,  7.50000000e-01)
( 8.25000000e-01,  1.00000000e+00)
( 4.87500000e-01,  1.00000000e+00)};\label{point:t3}

\addplot [red!50!black, mark=x, mark size=1.5, mark options={solid}, only marks]
coordinates {
( 7.50000000e-01,  0.00000000e+00)
( 1.00000000e+00,  0.00000000e+00)
( 1.25000000e+00,  0.00000000e+00)
( 1.50000000e+00,  0.00000000e+00)
( 1.00000000e+00,  2.50000000e-01)
( 1.25000000e+00,  2.50000000e-01)
( 1.50000000e+00,  2.50000000e-01)
( 1.25000000e+00,  5.00000000e-01)
( 1.50000000e+00,  5.00000000e-01)
( 1.50000000e+00,  7.50000000e-01)};\label{point:t4}

\addplot [black, mark=asterisk, mark size=1.5, mark options={sharp corners}, only marks]
coordinates {
( 4.20000000e-01,  1.50000000e-01)};\label{point:query}

\addplot [very thin, fill={rgb,255:red,204;green,255;blue,204}, forget plot]
coordinates {
(-1.00000000e+00,  1.00000000e+00)
(-1.00000000e+00,  0.00000000e+00)
( 5.00000000e-01,  0.00000000e+00)
( 1.50000000e+00,  0.00000000e+00)
( 1.50000000e+00,  1.00000000e+00)
( 1.50000000e-01,  1.00000000e+00)
(-1.00000000e+00,  1.00000000e+00)};

\addplot [black, very thin, forget plot]
coordinates {
( 5.00000000e-01,  0.00000000e+00)
(-1.00000000e+00,  1.00000000e+00)};

\addplot [black, very thin, forget plot]
coordinates {
( 5.00000000e-01,  0.00000000e+00)
( 1.50000000e-01,  1.00000000e+00)};

\addplot [black, very thin, forget plot]
coordinates {
( 5.00000000e-01,  0.00000000e+00)
( 1.50000000e+00,  1.00000000e+00)};

\node[]    at    (axis cs:-0.85, 0.2) {$K_1$};
\node[]    at    (axis cs:-0.3, 0.85) {$K_2$};
\node[]    at    (axis cs:0.6, 0.85) {$K_3$};
\node[]    at    (axis cs:1.35, 0.15) {$K_4$};
\nextgroupplot[axis equal image, axis lines=none, width=0.65\textwidth]
\addplot [magenta, mark=square*, mark size=1, mark options={solid, thin}, only marks, forget plot]
coordinates {
( 1.25000000e-01,  2.50000000e-01)};

\addplot [gray, mark=star, mark size=2, mark options={solid}, only marks, forget plot]
coordinates {
( 5.00000000e-01,  0.00000000e+00)};

\addplot [green!60!black, mark=diamond*, mark size=1.5, mark options={solid}, only marks, forget plot]
coordinates {
( 4.12500000e-01,  2.50000000e-01)};

\addplot [orange, mark=pentagon*, mark size=1.5, mark options={solid}, only marks, forget plot]
coordinates {
( 7.50000000e-01,  2.50000000e-01)};

\addplot [red!50!black, mark=x, mark size=1.5, mark options={solid}, only marks, forget plot]
coordinates {
( 7.50000000e-01,  0.00000000e+00)};

\addplot [black, mark=asterisk, mark size=1.5, mark options={sharp corners}, only marks, forget plot]
coordinates {
( 4.20000000e-01,  1.50000000e-01)};

\addplot [very thin, fill={rgb,255:red,204;green,255;blue,204}, forget plot]
coordinates {
(-1.00000000e+00,  1.00000000e+00)
(-1.00000000e+00,  0.00000000e+00)
( 5.00000000e-01,  0.00000000e+00)
( 1.50000000e+00,  0.00000000e+00)
( 1.50000000e+00,  1.00000000e+00)
( 1.50000000e-01,  1.00000000e+00)
(-1.00000000e+00,  1.00000000e+00)};

\addplot [black, very thin, forget plot]
coordinates {
( 5.00000000e-01,  0.00000000e+00)
(-1.00000000e+00,  1.00000000e+00)};

\addplot [black, very thin, forget plot]
coordinates {
( 5.00000000e-01,  0.00000000e+00)
( 1.50000000e-01,  1.00000000e+00)};

\addplot [black, very thin, forget plot]
coordinates {
( 5.00000000e-01,  0.00000000e+00)
( 1.50000000e+00,  1.00000000e+00)};

\addplot [red, densely dashdotted, very thin, forget plot]
coordinates {
( 4.20000000e-01,  1.50000000e-01)
( 1.25000000e-01,  2.50000000e-01)};

\addplot [red, densely dashdotted, very thin, forget plot]
coordinates {
( 4.20000000e-01,  1.50000000e-01)
( 4.12500000e-01,  2.50000000e-01)};

\addplot [red, densely dashdotted, very thin, forget plot]
coordinates {
( 4.20000000e-01,  1.50000000e-01)
( 5.00000000e-01,  0.00000000e+00)};

\addplot [red, densely dashdotted, very thin, forget plot]
coordinates {
( 4.20000000e-01,  1.50000000e-01)
( 7.50000000e-01,  2.50000000e-01)};

\addplot [red, densely dashdotted, very thin, forget plot]
coordinates {
( 4.20000000e-01,  1.50000000e-01)
( 7.50000000e-01,  0.00000000e+00)};

\node[]    at    (axis cs:-0.85, 0.2) {$K_1$};
\node[]    at    (axis cs:-0.3, 0.85) {$K_2$};
\node[]    at    (axis cs:0.6, 0.85) {$K_3$};
\node[]    at    (axis cs:1.35, 0.15) {$K_4$};
\nextgroupplot[axis equal image, axis lines=none, width=0.65\textwidth]
\addplot [black, mark=asterisk, mark size=1.5, mark options={sharp corners}, only marks, forget plot]
coordinates {
( 4.20000000e-01,  1.50000000e-01)};

\addplot [very thin, fill={rgb,255:red,204;green,255;blue,204}]
coordinates {
(-1.00000000e+00,  1.00000000e+00)
(-1.00000000e+00,  0.00000000e+00)
( 5.00000000e-01,  0.00000000e+00)
( 1.50000000e+00,  0.00000000e+00)
( 1.50000000e+00,  1.00000000e+00)
( 1.50000000e-01,  1.00000000e+00)
(-1.00000000e+00,  1.00000000e+00)};\label{point:in_polyg}

\addplot [black, very thin, forget plot]
coordinates {
( 5.00000000e-01,  0.00000000e+00)
(-1.00000000e+00,  1.00000000e+00)};

\addplot [black, very thin, fill=blue!50!white, opacity=0.6, forget plot]
coordinates {
( 5.00000000e-01,  0.00000000e+00)
(-1.00000000e+00,  1.00000000e+00)
( 1.50000000e-01,  1.00000000e+00)};

\addplot [black, very thin, forget plot]
coordinates {
( 5.00000000e-01,  0.00000000e+00)
( 1.50000000e-01,  1.00000000e+00)};

\addplot [black, very thin, forget plot]
coordinates {
( 5.00000000e-01,  0.00000000e+00)
( 1.50000000e+00,  1.00000000e+00)};

\node[]    at    (axis cs:-0.85, 0.2) {$K_1$};
\node[]    at    (axis cs:-0.3, 0.85) {$K_2$};
\node[]    at    (axis cs:0.6, 0.85) {$K_3$};
\node[]    at    (axis cs:1.35, 0.15) {$K_4$};
\end{groupplot}\end{tikzpicture}
	\caption{Illustration of search algorithm (Section~\ref{sec:mbp:pnteval})
	for element in which a point (\ref{point:query}) lies. \textit{Top}:
	The point distributions
	$\Dcal_{K_1}$ (\ref{point:t1},\ref{point:t12},\ref{point:t1234}),
	$\Dcal_{K_2}$ (\ref{point:t2},\ref{point:t12},\ref{point:t23},\ref{point:t1234}),
	$\Dcal_{K_3}$ (\ref{point:t3},\ref{point:t23},\ref{point:t34},\ref{point:t1234}),
	$\Dcal_{K_4}$ (\ref{point:t4},\ref{point:t34},\ref{point:t1234}).
	\textit{Middle}: The $k=5$ nearest points to the query point $r$
	(\ref{point:query}): $y_r^1$ (\ref{point:t23}),
	$y_r^2$ (\ref{point:t1234}),
	$y_r^3$ (\ref{point:t12}),
	$y_r^4$ (\ref{point:t34}),
	$y_r^5$ (\ref{point:t4}).
	The nearest points correspond to element sets
	$\Jcal_r^1=\{K_2,K_3\}$,
	$\Jcal_r^2=\{K_1,K_4\}$,
	$\Jcal_r^3=\emptyset$,
	$\Jcal_r^4=\emptyset$,
	$\Jcal_r^5=\emptyset$.
	\textit{Bottom}: The elements identified containing $r$ (\ref{point:query}):
	$\Kcal_r^1=\{K_2\}$,
	$\Kcal_r^2=\emptyset$,
	$\Kcal_r^3=\emptyset$,
	$\Kcal_r^4=\emptyset$,
	$\Kcal_r^5=\emptyset$.
	}
\label{fig:point:query}
\end{figure}

\section{Implicit shock tracking for shock-dominated flow over complex domains}
\label{sec:ist}
In this section we formulate the High-Order Implicit Shock Tracking method
(Section~\ref{sec:ist:form}) from a discontinuous Galerkin discretization
(Section~\ref{sec:ist:disc}) of a general system of conservation laws
posed on a fixed reference domain so domain mappings (nodal coordinates
after discretization) appear explicitly in the flux and source terms
(Section~\ref{sec:ist:govern}).

\subsection{Governing equations}
\label{sec:ist:govern}
Consider a general system of $m$ hyperbolic partial differential equations
posed in the domain $\Lambda \subset \Rbb^d$
\begin{equation} \label{eqn:claw}
  \nabla \cdot F(U) = S(U),
\end{equation}
where $x = (x_1, ..., x_{d}) \in \Lambda$ is the coordinate,
$U : \Lambda \rightarrow \Rbb^m$ is the conservative state
implicitly defined as the solution to (\ref{eqn:claw}),
$F : \Rbb^m \rightarrow \Rbb^{m \times d}$ with $F : W \mapsto F(W)$
is the physical flux function, $S : \Rbb^m \rightarrow \Rbb^m$ is the physical source term,
$(\nabla \cdot)$ is the divergence operator on the domain $\Lambda$ defined as
$(\nabla\cdot\psi)_i := \partial_{x_j} \psi_{ij}$ (summation implied on repeated
index), and $\partial \Lambda$ is the boundary of the spatial domain (with appropriate
boundary conditions prescribed). In general, the solution $U(x)$ may contain
discontinuities, in which case, the conservation law (\ref{eqn:claw}) holds
away from the discontinuities and the Rankine-Hugoniot conditions hold at
discontinuities.

Next, we write the physical domain $\Lambda$ as a transformation of a 
fixed reference domain $\Omega\subset\Rbb^d$ (the computational domain introduced
in Section~\ref{sec:mbp:msh}), i.e., $\Lambda = \Xi(\Omega)$,
where $\Xi : \Omega \rightarrow \Lambda$ with $\Xi : \bar{x} \mapsto \Xi(\bar{x})$
is any diffeomorphism from the reference domain to the physical domain. Then we recast the
conservation law (\ref{eqn:claw}) on the physical domain as a transformed conservation law
on the reference domain so the transformation appears explicitly in the flux and source
terms \cite{persson2009discontinuous,zahr2018shktrk}, i.e.,
\begin{equation} \label{eqn:trans1}
	\bar\nabla \cdot \bar{F}(\bar{U}; \partial_{\bar{x}} \Xi) = \bar{S}(\bar{U}; \det(\partial_{\bar{x}} \Xi)),
\end{equation}
where $\bar{U} : \Lambda \rightarrow \Rbb^m$ is the solution of the transformed
conservation law, $\bar{F} : \Rbb^m \times \Rbb^{d\times d} \rightarrow \Rbb^{m\times d}$
and $\bar{S} : \Rbb^m \times \Rbb \rightarrow \Rbb^m$ are the transformed flux function
and source term, respectively, and $\bar\nabla\cdot$ is the divergence operator in the
reference domain $\Omega$ defined as
$(\bar\nabla \cdot \psi)_i = \partial_{\bar{x}_j}\psi_{ij}$.
The reference domain quantities are defined in terms of the corresponding
physical domain quantities as \cite{zahr2018shktrk}
\begin{equation} \label{eqn:trans2}
	\bar{U}(\bar{x}) = U(\Xi(\bar{x})),
 	\qquad
	\bar{F} : (\bar{W}; \Theta) \mapsto (\det\Theta) \Fcal(\bar{W}) \Theta^{-T},
 	\qquad
	\bar{S} : (\bar{W}; q) \mapsto q S(\bar{W}).
\end{equation}

\subsection{Discontinuous Galerkin discretization}
\label{sec:ist:disc}
We discretize the transformed conservation law (\ref{eqn:trans1}) with a discontinuous
Galerkin method \cite{hesthaven_nodal_2008} with high-order piecewise polynomials spaces
used to approximate the state $\bar{U}$ and domain mapping $\Xi$.  To this end, let
$\hat\Omega_{h,q}$ represent a discretization of the reference domain $\Omega$ into
non-overlapping, potentially curved, computational elements (the computational
mesh introduced in Section~\ref{sec:mbp:msh}) and let $\Omega_{h,q}$ be the
surrogate domain (\ref{eqn:surdom}). To establish the finite-dimensional
DG formulation, we introduce the DG approximation (trial) space of discontinuous
piecewise polynomials associated with the mesh $\hat\Omega_{h,q}$
\begin{equation}
	\Vcal_{h,q}^p \coloneqq
 \left\{
	 v \in [L^2(\Omega_{h,q})]^m \suchthat
	 \left. v\right|_{\bar{K}} \in [\Pcal_p(\bar{K})]^m,~\forall \bar{K}\in\hat\Omega_{h,q}
 \right\},
\end{equation}
Furthermore, we define the space of globally continuous piecewise polynomials of
degree $q$ associated with the mesh $\hat\Omega_{h,q}$ as
\begin{equation}
	\Wcal_{h,q} \coloneqq
 \left\{
	 v \in C^0(\Omega_{h,q}) \suchthat \left. v\right|_{\bar{K}} \in \Pcal_q(\bar{K}),~\forall \bar{K}\in\hat\Omega_{h,q}
 \right\}
\end{equation}
and discretize the domain mapping with the corresponding vector-valued space
$\left[\Wcal_{h,q}\right]^d$. With these definitions, the DG variational problem
is: given $\Xi_h \in \left[\Wcal_{h,q}\right]^d$, find $\bar{U}_h\in\Vcal_{h,q}^p$
such that for all $\bar\psi_h \in \Vcal_{h,q}^{p'}$, we have
\begin{equation} \label{eqn:weak1}
	r_h^{p',p}(\bar\psi_h, \bar{U}_h; \partial_{\bar{x}}\Xi_h) = 0
\end{equation}
where $p' \geq p$ and the global residual function
$r_h^{p',p} : \Vcal_{h,q}^{p'}\times\Vcal_{h,q}^p\times[\Wcal_{h,q}]^d \rightarrow \Rbb$
is defined in terms of elemental residuals
$r_{\bar{K}}^{p',p} : \Vcal_{h,q}^{p'}\times\Vcal_{h,q}^p\times[\Wcal_{h,q}]^d \rightarrow \Rbb$
as
\begin{equation} \label{eqn:weak2}
r_h^{p', p} : (\bar\psi_h, \bar{W}_h; \Theta_h) \mapsto
	\sum_{\bar{K}\in\hat\Omega_{h,q}} r_{\bar{K}}^{p', p} (\bar\psi_h, \bar{W}_h; \Theta_h).
\end{equation}
The elemental residuals come directly from a standard DG formulation applied to
the transformed conservation law in (\ref{eqn:trans1})
\begin{equation} \label{eqn:weak3}
\begin{aligned}
 r_{\bar{K}}^{p',p} : (\bar\psi_h, \bar{W}_h; \Theta_h) \mapsto
	&\int_{\partial \bar{K}}\bar\psi_h \cdot \bar{\Hcal} ( W^+_h,  W^-_h, \bar{n}_h; \Theta_h  ) \, dS  \\
	- &\int_{\bar{K}} \bar{F} (W_h; \Theta_h) : \bar{\nabla} \bar\psi_h \, dV \\
	- &\int_{\bar{K}} \bar\psi_h \cdot \bar{S} (W_h; \det \Theta_h)) \, dV,
\end{aligned}
\end{equation}
where $\bar\eta_h : \partial\bar{K} \mapsto \Rbb^d$ is the outward unit normal to
an element $\bar{K} \in \hat\Omega_{h,q}$, $\bar{W}_h^+$ ($\bar{W}_h^-$) denotes the interior
(exterior) trace of $\bar{W}_h \in \Vcal_{h,q}^p$ to the element, and $\bar{\Hcal}$ is the transformed numerical (Roe) flux function \cite{roe1981approximate,zahr_implicit_2020}.

Finally, we introduce a basis for the test space ($\Vcal_{h,q}^{p'}$), trial space
($\Vcal_{h,q}^p$), and domain mapping space ($[\Wcal_{h,q}]^d$) to reduce the weak
formulation in (\ref{eqn:weak1})-(\ref{eqn:weak3}) to a system of nonlinear
algebraic equations. In the case where $p = p'$, we have
\begin{equation}
 \rbm : \Rbb^{N_u} \times \Rbb^{d N_x} \rightarrow \Rbb^{N_u}, \qquad
 \rbm : (\ubm, \xbm) \mapsto \rbm(\ubm,\xbm),
\end{equation}
where $N_u = \mathrm{dim}(\Vcal_{h,q}^p)$ and $N_x = \mathrm{dim}(\Wcal_{h,q})$,
which is the residual of a standard DG method. The algebraic state $\ubm\in\Rbb^{N_u}$
contains the coefficients of the flow solution in the chosen basis and
$\xbm\in\Rbb^{d N_x}$ contains the nodal coordinates of the physical mesh
because a Lagrangian basis is used for the continuous $\Wcal_{h,q}$ space.
Next, we define the algebraic enriched residual associated with a test space of
degree $p' > p$ ($p' = p+1$ in this work) as
\begin{equation}
 \Rbm : \Rbb^{N_u} \times \Rbb^{dN_x} \rightarrow \Rbb^{N_u'}, \qquad
 \Rbm : (\ubm, \xbm) \mapsto \Rbm(\ubm,\xbm),
\end{equation}
where $N_u' = \mathrm{dim}(\Vcal_{h,q}^{p'})$, which will be used to construct
the implicit shock tracking objective function.

\subsection{Implicit shock tracking formulation}
\label{sec:ist:form}
The HOIST method \cite{zahr_implicit_2020} simultaneously computes the discrete solution
of the conservation law and the nodal coordinates of the mesh that causes element
faces to align with discontinuities. This is achieved through a fully discrete, full
space PDE-constrained optimization formulation with the optimization variables taken
to be the discrete flow solution ($\ubm$) and nodal coordinates of the mesh ($\xbm$).
The nodal coordinates are constrained to ensure all nodes remain on their original
boundaries using the mesh-based parametrization of Section~\ref{sec:mbp:msh}. This is
accomplished by requiring the nodal coordinates be in the image of the mesh-based
parametrization $\phibold$, i.e., there exists $\ybm\in\Ycal_{h',q'}$ such that
$\xbm = \phibold(\ybm)$. This leads to the following formulation of the HOIST method
\begin{equation} \label{eqn:pde-opt}
    (\ubm^\star,\ybm^\star) \coloneqq
    \argmin_{\ubm\in\Rbb^{N_u},\ybm\in\Rbb^{N_y}}
	f(\ubm,\ybm) \quad \text{subject to:} \quad \cbm(\ubm,\ybm) = \zerobold,
\end{equation}
where $\func{f}{\Rbb^{N_u}\times\Rbb^{N_y}}{\Rbb}$ is the objective
function, $\func{\cbm}{\Rbb^{N_u}\times\Rbb^{N_y}}{\Rbb^{N_u}}$ is
the constraint function, and the nodal coordinates of the aligned mesh are
$\xbm^\star = \phibold(\ybm^\star)$. The objective function
is composed of two terms as
\begin{equation}\label{eqn:obj0}
 f : (\ubm,\ybm) \mapsto f_\text{err}(\ubm,\ybm) + \kappa^2 f_\text{msh}(\ybm),
\end{equation}
which balances alignment of the mesh with non-smooth features and the quality
of the elements and $\kappa\in\Rbb_{\geq 0}$ is the mesh penalty parameter.
The mesh alignment term,
$\func{f_\text{err}}{\Rbb^{N_u}\times\Rbb^{N_y}}{\Rbb}$, is taken
to be the norm of the enriched DG residual
\begin{equation}\label{eqn:obj1}
 f_\text{err} : (\ubm,\ybm) \mapsto \frac{1}{2}\norm{\Rbm(\ubm,\phibold(\ybm))}_2^2.
\end{equation}
To ensure elements of the
discontinuity-aligned mesh are high-quality, we define the mesh distortion
term, $\func{f_\text{msh}}{\Rbb^{N_y}}{\Rbb}$, as
\begin{equation}\label{eqn:obj2}
 f_\text{msh} : \ybm \mapsto \frac{1}{2}\norm{\Rbm_\text{msh}(\phibold(\ybm))}_2^2,
\end{equation}
where $\func{\Rbm_\text{msh}}{\Rbb^{N_y}}{\Rbb^{|\hat\Omega_{h,q}|}}$ is the
elementwise mesh distortion with respect to an ideal element
\cite{zahr_implicit_2020,knupp_algebraic_2001,roca_defining_2012}.
Finally, the constraint is the standard DG residual to ensure the HOIST
method inherits the desirable properties of DG, i.e.,
\begin{equation}
	\cbm : (\ubm,\ybm) \mapsto \rbm(\ubm,\phibold(\ybm)).
\end{equation}

The constrained optimization problem in (\ref{eqn:pde-opt}) is
solved using the sequential quadratic programming (SQP) method
in \cite{zahr_implicit_2020,2022_huang_shktrk}.
This gradient-based optimizer requires the derivatives
of the objective and constraint with respect to both $\ubm$ and $\ybm$.
The $\ubm$ derivatives are straightforward and described
in \cite{zahr_implicit_2020,2022_huang_shktrk},
whereas the $\ybm$ derivatives involve the Jacobian of the mesh-based
parametrization $\phibold$. The $\ybm$-gradient of the objective function,
$\nabla_\ybm f: \Rbb^{N_u}\times\Rbb^{N_y} \mapsto \Rbb^{N_y}$,
takes the form
\begin{equation}
	\nabla_\ybm f : (\ubm,\ybm) \mapsto
	\pder{\phibold}{\ybm}(\ybm)^T\pder{\Rbm}{\xbm}(\ybm,\phibold(\ybm))^T
	\Rbm(\ubm,\phibold(\ybm))
\end{equation}
and the $\ybm$-Jacobian of the constraint function, $\pder{\cbm}{\ybm}: \Rbb^{N_u}\times\Rbb^{N_y} \mapsto \Rbb^{N_u\times N_y}$, is
\begin{equation}
	\pder{\cbm}{\ybm}(\ubm,\ybm) \mapsto
	\pder{\rbm}{\xbm}(\ubm,\phibold(\ybm))\pder{\phibold}{\ybm}(\ybm),
\end{equation}
where $\pder{\phibold}{\ybm}$ is the Jacobian of the mesh-based parametrization defined
in (\ref{eqn:mbp_msh_deriv}).

\begin{remark}
	The condition $\ybm\in\Ycal_{h',q'}$ was relaxed to $\ybm\in\Rbb^{N_y}$
	in (\ref{eqn:pde-opt}) because any $\ybm\not\in\Ycal_{h',q'}$ would
	lead to a mesh $\xbm=\phibold(\ybm)$ with inverted elements, which would
	cause $f_\mathrm{msh}(\xbm)$ to blow up and be rejected by the optimizer.
\end{remark}

\section{Numerical experiments}
\label{sec:numexp}
In this section we provide two concrete examples of the mesh-based parametrization
that show the surrogate objects being parametrized rapidly approach the true
object (Section~\ref{sec:numexp:surf}). We also provide two examples that
demonstrate mesh-based parametrizations integrate cleanly into implicit
shock tracking, effectively lead to shock-aligned grids even when the
shocks intersect the boundary, and do not impede convergence of the
optimization solver (Section~\ref{sec:numexp:advec}-\ref{sec:numexp:euler}).
Finally, we show mesh-based parametrizations can be used to easily slide
nodes along complex geometries and surface intersections using a relevant
hypersonic vehicle, the sliced cone flap (Section~\ref{sec:numexp:cone}).

\subsection{Surface approximation quality of mesh-based parametrizations}
\label{sec:numexp:surf}
In this section we demonstrate the accuracy of mesh-based parametrization and
its derivative with respect to known geometries: a two-dimensional Gaussian
bump (Section~\ref{sec:numexp:surf:gaussian}) and quarter sphere
(Section~\ref{sec:numexp:surf:sphere}). Despite being a parametrization
of a globally $C^0$ surrogate surface, the mesh-based parametrization
accurately represent the underlying surfaces and its normals, especially
when the geometry mesh is refined (small $h'$ or large $q'$).

Let $\Scal\subset\Rbb^d$ be the true surface of interest and
$\Mcal : \Rcal \rightarrow \Scal$ be a parametrization of the surface
with parameter domain $\Rcal\subset\Rbb^{d'}$. We quantify the error
in the mesh-based parametrization $\Mcal_{h',q'}$ and its derivative
$M_{h',q'}$ using the normalized error metrics,
$E : \Rcal_{h',q'} \rightarrow \Rbb_{\ge 0}$ and 
$E_\partial : \Rcal_{h',q'} \rightarrow \Rbb_{\ge 0}$,
defined as
\begin{equation}
	E : r \mapsto \frac{\norm{\Mcal_{h',q'}(r) - \Mcal(r)}}{\sup_{r'\in\Rcal} \norm{\Mcal(r')}}, \qquad
	E_\partial : r \mapsto \frac{\norm{M_{h',q'}(r) - \pder{\Mcal(r)}{r}}}{\sup_{r'\in\Rcal} \norm{\pder{\Mcal}{r}(r')}}.
\end{equation}

\subsubsection{Two-dimensional Gaussian bump}
\label{sec:numexp:surf:gaussian}
\begin{figure}
  \centering
  \begin{tikzpicture}
\begin{axis}[
axis y line*=left,
axis x line*=bottom,
ytick={-0.1, 0.0625},
width=0.7\textwidth,
height=0.2\textwidth,
xlabel=$r_1$]
\addplot [domain=-1.5:1.5, samples=201, black]
{
(1/16)*exp(-(25)*x^2)};\label{gauss:exact}

\addplot [mark=*, mark size=1, mark options={solid, thin}, blue, densely dashed]
coordinates {
(-1.50000000e+00,  2.32710195e-26)
(-1.25000000e+00,  6.78034538e-19)
(-1.00000000e+00,  8.67996492e-13)
(-7.50000000e-01,  4.88218088e-08)
(-5.00000000e-01,  1.20653384e-04)
(-2.50000000e-01,  1.31007117e-02)
( 0.00000000e+00,  6.25000000e-02)
( 2.50000000e-01,  1.31007117e-02)
( 5.00000000e-01,  1.20653384e-04)
( 7.50000000e-01,  4.88218088e-08)
( 1.00000000e+00,  8.67996492e-13)
( 1.25000000e+00,  6.78034519e-19)
( 1.50000000e+00,  0.00000000e+00)};\label{gauss:n12_p1}

\end{axis}
\end{tikzpicture}
  \caption{The true surface $\Scal$ (\ref{gauss:exact}) for the Gaussian bump
	and a surrogate surface $\Scal_{h',q'}$ with $h'=0.25$, $q'=1$ (\ref{gauss:n12_p1}).}
  \label{fig:gauss-geom}
\end{figure}
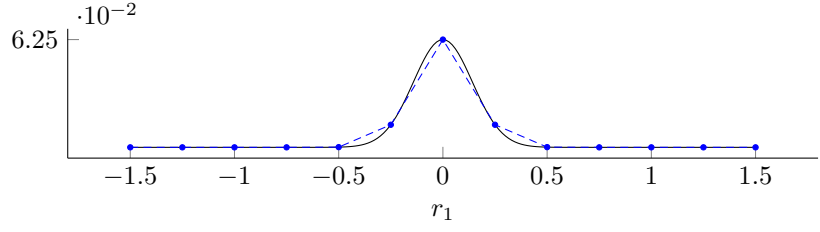

Consider the following surface (curve) $\Scal$ (Figure~\ref{fig:gauss-geom})
embedded in a two-dimensional space $d = 2$ ($d' = d-1 = 1$)
\begin{equation}
	\Scal = \left\{\Mcal(r)\in\Rbb^2 \suchthat r\in\Rcal\right\},
\end{equation}
where $\Rcal \coloneqq (-1.5, 1.5)$ (parameter domain), $\Mcal : r \mapsto (r, f(r))$,
and $f : \Rcal \rightarrow (0, a)$, $f : r \mapsto a\exp{(-br^2)}$ defines
the curve, and $a=1/16$ and $b=25$. Because $f$ is Gaussian, the surface $\Scal$
will not be exactly represented by a piecewise polynomial space so the mesh-based
parametrization will incur a discretization error. To study the behavior of the
mesh-based parametrization under refinement, we define a mesh-based parametrization
$\Mcal_{h',q'}$ (Figure~\ref{fig:gauss-geom}) and its derivative $M_{h',q'}$ for
three grids,  $h' = 3 / N$ for $N = 12, 120, 1200$ and three polynomial degrees
$q = 1, 2, 6$. For this problem, we take $\Pi_{h',q'} : x \mapsto x_1$,
which leads to the parameter domain $\Rcal_{h',q'} = \Rcal$.

\begin{figure}
    \centering
    \input{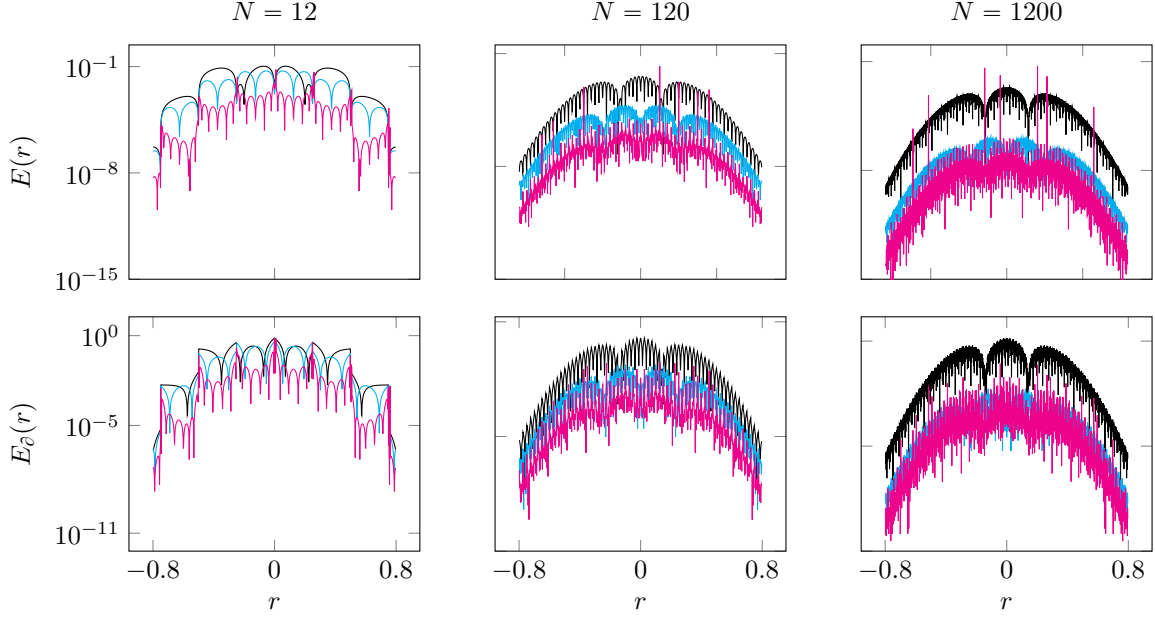}
	\caption{
		The error in the mesh-based parametrization (\textit{top}) and
		its derivative (\textit{bottom}) of the Gaussian bump for three
		$h$-refinement levels (\textit{columns}) for polynomial degrees
		$q' = 1$ (\ref{p1:gauss_bump}), $q' = 2$ (\ref{p2:gauss_bump}),
		and $q' = 6$ (\ref{p6:gauss_bump}). The $r$ limits are restricted
		to $(-0.8,0.8)$ because the errors in the near-constant region are
		negligible.
	}
    \label{fig:gauss_param_diff_demo}
\end{figure}

The mesh-based parametrization show clear convergence as $h'$ decreases or $q'$ increases
and, as expected,  the error decreases more rapidly with $h'$ for the larger
polynomial degrees (Figure~\ref{fig:gauss_param_diff_demo}).
At the finest resolution, $N = 1200$ and $q' = 6$, the error in the mesh-based
parametrization and its derivative are $\Ocal(10^{-8})$ and $\Ocal(10^{-5})$,
respectively, which shows the mesh-based parametrization can be made effectively
indistinguishable from a parametrization of the true geometry. We will show in
Section~\ref{sec:numexp:advec} that implicit shock tracking
is still effective when much coarser geometry grids are used.

\begin{remark}
	The mapping $\Pi_{h',q'} : x \mapsto x_1$ used for this problem is
	obtained from the approach in~\ref{sec:surfproj:surf} with $\hat{x} = 0$
	and $n = (0, 1)$.
\end{remark}

\subsubsection{Three-dimensional quarter sphere}
\label{sec:numexp:surf:sphere}
\begin{figure}
    \centering
    \begin{tikzpicture}
\begin{groupplot} [
group style={group size = 2 by 1, horizontal sep = 1cm, vertical sep = 1cm}]
\nextgroupplot[axis equal image, view={135}{30}, xlabel=$x_2$, ylabel=$x_1$, zlabel=$x_3$, width=0.57\textwidth]
\addplot3 [surf, shader=flat, domain=0:1.5708, y domain=0:3.1416, samples=25, samples y=25, color=gray!50, opacity=0.7, forget plot]
(
{cos(\x r)*cos(\y r)}, {cos(\x r)*sin(\y r)}, {sin(\x r)});

\nextgroupplot[axis equal image, axis y line=none, axis x line=none, width=0.42\textwidth]
\addplot []
graphics [xmin=0.0,xmax=1.1155,ymin=0,ymax=0.917] { 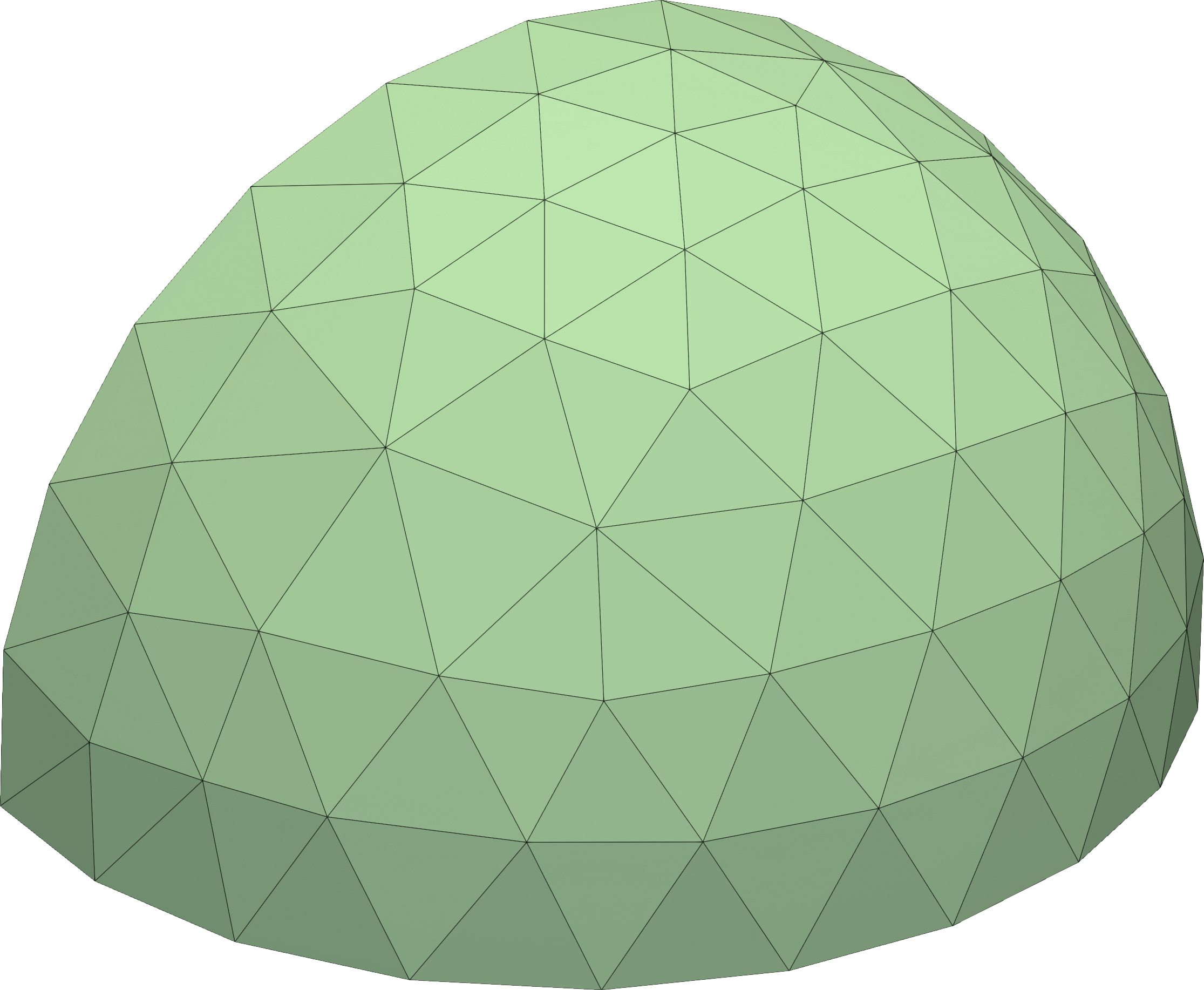};

\end{groupplot}\end{tikzpicture}
	\caption{The true surface $\Scal$ (\textit{left}) of the quarter sphere
	and a surrogate surface $\Scal_{h',q'}$ for $h'\approx 0.15$, $q'=1$
	(\textit{right}).}
        \label{fig:gauss-sphere}
\end{figure}

\begin{figure}
        \centering
        \begin{tikzpicture}
\begin{groupplot} [
group style={group size = 4 by 1, horizontal sep = 1cm, vertical sep = 1.5cm}]
\nextgroupplot[axis equal image, title={(a)}, xlabel={$r_1$}, ytick={-1,0,1}, xtick={0,0.5,1}, ylabel={$r_2$}, ymin=-1, ymax=1, xmin=0, xmax=1, width=0.45\textwidth]
\addplot []
graphics [xmin=0.0,xmax=1.0,ymin=-1.0,ymax=1.0] { 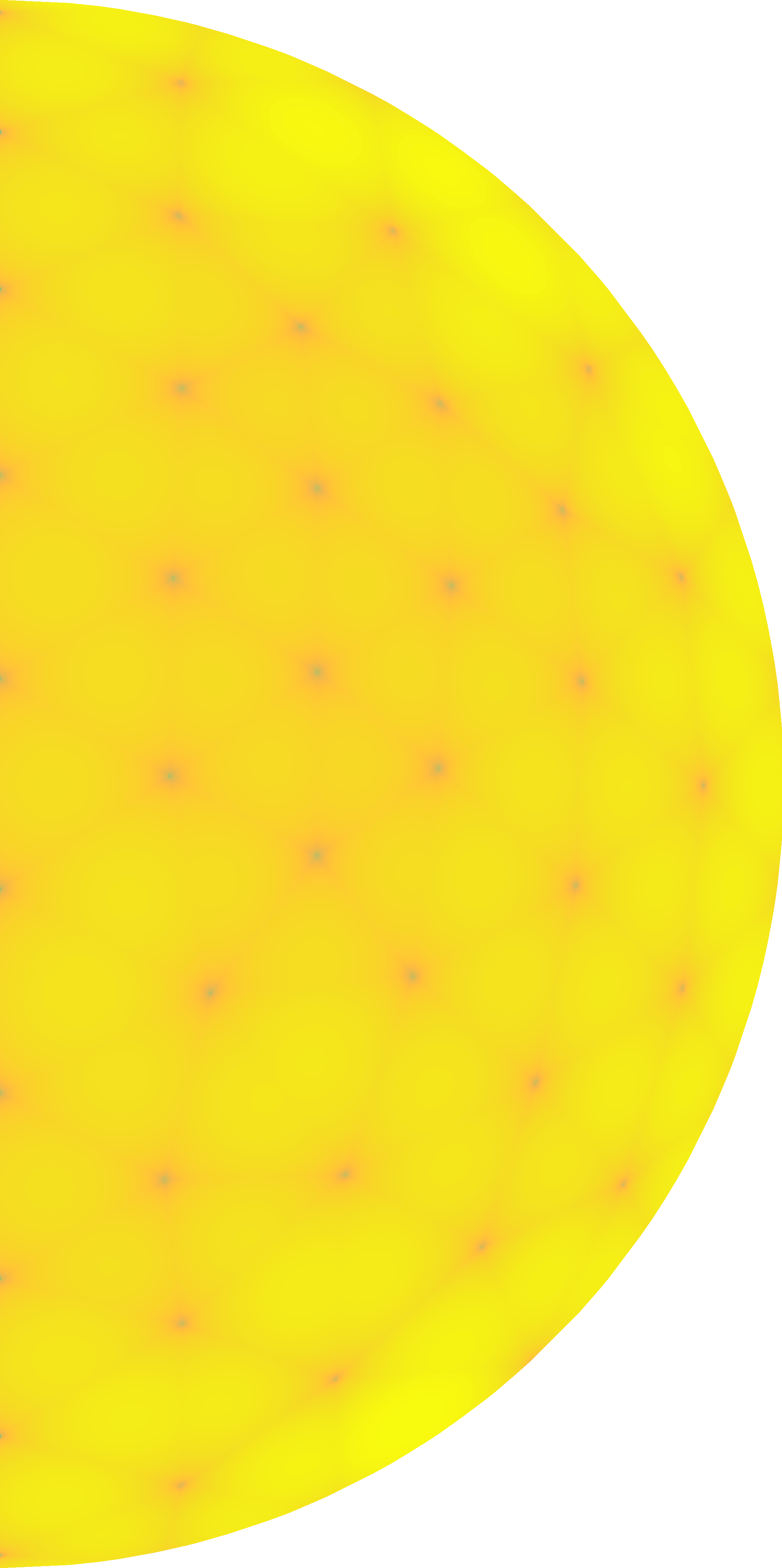};

\nextgroupplot[axis equal image, title={(b)}, yticklabels={,,}, xtick={0,0.5,1}, xlabel={$r_1$}, ymin=-1, ymax=1, xmin=0, xmax=1, width=0.45\textwidth]
\addplot []
graphics [xmin=0.0,xmax=1.0,ymin=-1.0,ymax=1.0] { 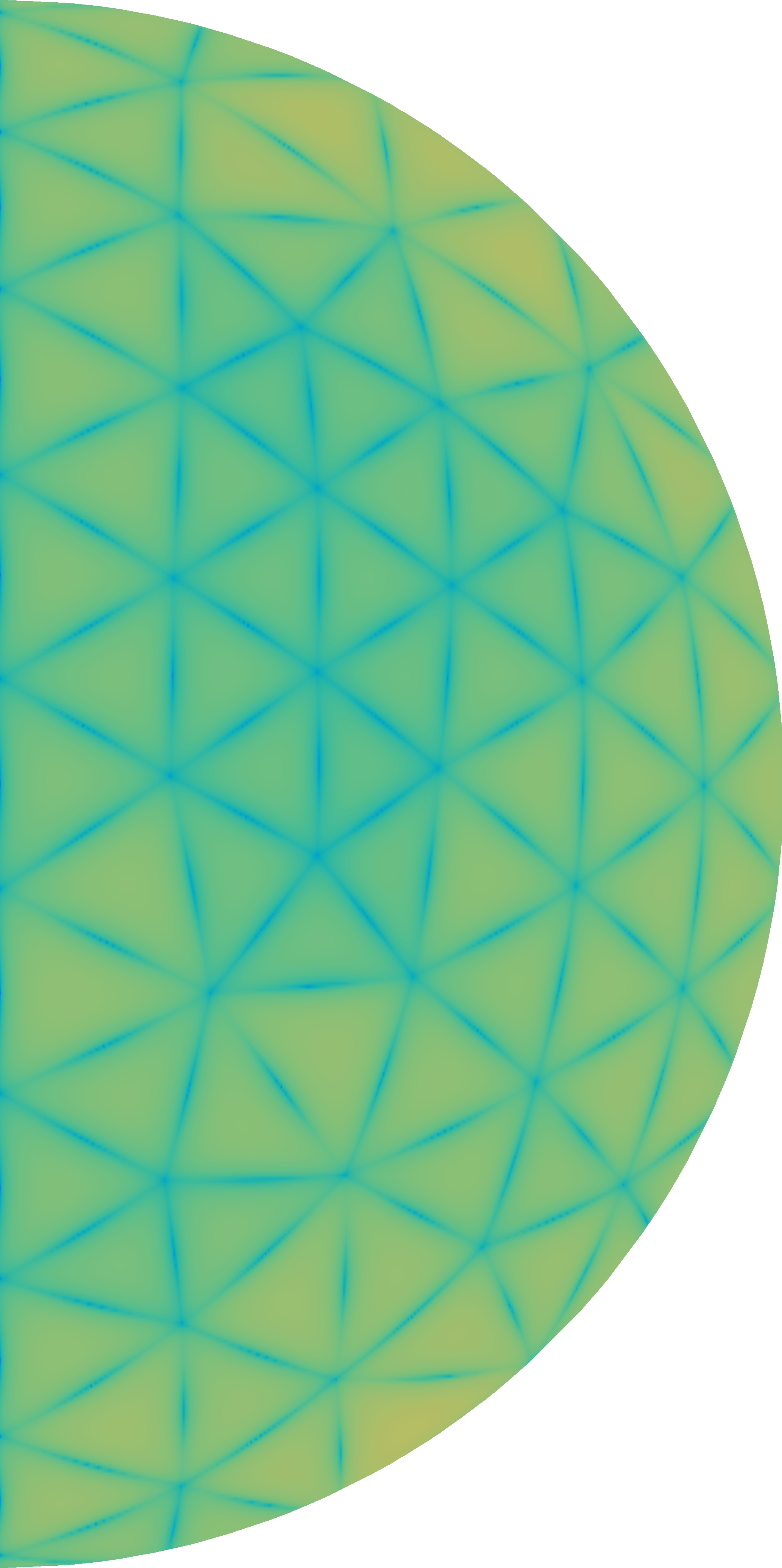};

\nextgroupplot[axis equal image, title={(c)}, yticklabels={,,}, xtick={0,0.5,1}, xlabel={$r_1$}, ymin=-1, ymax=1, xmin=0, xmax=1, width=0.45\textwidth]
\addplot []
graphics [xmin=0.0,xmax=1.0,ymin=-1.0,ymax=1.0] { 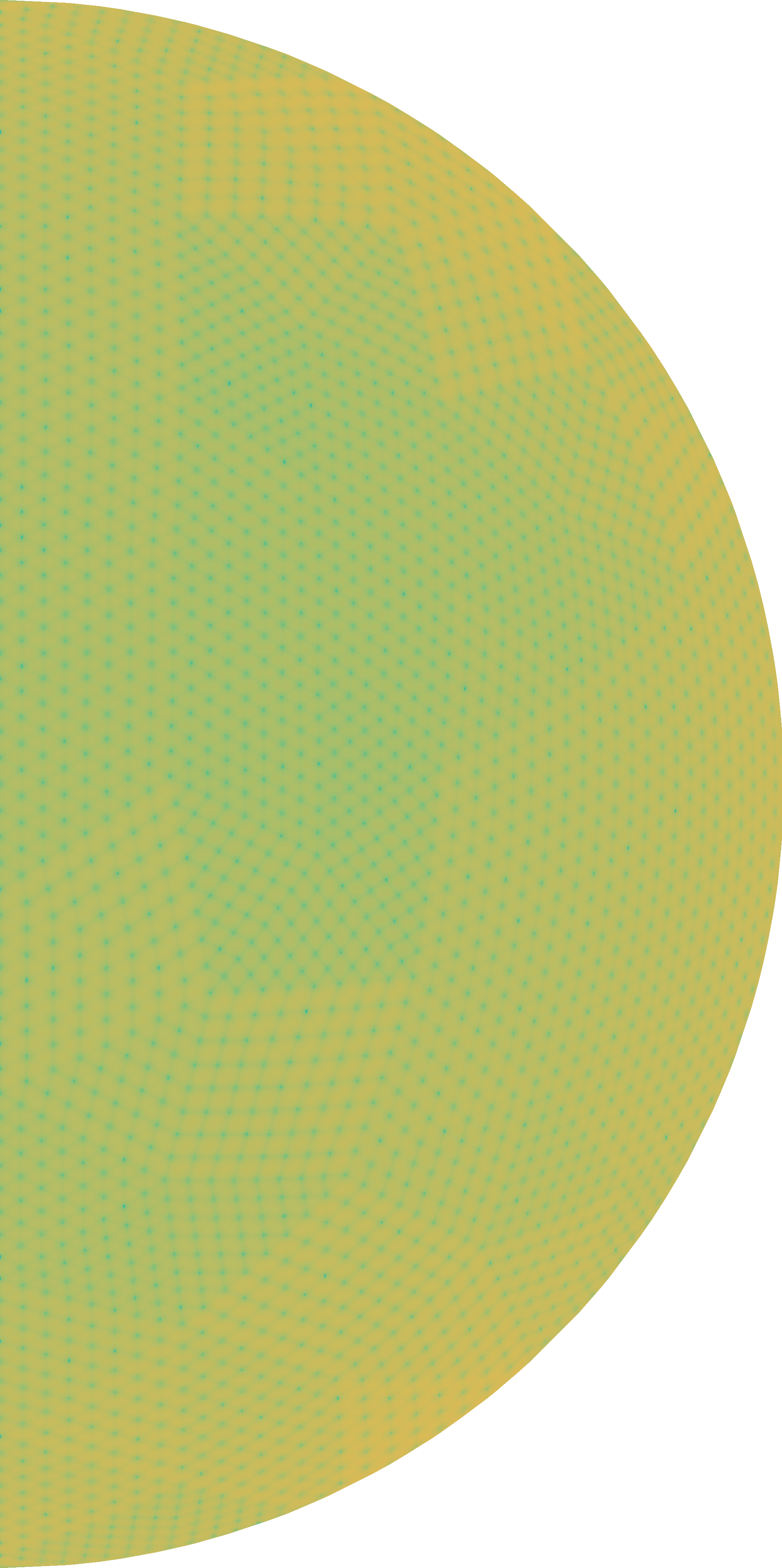};

\nextgroupplot[axis equal image, title={(d)}, yticklabels={,,}, xtick={0,0.5,1}, xlabel={$r_1$}, ymin=-1, ymax=1, xmin=0, xmax=1, width=0.45\textwidth]
\addplot []
graphics [xmin=0.0,xmax=1.0,ymin=-1.0,ymax=1.0] { 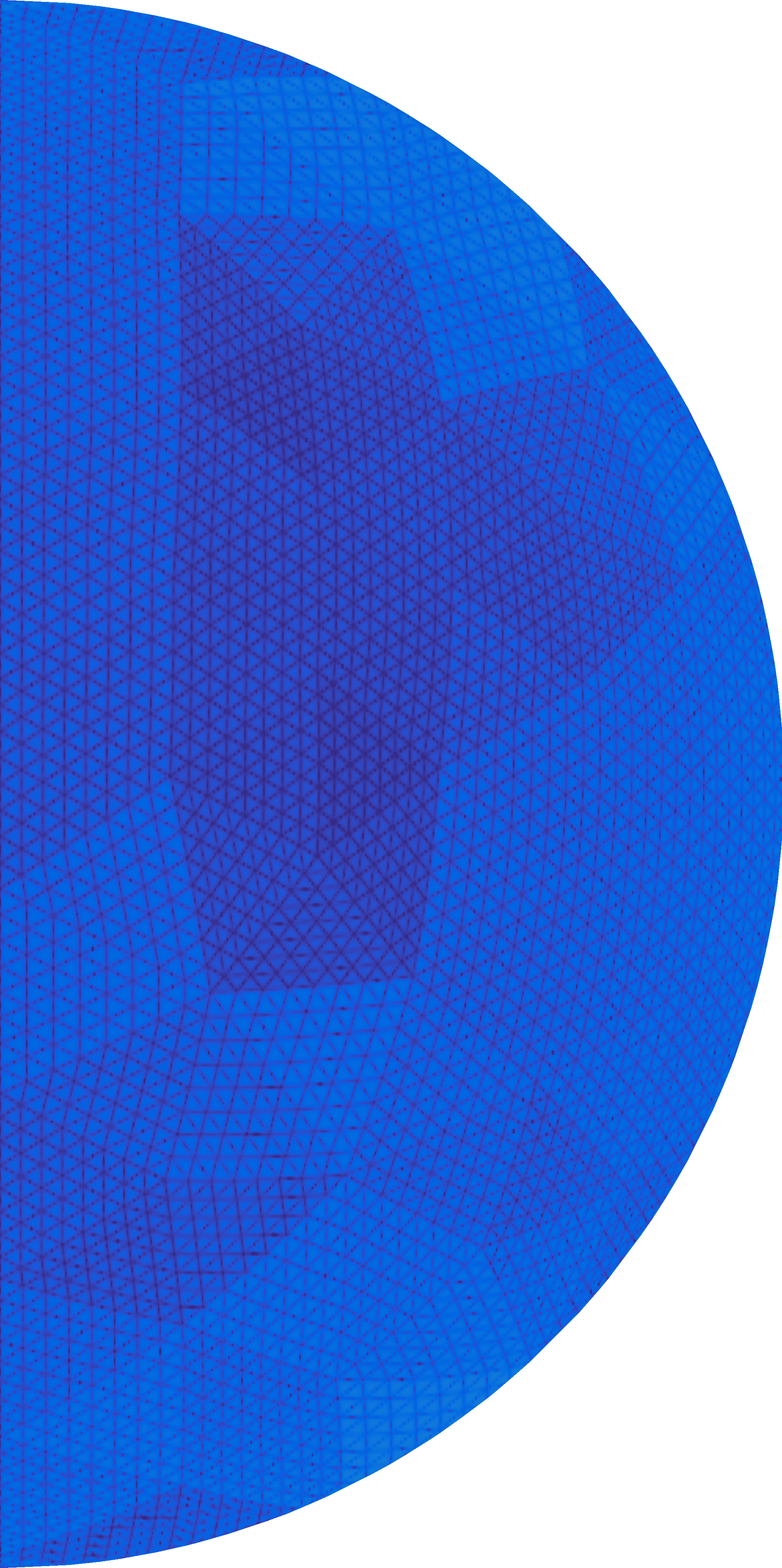};

\end{groupplot}\end{tikzpicture}
        \colorbarMatlabParula{-9}{-7.1385325}{-5.277065}{-3.4155975}{-1.55413}
	\caption{The logarithm (base 10) of the error in the mesh-based
	         parametrization ($\log_{10} E(r)$) for two refinement levels
		 ($N = 134, 8576$) and polynomial degrees ($q' = 1, 2$).
	         Legend: (a) $N=134$, $q'=1$, (b) $N=134$, $q'=2$, (c) $N=8576$, $q'=1$,
	         (d) $N=8576$, $q'=2$.
        }
        \label{fig:sphere_diff_demo}
    \end{figure}

\begin{figure}
        \centering
        \begin{tikzpicture}
\begin{groupplot} [
group style={group size = 4 by 1, horizontal sep = 1cm, vertical sep = 1.5cm}]
\nextgroupplot[axis equal image, title={(a)}, xtick={0,0.5,1}, xlabel={$r_1$}, ytick={-1,0,1}, ylabel={$r_2$}, ymin=-1, ymax=1, xmin=0, xmax=1, width=0.45\textwidth]
\addplot []
graphics [xmin=0.0,xmax=1.0,ymin=-1.0,ymax=1.0] { 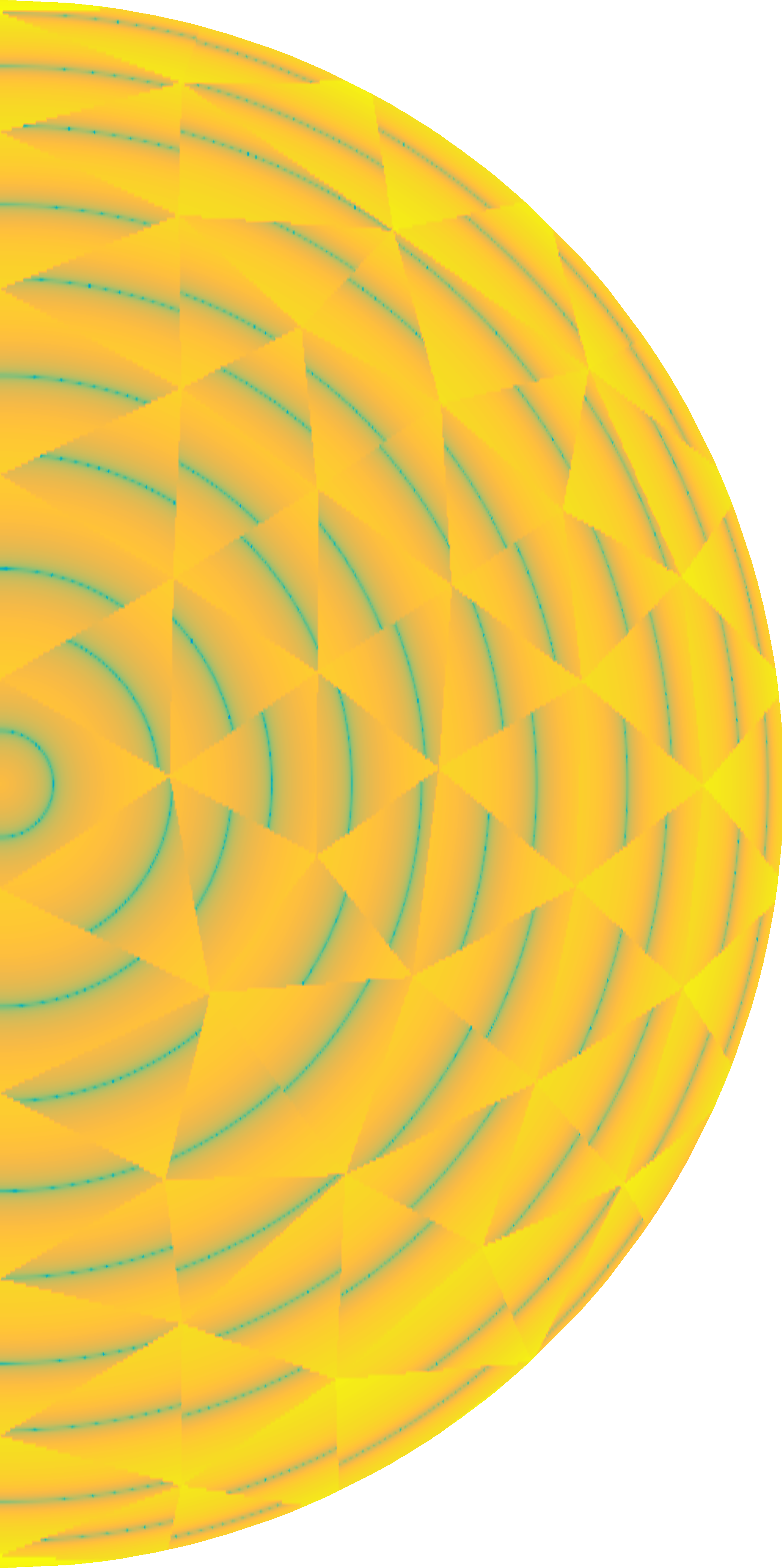};

\nextgroupplot[axis equal image, title={(b)}, yticklabels={,,}, xtick={0,0.5,1}, xlabel={$r_1$}, ymin=-1, ymax=1, xmin=0, xmax=1, width=0.45\textwidth]
\addplot []
graphics [xmin=0.0,xmax=1.0,ymin=-1.0,ymax=1.0] { 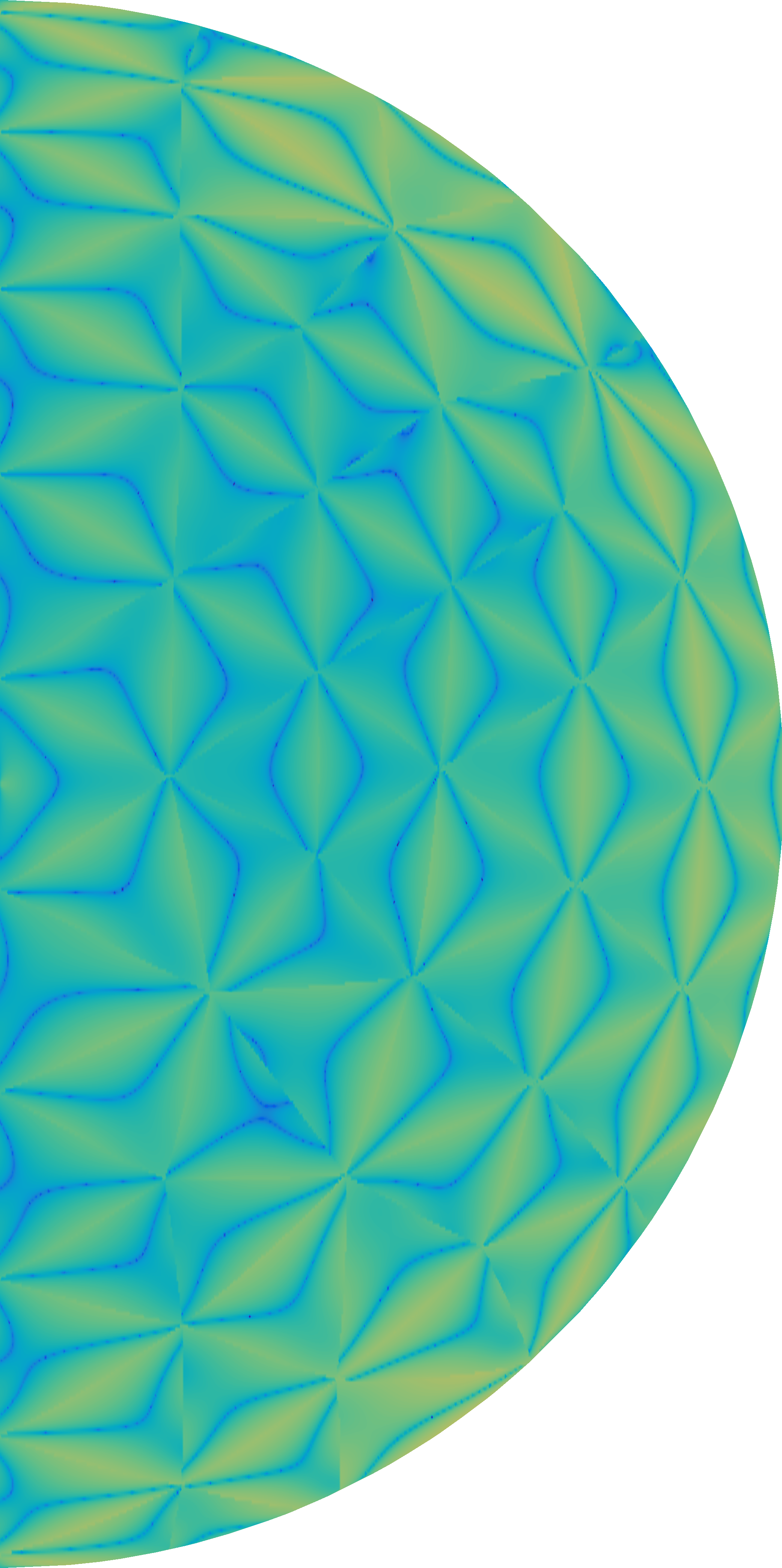};

\nextgroupplot[axis equal image, title={(c)}, yticklabels={,,}, xtick={0,0.5,1}, xlabel={$r_1$}, ymin=-1, ymax=1, xmin=0, xmax=1, width=0.45\textwidth]
\addplot []
graphics [xmin=0.0,xmax=1.0,ymin=-1.0,ymax=1.0] { 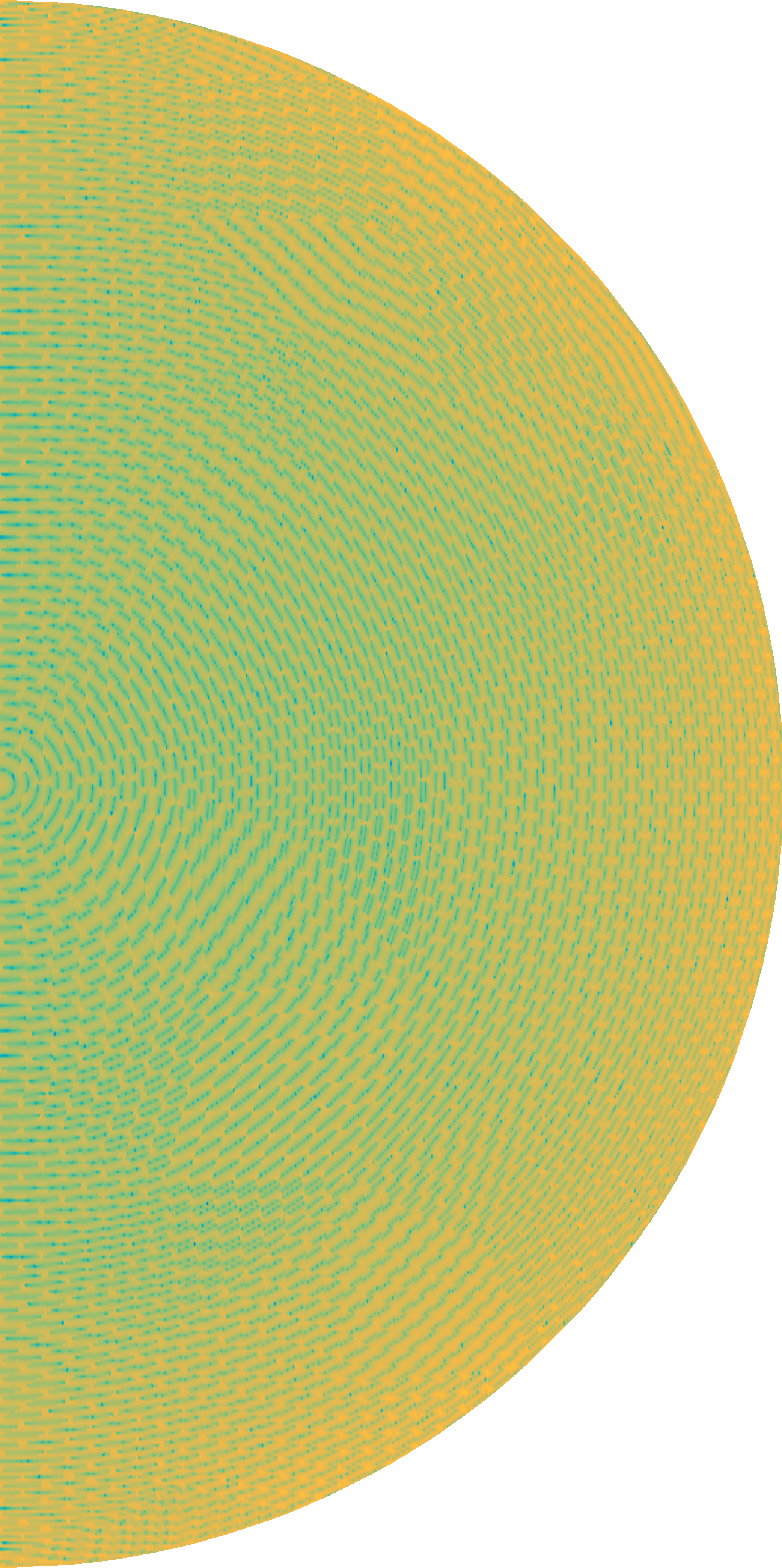};

\nextgroupplot[axis equal image, title={(d)}, yticklabels={,,}, xtick={0,0.5,1}, xlabel={$r_1$}, ymin=-1, ymax=1, xmin=0, xmax=1, width=0.45\textwidth]
\addplot []
graphics [xmin=0.0,xmax=1.0,ymin=-1.0,ymax=1.0] { 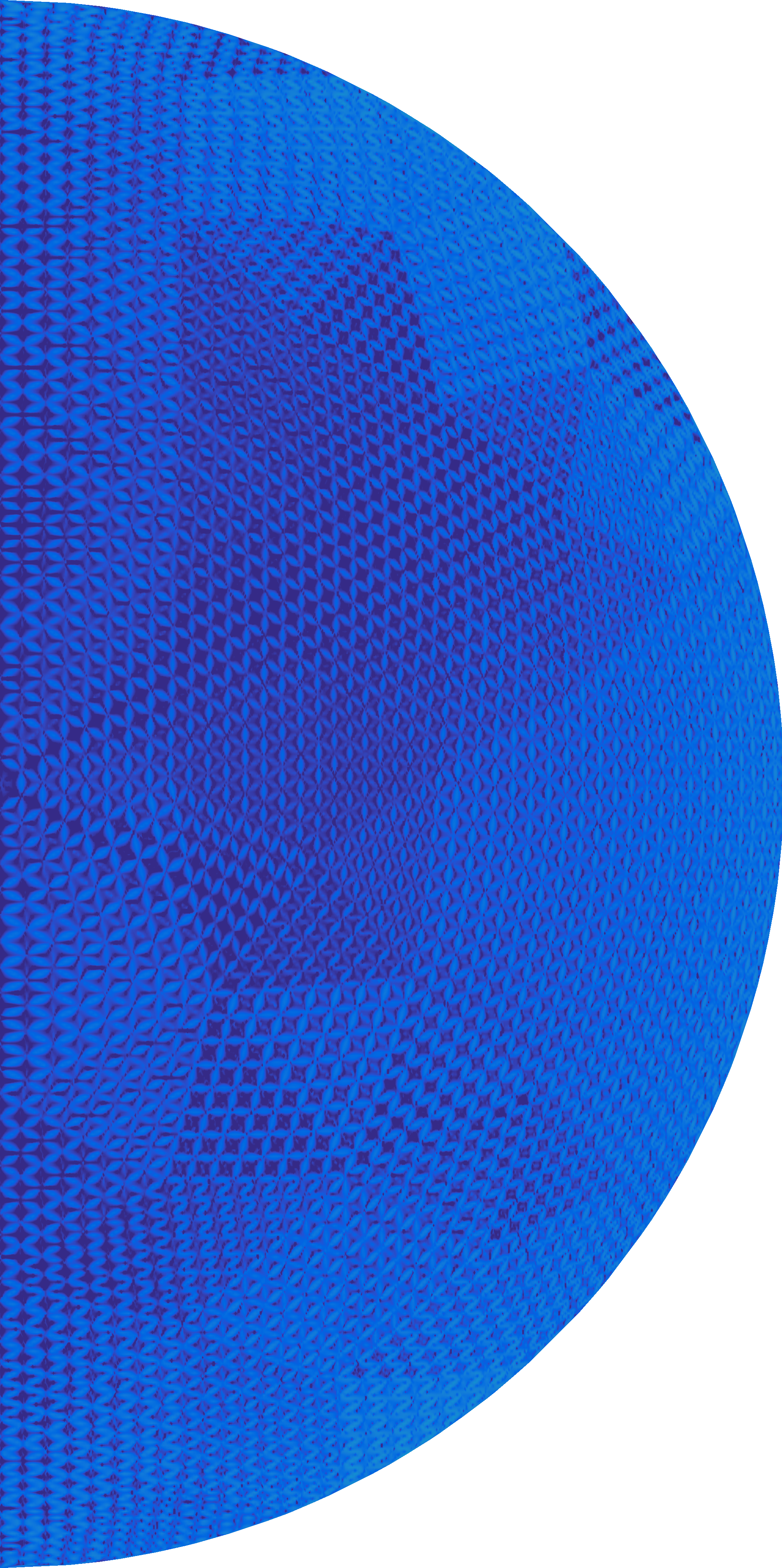};

\end{groupplot}\end{tikzpicture}
        \colorbarMatlabParula{-7}{-5.35741263}{-3.71482527}{-2.0722379}{-0.42965054}
	\caption{The logarithm (base 10) of the error in the mesh-based
         parametrization derivative ($\log_{10} E_\partial(r)$) for two
 	 refinement levels ($N = 134, 8576$) and polynomial degrees ($q' = 1, 2$).
	 Legend: (a) $N=134$, $q'=1$, (b) $N=134$, $q'=2$, (c) $N=8576$, $q'=1$,
	         (d) $N=8576$, $q'=2$.
	}
        \label{fig:sphere_diff_prime_demo}
    \end{figure}

Consider the following surface $\Scal$ (Figure~\ref{fig:gauss-sphere})
embedded in a three-dimensional space $d = 3$ ($d' = d-1 = 2$)
\begin{equation}
	\Scal = \left\{\Mcal(r)\in\Rbb^2 \suchthat r\in\Rcal\right\},
\end{equation}
where $\Rcal = \{r\in\Rbb^2 \mid r_1^2+r_2^2\leq 1,~r_1\geq 0\}$ (parameter domain),
$\Mcal : r \mapsto (r_1, r_2, f(r_3))$, and
$f : \Rcal \rightarrow (0, 1)$, $f : r \mapsto \sqrt{1 - r_1^2 - r_2^2}$ defines
the surface. Similar to Section~\ref{sec:numexp:surf:gaussian}, the surface $\Scal$
will not be exactly represented by a piecewise polynomial space so the mesh-based
parametrization will incur a discretization error. To study the behavior of the
mesh-based parametrization under refinement, we define a mesh-based parametrization
$\Mcal_{h',q'}$ (Figure~\ref{fig:gauss-sphere}) and its derivative $M_{h',q'}$ for
two grids,  $h' = \sqrt{\pi/N}$ for $N = 134, 8576$ and two polynomial degrees $q' = 1, 2$.
For this problem, we take $\Pi_{h',q'} : x \mapsto (x_1,x_2)$, which leads to
$\Rcal_{h',q'}$ being a piecewise polynomial approximation to $\Rcal$.
The error in the mesh-based parametrization
(Figure~\ref{fig:sphere_diff_demo}) decreases by two orders of magnitude when
the polynomial degree is increased from $q' = 1$ to $q' = 2$ or each surface triangle
is split into $64$ triangles, which demonstrates the effectivity of polynomial refinement
over $h$-refinement. The benefit of polynomial refinement is more pronounced for
mesh-based parametrization derivative (Figure \ref{fig:sphere_diff_prime_demo}) as the error decreases by
three orders of magnitude when the polynomial degree in increased from $q'=1$ to
$q'=2$ but just over one order of magnitude by splitting each surface triangle into
$64$ triangles. Based on this observation we opt for geometry meshes with larger
polynomial degrees when integrated into implicit shock tracking (Sections~\ref{sec:numexp:advec}-\ref{sec:numexp:euler})
rather than using more elements to improve the geometry resolution.

\begin{remark}
	The mapping $\Pi_{h',q'} : x \mapsto (x_1, x_2)$ used for this problem is
        obtained from the approach in~\ref{sec:surfproj:surf} with $\hat{x} = 0$
        and $n = (0, 0, 1)$.
\end{remark}

\subsection{Linear advection: straight-sided shock intersecting curved boundary}
\label{sec:numexp:advec}
Consider steady, linear advection of a scalar quantity $U : \Omega \mapsto \Rbb$ 
through a domain $\Omega\subset\Rbb^d$, which is a conservation law
of the form (\ref{eqn:claw}) with $m=1$ and
\begin{equation}
	F : W \mapsto W \beta^T, \quad S : W \mapsto 0,
\end{equation}
where $\beta:\Omega \mapsto \Rbb^{d}$ is the flow direction, $U(x)$
is implicitly defined as the solution of (\ref{eqn:claw}) with boundary
condition $U(x) = U_\infty(x)$ for $x \in \Gamma_\mathrm{in}$,
$\Gamma_\mathrm{in} = \{x \in \partial\Omega ~\mid~\beta(x)\cdot n(x) < 0\}$
is the inflow boundary, and $U_\infty : \partial\Omega \rightarrow \Rbb$
is the boundary condition. In this problem, we consider a two-dimensional
domain ($d=2$) defined as
$\Omega \coloneqq \left\{ x\in\Rbb^2 \suchthat x_1 \in (-1,1), x_2 \in (0,f(x_1))\right\}$,
where $f : (-1, 1) \rightarrow \Rbb$ is a polynomial of degree seven, defined as
$f : s \mapsto 1+0.07(s^2-1)(s^5-0.1s^4+4.9s^3+10s^2-7.8s-2.8)$.
We also define a constant velocity
field $\beta : x \mapsto (0.25, 1)$ with boundary condition $U_\infty : x \mapsto H(x)$,
where $H : \Rbb \rightarrow \{0,1\}$ is the Heaviside function, which leads to a solution
with discontinuity along the line $\left\{(s/4, s) \suchthat s\in\Rbb_{\geq 0}\right\}$ with
constant solution on either side. All four boundaries (including the planar ones) and
their intersections (in $d=2$, points at the intersection of two boundaries are fixed)
are parametrized using mesh-based parametrization (\ref{eqn:mbp_msh}) with
$h' = \ell_i/4$ and $q' = 6$, where $\ell_i$ is the length of boundary $i$. The
mesh-based parametrization $\phibold$ is slightly modified to fix the node at
$(0,0)$ to ensure the discontinuous boundary condition is represented correctly,
i.e., $\phibold_I(\ybm_I) = \Xbm_I$ where $I\in\{1,\dots,N_x\}$ such that
$\Xbm_I = (0,0)$. The domain is discretized into $24$ triangular elements
with $q = 6$ domain approximation and $p = 0$ solution approximation.
The DG discretization uses the smoothed Riemann numerical flux described
in \cite{zahr2020react, zahr2018shktrk}.

\begin{figure}
	\centering
	\begin{tikzpicture}
\begin{groupplot} [
group style={group size = 2 by 1, horizontal sep = 1cm, vertical sep = 1.50cm}]
\nextgroupplot[axis equal image, axis line style={gray}, axis x line*=bottom, width=0.49\textwidth, xtick={-1.0, 0, 1.0}, ytick={0.0, 1.0}, xmin=-1, xmax=1, ymin=0, ymax=1.26]
\addplot []
graphics [xmin=-1,xmax=1,ymin=0,ymax=1.26] { 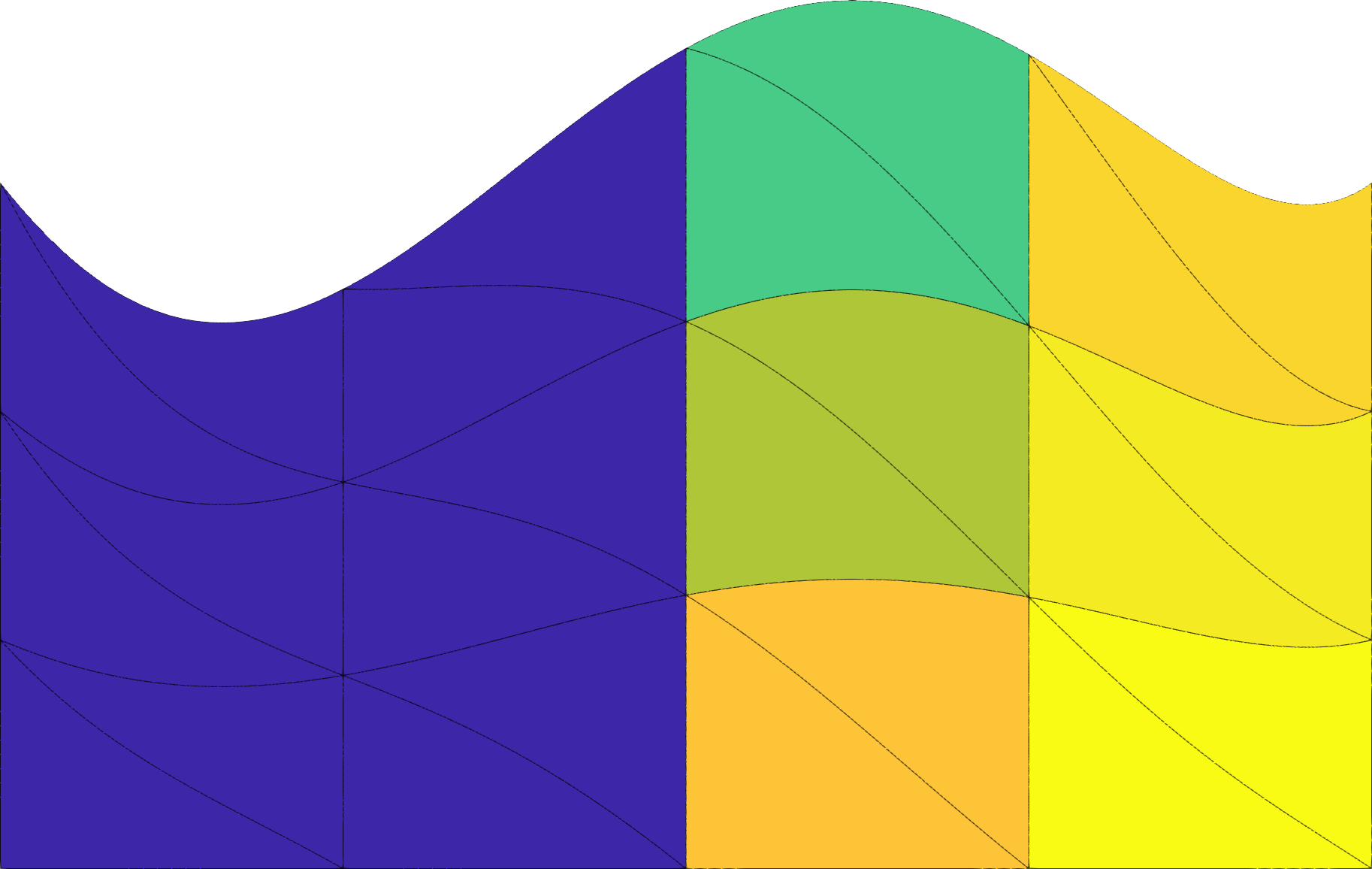};

\addplot [domain=0:0.31479, color=purple, thick]
{
(1/0.25)*x};\label{line:wavy:discont_surf}

\nextgroupplot[axis equal image, axis line style={gray}, axis x line*=bottom, yticklabels={,,}, width=0.49\textwidth, xtick={-1.0, 0, 1.0}, xmin=-1, xmax=1, ymin=0, ymax=1.26]
\addplot []
graphics [xmin=-1,xmax=1,ymin=0,ymax=1.26] { 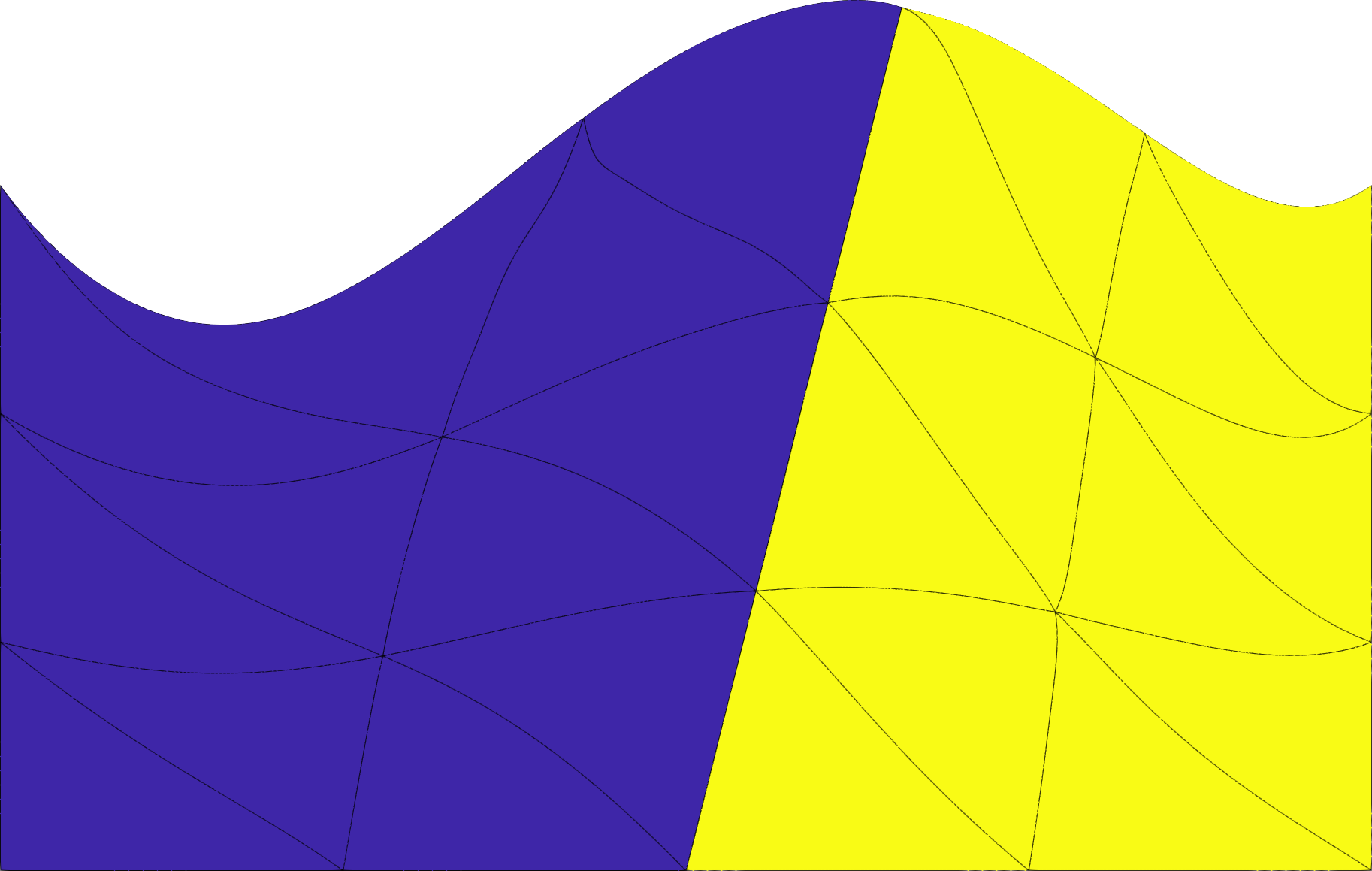};

\addplot [domain=0:0.31479, color=purple, thick, forget plot]
{
(1/0.25)*x};

\end{groupplot}\end{tikzpicture}
	\caption{
		Initial $p=0$ solution \textit{(left)} and converged HOIST
		solution \textit{(right)} of the linear advection problem
		with the true discontinuity surface highlighted
		(\ref{line:wavy:discont_surf}).
	}
	\label{fig:advec_wavy}
\end{figure}

\begin{figure}
	\centering
	\input{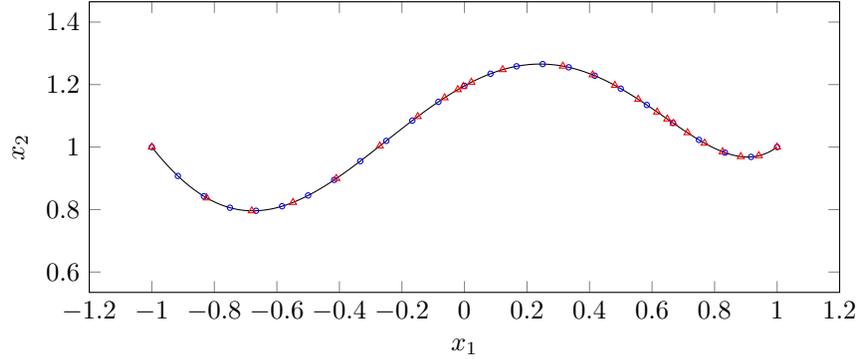}
	\caption{
		The surrogate surface of the upper boundary for the linear
		advection problem (\ref{fig:wavy_nd_exact}) with the initial
		(\ref{fig:wavy_nd_0}) and final (\ref{fig:wavy_nd_1}) positions
		of the high-order nodes on this boundary.
	}
  	\label{fig:advec_wavy_nd}
\end{figure}

\begin{figure}
	\centering
	\begin{tikzpicture}
\begin{axis}[
ymode=log,
xtick={0,15,33},
height=0.33\textwidth,
width=0.7\textwidth,
xlabel=Iteration]
\addplot [black, mark=*, mark size=1, mark options={solid, thin}]
coordinates {
( 0.00000000e+00,  3.40899501e-03)
( 1.03030303e+00,  1.99187657e-03)
( 2.06060606e+00,  1.71664736e-03)
( 3.09090909e+00,  8.84320236e-04)
( 4.12121212e+00,  8.65305085e-04)
( 5.15151515e+00,  7.56871241e-04)
( 6.18181818e+00,  7.26880624e-04)
( 7.21212121e+00,  6.81551924e-04)
( 8.24242424e+00,  3.04066988e-04)
( 9.27272727e+00,  1.71472164e-04)
( 1.03030303e+01,  2.69359405e-05)
( 1.13333333e+01,  4.62838689e-06)
( 1.23636364e+01,  8.41852344e-07)
( 1.33939394e+01,  1.99941742e-07)
( 1.44242424e+01,  4.38258738e-08)
( 1.54545455e+01,  1.32323617e-08)
( 1.64848485e+01,  4.98207696e-09)
( 1.75151515e+01,  2.46524442e-09)
( 1.85454545e+01,  1.50324340e-09)
( 1.95757576e+01,  1.04225016e-09)
( 2.06060606e+01,  7.79122451e-10)
( 2.16363636e+01,  6.10443819e-10)
( 2.26666667e+01,  4.93950651e-10)
( 2.36969697e+01,  4.09382834e-10)
( 2.47272727e+01,  3.45758281e-10)
( 2.57575758e+01,  2.96570795e-10)
( 2.67878788e+01,  2.57710419e-10)
( 2.78181818e+01,  2.26454024e-10)
( 2.88484848e+01,  2.00929634e-10)
( 2.98787879e+01,  1.79811866e-10)
( 3.09090909e+01,  1.62139224e-10)
( 3.19393939e+01,  1.47199677e-10)
( 3.29696970e+01,  1.34456481e-10)
( 3.40000000e+01,  1.23498678e-10)};\label{adv_res_dg}

\addplot [red, mark=triangle*, mark size=1.5, mark options={solid, thin}]
coordinates {
( 0.00000000e+00,  4.13821161e-02)
( 1.03030303e+00,  3.25052734e-02)
( 2.06060606e+00,  2.99331813e-02)
( 3.09090909e+00,  2.19836802e-02)
( 4.12121212e+00,  2.18148436e-02)
( 5.15151515e+00,  2.07159966e-02)
( 6.18181818e+00,  2.01005081e-02)
( 7.21212121e+00,  1.90442626e-02)
( 8.24242424e+00,  1.25559923e-02)
( 9.27272727e+00,  8.94026040e-03)
( 1.03030303e+01,  3.13791003e-03)
( 1.13333333e+01,  1.21722472e-03)
( 1.23636364e+01,  5.21667835e-04)
( 1.33939394e+01,  2.52849460e-04)
( 1.44242424e+01,  1.19828939e-04)
( 1.54545455e+01,  6.17392615e-05)
( 1.64848485e+01,  3.61975912e-05)
( 1.75151515e+01,  2.53854670e-05)
( 1.85454545e+01,  1.94378965e-05)
( 1.95757576e+01,  1.57436539e-05)
( 2.06060606e+01,  1.32779047e-05)
( 2.16363636e+01,  1.15218584e-05)
( 2.26666667e+01,  1.02023037e-05)
( 2.36969697e+01,  9.16830959e-06)
( 2.47272727e+01,  8.33187800e-06)
( 2.57575758e+01,  7.63874502e-06)
( 2.67878788e+01,  7.05359522e-06)
( 2.78181818e+01,  6.55230594e-06)
( 2.88484848e+01,  6.11770810e-06)
( 2.98787879e+01,  5.73716193e-06)
( 3.09090909e+01,  5.40110444e-06)
( 3.19393939e+01,  5.10213863e-06)
( 3.29696970e+01,  4.83443834e-06)
( 3.40000000e+01,  4.59334476e-06)};\label{adv_en_res_dg}

\end{axis}
\end{tikzpicture}
	\caption{
		Convergence of the standard (\ref{eul_res_dg}) and enriched
		(\ref{eul_en_res_dg}) DG residual as a
		function of the HOIST SQP iterations.
	}
	\label{fig:advec_conv}
\end{figure}
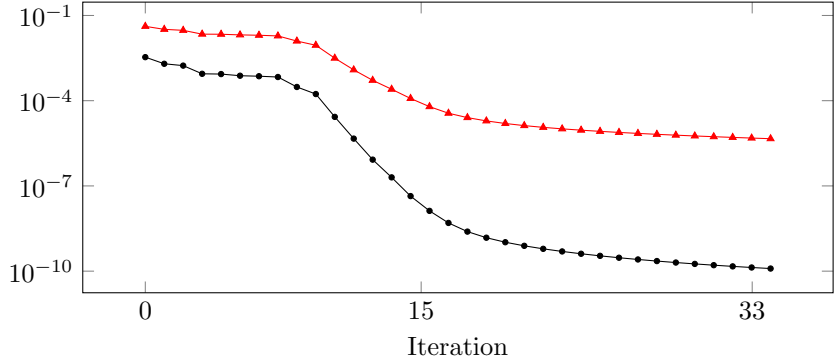

Starting from a shock-agnostic mesh and a first-order finite volume solution,
the HOIST method converges to a mesh with element faces aligned with the
discontinuity and an accurate flow solution (Figure~\ref{fig:advec_wavy}).
Nodes on the top boundary slide perfectly along the top surrogate surface
using the mesh-based parametrization $\phibold$ (Figure~\ref{fig:advec_wavy_nd})
constructed using the framework in Section~\ref{sec:mbp} (without using the
analytical expression of the boundary, only the geometry mesh, $\hat\Omega_{h',q'}$).
In only $k=33$ iterations, the residual reaches decreases nearly eight
orders of magnitude and the enriched residual decreases four orders of
magnitude (Figure~\ref{fig:advec_conv}).


\subsection{Inviscid, transonic flow over RAE2822 airfoil}
\label{sec:numexp:euler}
In this section we consider steady fluid flow in an inviscid, compressible fluid through a
domain $\Omega \subset \Rbb^{d}$ governed by the Euler equations of gasdynamics
(\ref{eqn:claw}) with
\begin{equation} \label{eqn:euler}
	U = \begin{bmatrix} \rho \\ \rho v \\ \rho E\end{bmatrix}, \qquad
		F : U \mapsto \begin{bmatrix} \rho v^T \\ \rho vv^T + P I_{d\times d} \\ (\rho E+P)v^T \end{bmatrix}, \qquad
	S : U \mapsto \begin{bmatrix} 0 \\ 0_d \\ 0 \end{bmatrix}.
\end{equation}
The density $\rho : \Omega \mapsto \Rbb_{>0}$, velocity $v_{i}: \Omega \mapsto \Rbb$ 
and the energy $E : \Omega \mapsto \Rbb_{>0}$ are defined implicitly as the solution of 
(\ref{eqn:claw}) and (\refeq{eqn:euler}). For a calorically perfect fluid, the pressure
 $P:x \mapsto \Rbb_{>0}$ is related to the energy by the ideal gas law
\begin{equation}
 P = (\gamma-1)\left(\rho E - \frac{\rho v_i v_i}{2}\right),
\end{equation}
where $\gamma \in \Rbb_{>0}$ is the ratio of specific heats.

In this problem, we consider transonic flow at Mach number $M = 0.82$ and
angle of attack $\alpha = 0.92^\circ$ over the RAE2822 airfoil with chord
length $c = 1$ and $\gamma=1.4$. The left and right boundaries are two chords
lengths from the airfoil and the top and bottom boundaries are three chord lengths
away. The farfield conditions $\rho_{\infty} = 1$, $P_{\infty}=1/\gamma$ are
enforced using Roe's numerical flux, i.e., solving an approximate Riemann problem
between the interior and boundary state. A no-penetration ($v \cdot n = 0$) is enforced
on the airfoil surfaces. The conservation law is discretized using DG on a mesh
($\hat\Omega_{h,q}$) consisting of 755 quadratic ($q=2$) triangular elements with
quadratic ($p = 2$) flow approximation. We use the Roe flux with Harten-Hyman entropy
fix \cite{harten1983self}. The boundary-preserving parametrization $\phibold$ is
constructed via mesh-based parametrization of the upper and lower surfaces of the airfoil
and fixed nodes on the farfield boundaries. Under this parametrization, the nodes on the
leading and trailing edge remain fixed because these points represent the intersection of
two surfaces (i.e., the upper and lower airfoil) in two dimensions
(Section~\ref{sec:mbp:msh}). The geometry mesh of the airfoil surfaces
is extracted from the computational mesh ($h' = h$, $q' = q$).

\begin{figure}
    \centering
    \begin{tikzpicture}

\begin{groupplot}[
  group style={
      group size=3 by 2,
      horizontal sep=0.5cm,
      vertical sep=0.5cm
  },
  width=0.42\textwidth,
  axis equal image,
  axis line style={gray},
  axis x line*=bottom,
  axis y line*=left,
  xtick = {-1.0, 0, 2.0},
  ytick = {-1.5, 1.5},
  xmin=-1, xmax=2,
  ymin=-1.5, ymax=1.5
]
\nextgroupplot[xlabel={}, xticklabels={,,}]
\addplot graphics [xmin=-1.0, xmax=2, ymin=-1.5, ymax=1.5] {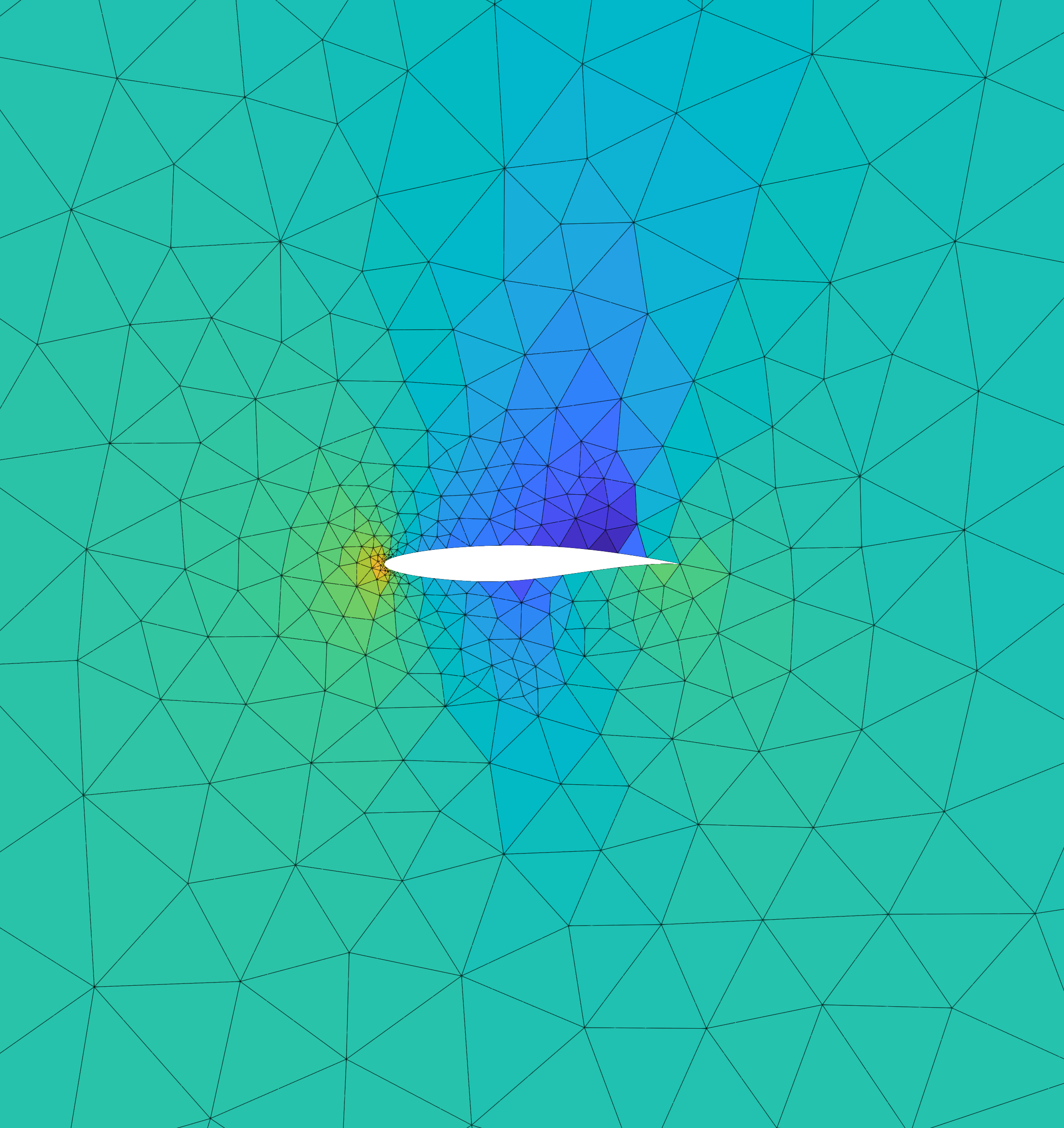};
\nextgroupplot[xlabel={}, ylabel={}, xticklabels={,,}, yticklabels={,,}]
\addplot graphics [xmin=-1.0, xmax=2, ymin=-1.5, ymax=1.5] {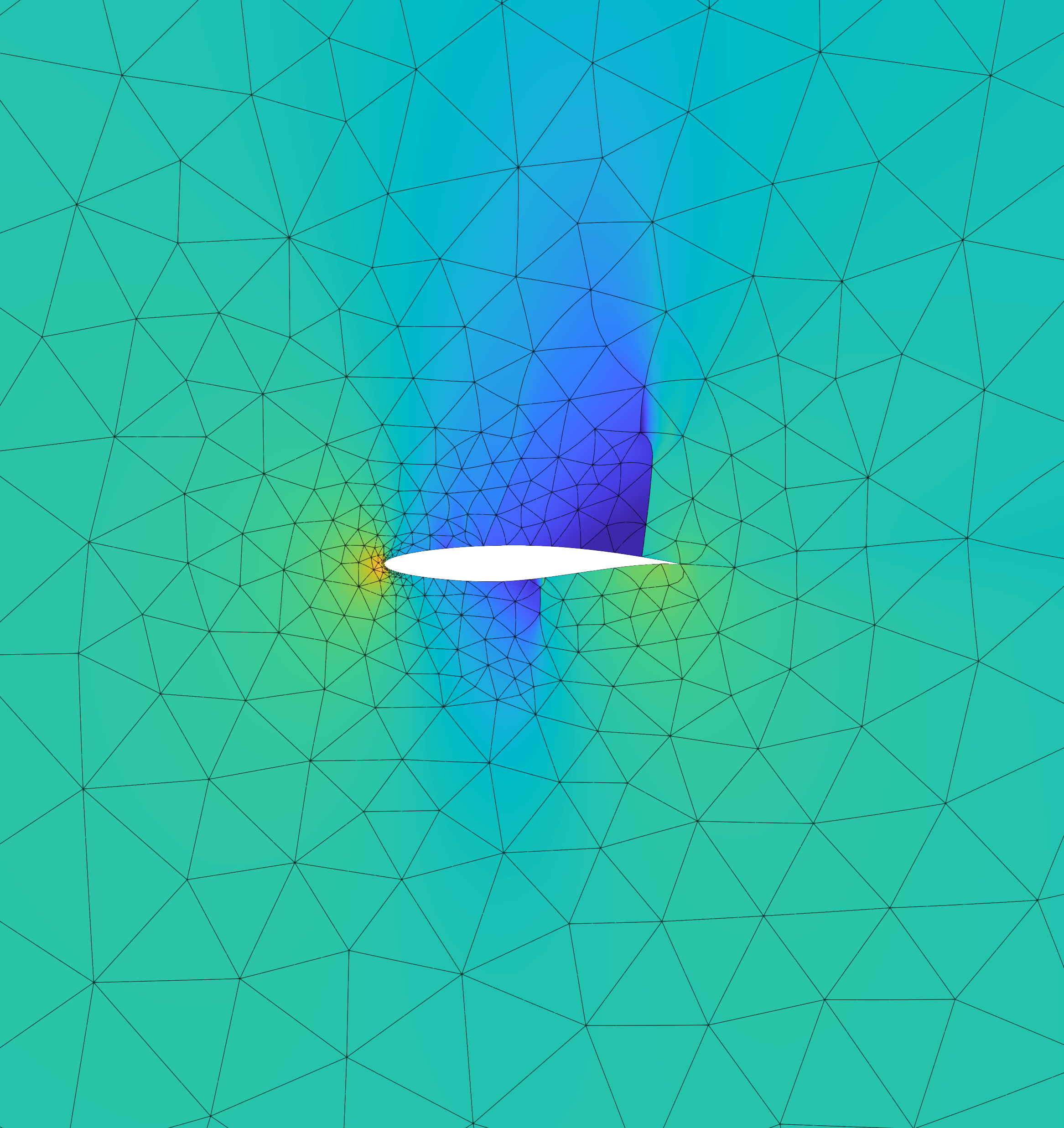};

\nextgroupplot[xlabel={}, ylabel={}, xticklabels={,,}, yticklabels={,,}]
\addplot graphics [xmin=-1.0, xmax=2.0, ymin=-1.5, ymax=1.5] {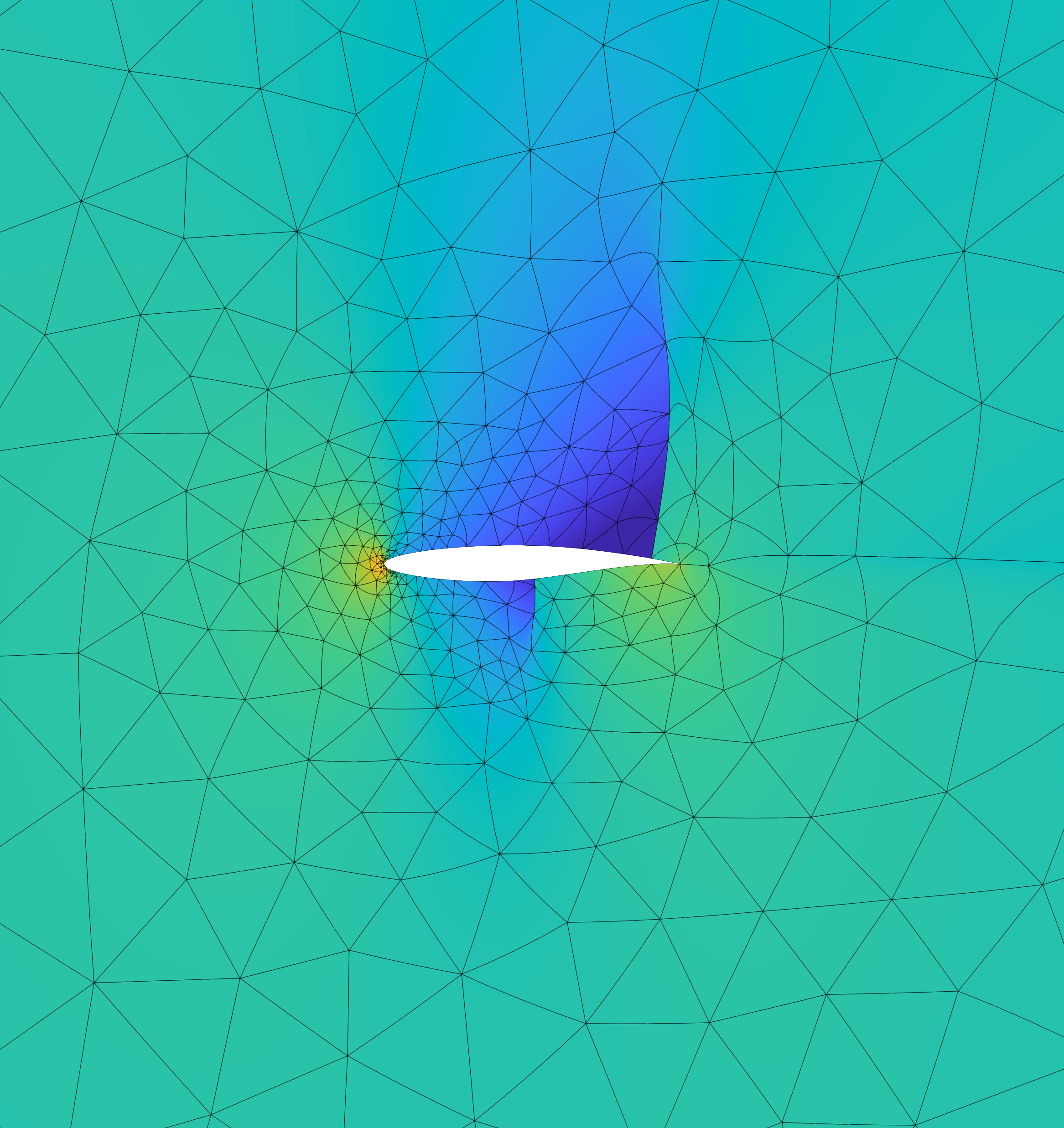};
\nextgroupplot[]
\addplot graphics [xmin=-1.0, xmax=2.0, ymin=-1.5, ymax=1.5] {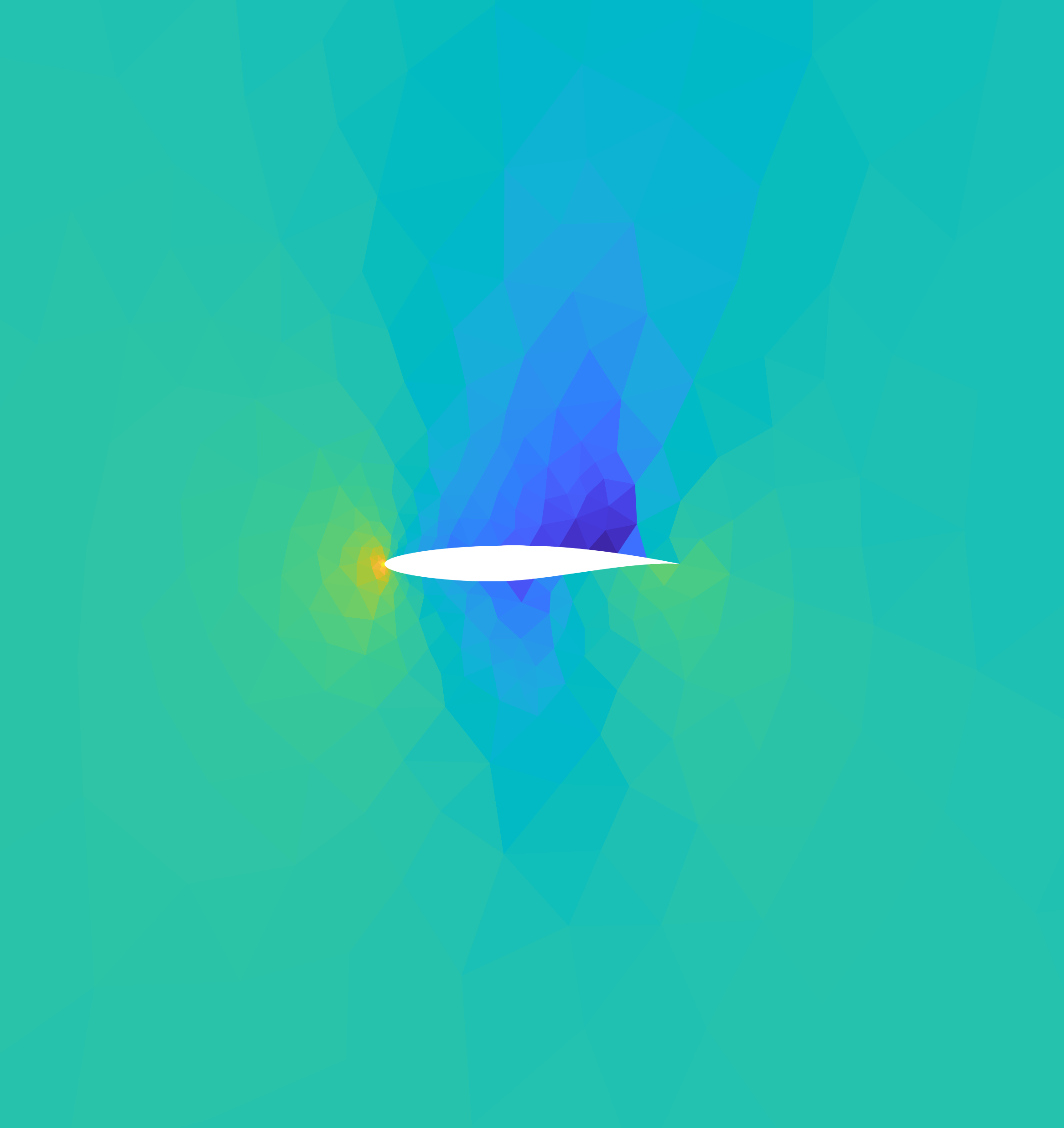};
\nextgroupplot[ylabel={}, yticklabels={,,}]
\addplot graphics [xmin=-1.0, xmax=2.0, ymin=-1.5, ymax=1.5] {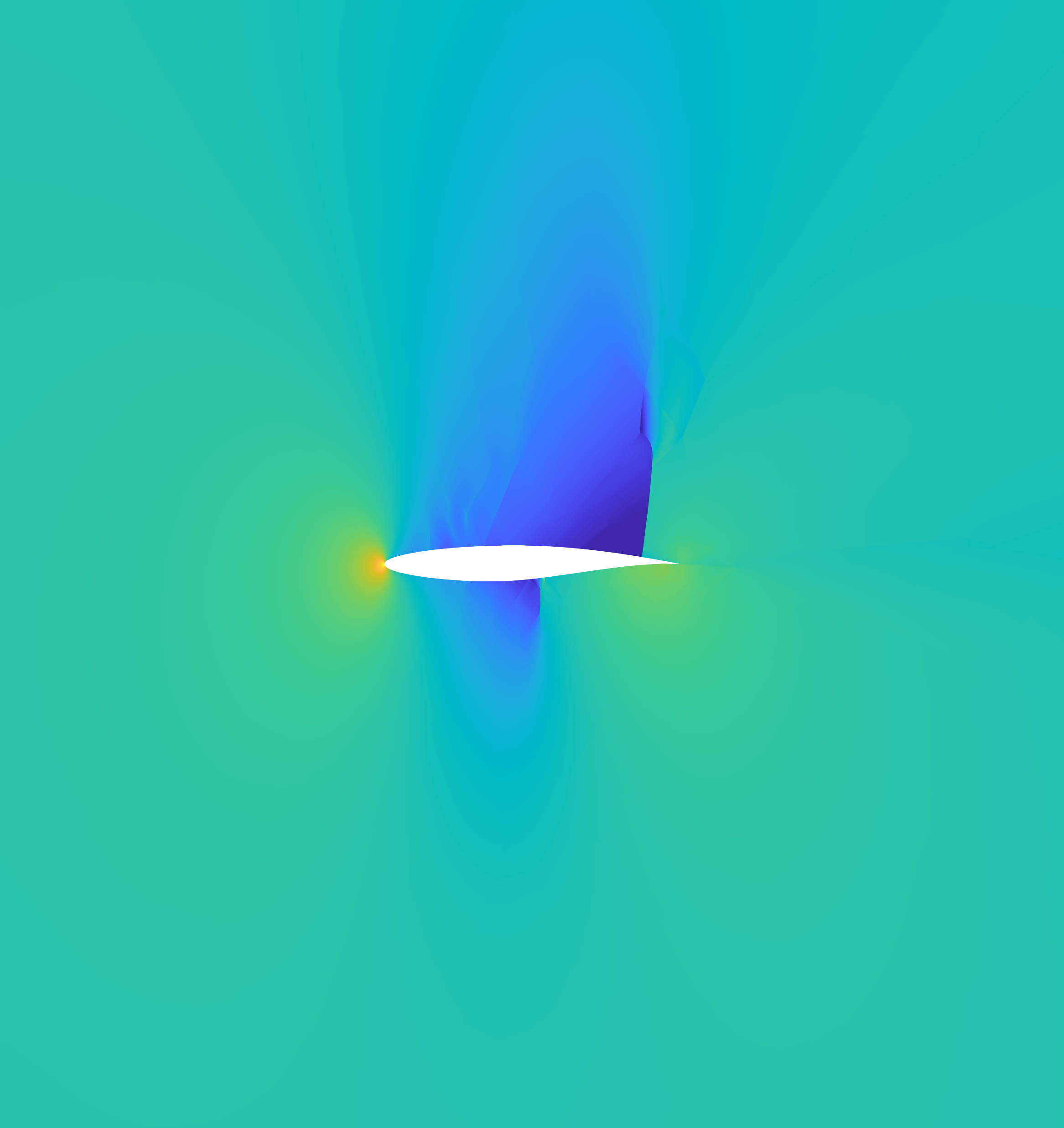};

\nextgroupplot[ylabel={}, yticklabels={,,}]
\addplot graphics [xmin=-1.0, xmax=2, ymin=-1.5, ymax=1.5] {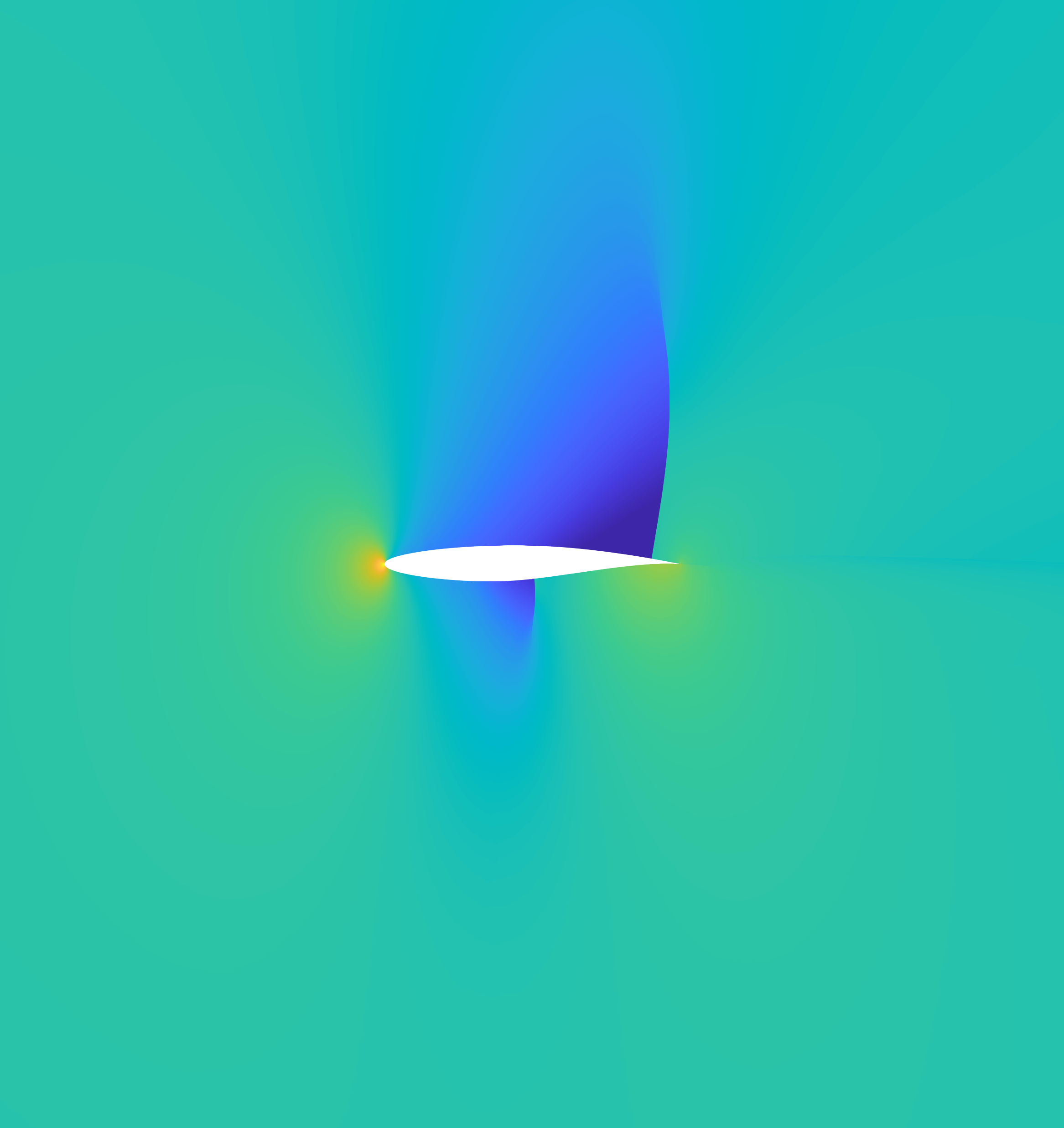};
\end{groupplot}
\end{tikzpicture}
    \colorbarMatlabParula{6.2e-1}{0.8}{1}{1.2}{1.4}
    \caption{
        Initial $p=0$ solution \textit{(left)}, an unconverged SQP iteration
	($k=7$) \textit{(middle)}, and converged HOIST
        solution \textit{(right)} of the RAE2822 problem.
    }
    \label{fig:rae_sols}
\end{figure}

\begin{figure}
    \centering
    \input{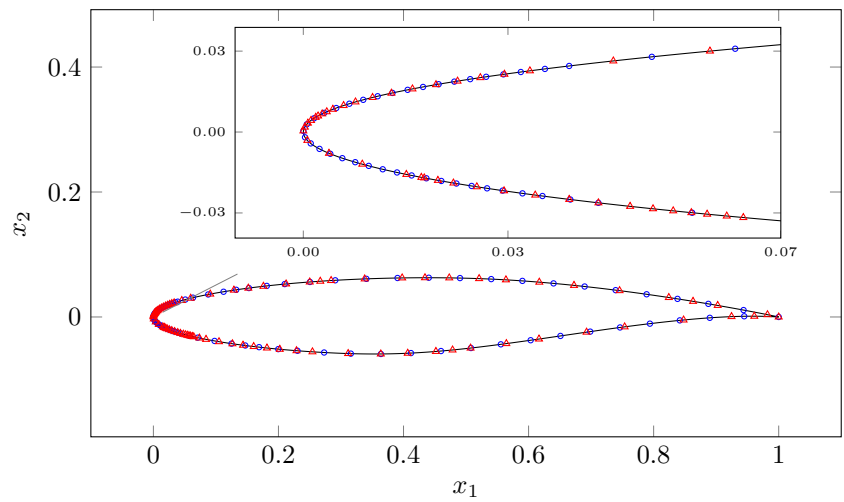}
    \caption{
	    The surrogate surface of the RAE2822 airfoil (\ref{fig:rae_nd_exact})
	    and the surface nodes of the computational mesh in their initial
	    (\ref{fig:rae_nd_0}) and final (\ref{fig:rae_nd_1}) positions
	    after $k=100$ HOIST iterations.
    }
    \label{fig:rae_nd_motion}
\end{figure}

\begin{figure}
    \centering
    \begin{tikzpicture}
\begin{axis}[
ymode=log,
xtick={0,25,50,75,100},
height=0.33\textwidth,
width=0.7\textwidth,
xlabel=Iteration]
\addplot [black, mark repeat=10, mark=*, mark size=1, mark options={solid, thin}]
coordinates {
( 0.00000000e+00,  1.03734523e-03)
( 1.01000000e+00,  5.49327604e-04)
( 2.02000000e+00,  3.40960444e-04)
( 3.03000000e+00,  2.79164994e-04)
( 4.04000000e+00,  1.97871562e-04)
( 5.05000000e+00,  9.45583293e-05)
( 6.06000000e+00,  7.10978637e-05)
( 7.07000000e+00,  3.82236700e-05)
( 8.08000000e+00,  2.92221846e-05)
( 9.09000000e+00,  2.76176312e-05)
( 1.01000000e+01,  9.70914600e-04)
( 1.11100000e+01,  2.53308295e-04)
( 1.21200000e+01,  2.88171212e-05)
( 1.31300000e+01,  1.64084125e-05)
( 1.41400000e+01,  1.09781703e-05)
( 1.51500000e+01,  9.00560119e-06)
( 1.61600000e+01,  6.54170266e-06)
( 1.71700000e+01,  5.42203349e-06)
( 1.81800000e+01,  4.75430104e-06)
( 1.91900000e+01,  4.30247537e-06)
( 2.02000000e+01,  3.96755591e-06)
( 2.12100000e+01,  3.85433682e-06)
( 2.22200000e+01,  3.68265025e-06)
( 2.32300000e+01,  3.55750060e-06)
( 2.42400000e+01,  3.45356161e-06)
( 2.52500000e+01,  3.37099553e-06)
( 2.62600000e+01,  3.29693883e-06)
( 2.72700000e+01,  3.23390091e-06)
( 2.82800000e+01,  3.22125446e-06)
( 2.92900000e+01,  3.21262778e-06)
( 3.03000000e+01,  3.21057317e-06)
( 3.13100000e+01,  3.14897432e-06)
( 3.23200000e+01,  3.13209160e-06)
( 3.33300000e+01,  3.13153328e-06)
( 3.43400000e+01,  3.12430415e-06)
( 3.53500000e+01,  3.11024641e-06)
( 3.63600000e+01,  3.09695116e-06)
( 3.73700000e+01,  3.04708407e-06)
( 3.83800000e+01,  3.03460793e-06)
( 3.93900000e+01,  3.01152180e-06)
( 4.04000000e+01,  7.03916859e-04)
( 4.14100000e+01,  1.77570695e-04)
( 4.24200000e+01,  1.40268715e-05)
( 4.34300000e+01,  6.79887632e-06)
( 4.44400000e+01,  5.09458886e-06)
( 4.54500000e+01,  3.64131584e-06)
( 4.64600000e+01,  3.19859695e-06)
( 4.74700000e+01,  3.12260930e-06)
( 4.84800000e+01,  2.96253971e-06)
( 4.94900000e+01,  2.95783226e-06)
( 5.05000000e+01,  2.91093845e-06)
( 5.15100000e+01,  2.90834159e-06)
( 5.25200000e+01,  2.89109212e-06)
( 5.35300000e+01,  2.88403317e-06)
( 5.45400000e+01,  2.86215049e-06)
( 5.55500000e+01,  2.85058662e-06)
( 5.65600000e+01,  2.84903580e-06)
( 5.75700000e+01,  2.84741538e-06)
( 5.85800000e+01,  2.84698820e-06)
( 5.95900000e+01,  2.84623078e-06)
( 6.06000000e+01,  2.84344131e-06)
( 6.16100000e+01,  2.83820998e-06)
( 6.26200000e+01,  2.81907458e-06)
( 6.36300000e+01,  2.81728800e-06)
( 6.46400000e+01,  2.81728800e-06)
( 6.56500000e+01,  2.81728800e-06)
( 6.66600000e+01,  2.81728800e-06)
( 6.76700000e+01,  2.81728800e-06)
( 6.86800000e+01,  2.81728800e-06)
( 6.96900000e+01,  2.81728800e-06)
( 7.07000000e+01,  2.81728800e-06)
( 7.17100000e+01,  2.81728800e-06)
( 7.27200000e+01,  2.81728800e-06)
( 7.37300000e+01,  2.81728800e-06)
( 7.47400000e+01,  2.81728800e-06)
( 7.57500000e+01,  2.81728800e-06)
( 7.67600000e+01,  2.81728800e-06)
( 7.77700000e+01,  2.81728800e-06)
( 7.87800000e+01,  2.81728800e-06)
( 7.97900000e+01,  2.81728800e-06)
( 8.08000000e+01,  2.81728800e-06)
( 8.18100000e+01,  2.81728800e-06)
( 8.28200000e+01,  2.81728800e-06)
( 8.38300000e+01,  2.81728800e-06)
( 8.48400000e+01,  2.81728800e-06)
( 8.58500000e+01,  2.81728800e-06)
( 8.68600000e+01,  2.81728800e-06)
( 8.78700000e+01,  2.81728800e-06)
( 8.88800000e+01,  2.81728800e-06)
( 8.98900000e+01,  2.81728800e-06)
( 9.09000000e+01,  2.81728800e-06)
( 9.19100000e+01,  2.81728800e-06)
( 9.29200000e+01,  2.81728800e-06)
( 9.39300000e+01,  2.81728800e-06)
( 9.49400000e+01,  2.81728800e-06)
( 9.59500000e+01,  2.81728800e-06)
( 9.69600000e+01,  2.81728800e-06)
( 9.79700000e+01,  2.81728800e-06)
( 9.89800000e+01,  2.81728800e-06)
( 9.99900000e+01,  2.81728800e-06)
( 1.01000000e+02,  2.81728800e-06)};\label{eul_res_dg}

\addplot [red, mark repeat=10, mark=triangle*, mark size=1.5, mark options={solid, thin}]
coordinates {
( 0.00000000e+00,  6.96045871e-03)
( 1.01000000e+00,  5.01394889e-03)
( 2.02000000e+00,  3.85076778e-03)
( 3.03000000e+00,  3.43026633e-03)
( 4.04000000e+00,  3.02015484e-03)
( 5.05000000e+00,  1.57629962e-03)
( 6.06000000e+00,  1.37265487e-03)
( 7.07000000e+00,  1.24055797e-03)
( 8.08000000e+00,  1.24140978e-03)
( 9.09000000e+00,  1.21871246e-03)
( 1.01000000e+01,  1.00060419e-02)
( 1.11100000e+01,  4.92932567e-03)
( 1.21200000e+01,  1.41471063e-03)
( 1.31300000e+01,  1.05775426e-03)
( 1.41400000e+01,  6.01172501e-04)
( 1.51500000e+01,  4.99606488e-04)
( 1.61600000e+01,  3.93705936e-04)
( 1.71700000e+01,  3.38075953e-04)
( 1.81800000e+01,  3.18898426e-04)
( 1.91900000e+01,  3.17142671e-04)
( 2.02000000e+01,  3.26735859e-04)
( 2.12100000e+01,  3.29197153e-04)
( 2.22200000e+01,  3.30356809e-04)
( 2.32300000e+01,  3.27722206e-04)
( 2.42400000e+01,  3.20955626e-04)
( 2.52500000e+01,  3.12839547e-04)
( 2.62600000e+01,  3.01689625e-04)
( 2.72700000e+01,  2.89246171e-04)
( 2.82800000e+01,  2.84923254e-04)
( 2.92900000e+01,  2.82802521e-04)
( 3.03000000e+01,  2.82260192e-04)
( 3.13100000e+01,  2.66082730e-04)
( 3.23200000e+01,  2.61426410e-04)
( 3.33300000e+01,  2.61273715e-04)
( 3.43400000e+01,  2.58915289e-04)
( 3.53500000e+01,  2.54166740e-04)
( 3.63600000e+01,  2.49363302e-04)
( 3.73700000e+01,  2.30620476e-04)
( 3.83800000e+01,  2.25475965e-04)
( 3.93900000e+01,  2.15542640e-04)
( 4.04000000e+01,  5.05211124e-03)
( 4.14100000e+01,  2.52698640e-03)
( 4.24200000e+01,  7.71586988e-04)
( 4.34300000e+01,  6.85391909e-04)
( 4.44400000e+01,  5.21951858e-04)
( 4.54500000e+01,  3.14669141e-04)
( 4.64600000e+01,  2.68625704e-04)
( 4.74700000e+01,  2.53113115e-04)
( 4.84800000e+01,  2.11298411e-04)
( 4.94900000e+01,  2.10157860e-04)
( 5.05000000e+01,  1.93502301e-04)
( 5.15100000e+01,  1.93391217e-04)
( 5.25200000e+01,  1.92540996e-04)
( 5.35300000e+01,  1.92250378e-04)
( 5.45400000e+01,  1.91281774e-04)
( 5.55500000e+01,  1.91145666e-04)
( 5.65600000e+01,  1.91144885e-04)
( 5.75700000e+01,  1.91147231e-04)
( 5.85800000e+01,  1.91149899e-04)
( 5.95900000e+01,  1.91138333e-04)
( 6.06000000e+01,  1.91142312e-04)
( 6.16100000e+01,  1.91155134e-04)
( 6.26200000e+01,  1.91201959e-04)
( 6.36300000e+01,  1.91208382e-04)
( 6.46400000e+01,  1.91208382e-04)
( 6.56500000e+01,  1.91208382e-04)
( 6.66600000e+01,  1.91208382e-04)
( 6.76700000e+01,  1.91208382e-04)
( 6.86800000e+01,  1.91208382e-04)
( 6.96900000e+01,  1.91208382e-04)
( 7.07000000e+01,  1.91208382e-04)
( 7.17100000e+01,  1.91208382e-04)
( 7.27200000e+01,  1.91208382e-04)
( 7.37300000e+01,  1.91208382e-04)
( 7.47400000e+01,  1.91208382e-04)
( 7.57500000e+01,  1.91208382e-04)
( 7.67600000e+01,  1.91208382e-04)
( 7.77700000e+01,  1.91208382e-04)
( 7.87800000e+01,  1.91208382e-04)
( 7.97900000e+01,  1.91208382e-04)
( 8.08000000e+01,  1.91208382e-04)
( 8.18100000e+01,  1.91208382e-04)
( 8.28200000e+01,  1.91208382e-04)
( 8.38300000e+01,  1.91208382e-04)
( 8.48400000e+01,  1.91208382e-04)
( 8.58500000e+01,  1.91208382e-04)
( 8.68600000e+01,  1.91208382e-04)
( 8.78700000e+01,  1.91208382e-04)
( 8.88800000e+01,  1.91208382e-04)
( 8.98900000e+01,  1.91208382e-04)
( 9.09000000e+01,  1.91208382e-04)
( 9.19100000e+01,  1.91208382e-04)
( 9.29200000e+01,  1.91208382e-04)
( 9.39300000e+01,  1.91208382e-04)
( 9.49400000e+01,  1.91208382e-04)
( 9.59500000e+01,  1.91208382e-04)
( 9.69600000e+01,  1.91208382e-04)
( 9.79700000e+01,  1.91208382e-04)
( 9.89800000e+01,  1.91208382e-04)
( 9.99900000e+01,  1.91208382e-04)
( 1.01000000e+02,  1.91208382e-04)};\label{eul_en_res_dg}

\end{axis}
\end{tikzpicture}
    \caption{
    	Convergence of the standard (\ref{eul_res_dg}) and enriched
  	(\ref{eul_en_res_dg}) DG residual as a
   	function of the HOIST SQP iterations.
    }
    \label{fig:rae_conv}
\end{figure}
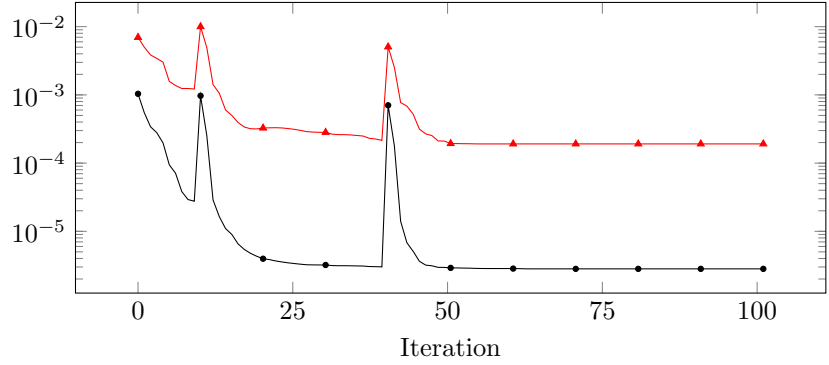

Starting from a shock-agnostic mesh and first-order finite volume solution, the
HOIST method aligns the mesh with both shocks, the strong shock on the upper
surface and the weak shock on the lower surface (Figure~\ref{fig:rae_sols}).
Even though there is no analytical expression for the RAE2822 surfaces, the
nodes slide along the top and bottom surrogate surfaces of the airfoil in such a way
that preserves the geometry of the airfoil using mesh-based parametrization
(Figure~\ref{fig:rae_nd_motion}). In only $k=100$ iterations, the residual
decreases almost three orders of magnitude and the enriched residual decreases
almost two orders of magnitude (Figure~\ref{fig:rae_conv}). This problem
demonstrates the ability of the HOIST method to track multiple shocks of
different strengths, and the effectivity of mesh-based parametrization in
pinning nodes to their original surfaces, even for geometries without analytical
expressions (only a geometry mesh of the surfaces and their intersections is
required).


\subsection{Mesh-based parametrization of the sliced cone flap geometry}
\label{sec:numexp:cone}
\begin{figure}
	\centering
  \begin{tikzpicture}
\begin{axis}[
axis lines = none,
axis equal image,
name=plot1,
scaled ticks=false,
xticklabels={,,},
yticklabels={,,},
ymin=0.0,
ymax=0.94,
xmin=0.0,
xmax=3.823,
width=0.49\textwidth]
\addplot []
graphics [xmin=0.0,xmax=3.823,ymin=0.0,ymax=0.94] { 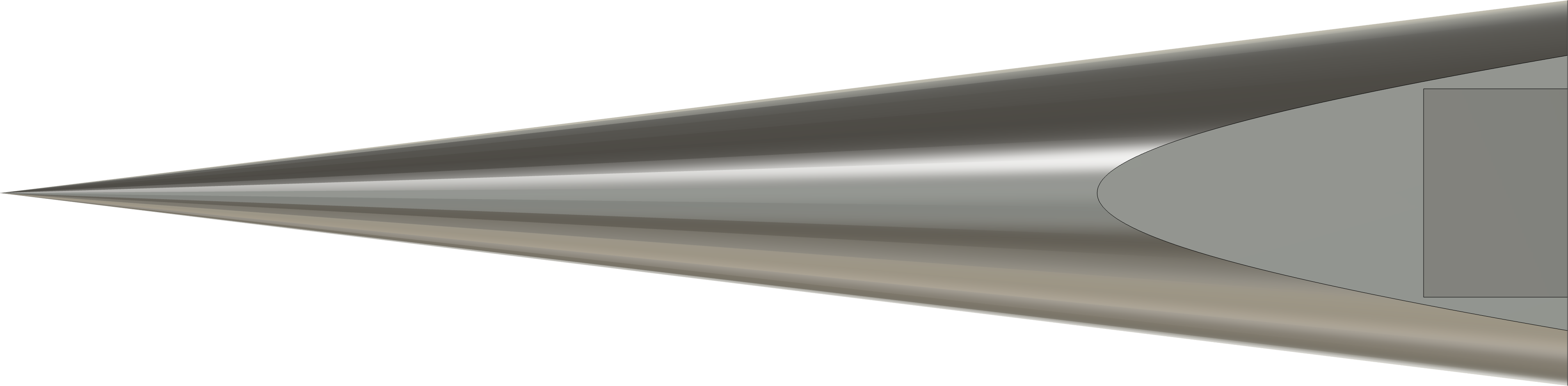};

\end{axis}
\begin{axis}[
axis lines = none,
axis equal image,
name=plot2,
at={($(plot1.south west)+(0,-1cm)$)},
anchor=north west,
scaled ticks=false,
xticklabels={,,},
yticklabels={,,},
ymin=0.0,
ymax=1.002,
xmin=0.0,
xmax=3.823,
width=0.49\textwidth]
\addplot []
graphics [xmin=0.0,xmax=3.823,ymin=0.0,ymax=1.002] { 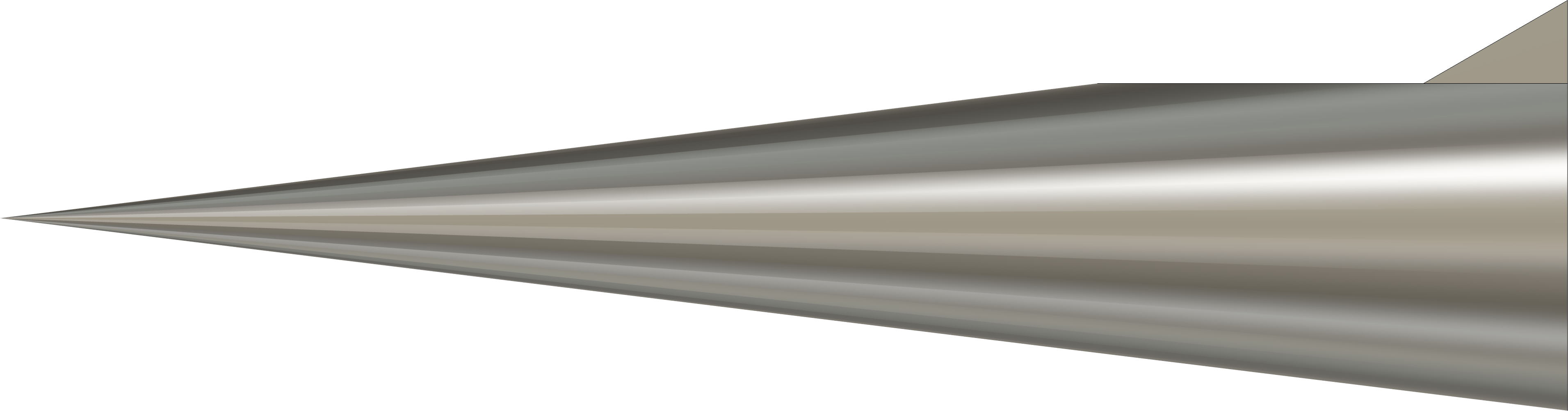};

\end{axis}
\begin{axis}[
axis lines=none,
axis equal image,
name=plot3,
at={($(plot1.north east)!0.0!(plot2.south east)+(1.5cm,0)$)},
anchor=north west,
scaled ticks=false,
xticklabels={,,},
yticklabels={,,},
xmin=0.0,
xmax=2.867,
ymin=0,
ymax=1.982,
width=0.49\textwidth]
\addplot []
graphics [xmin=0.0,xmax=2.867,ymin=0.0,ymax=1.982] { 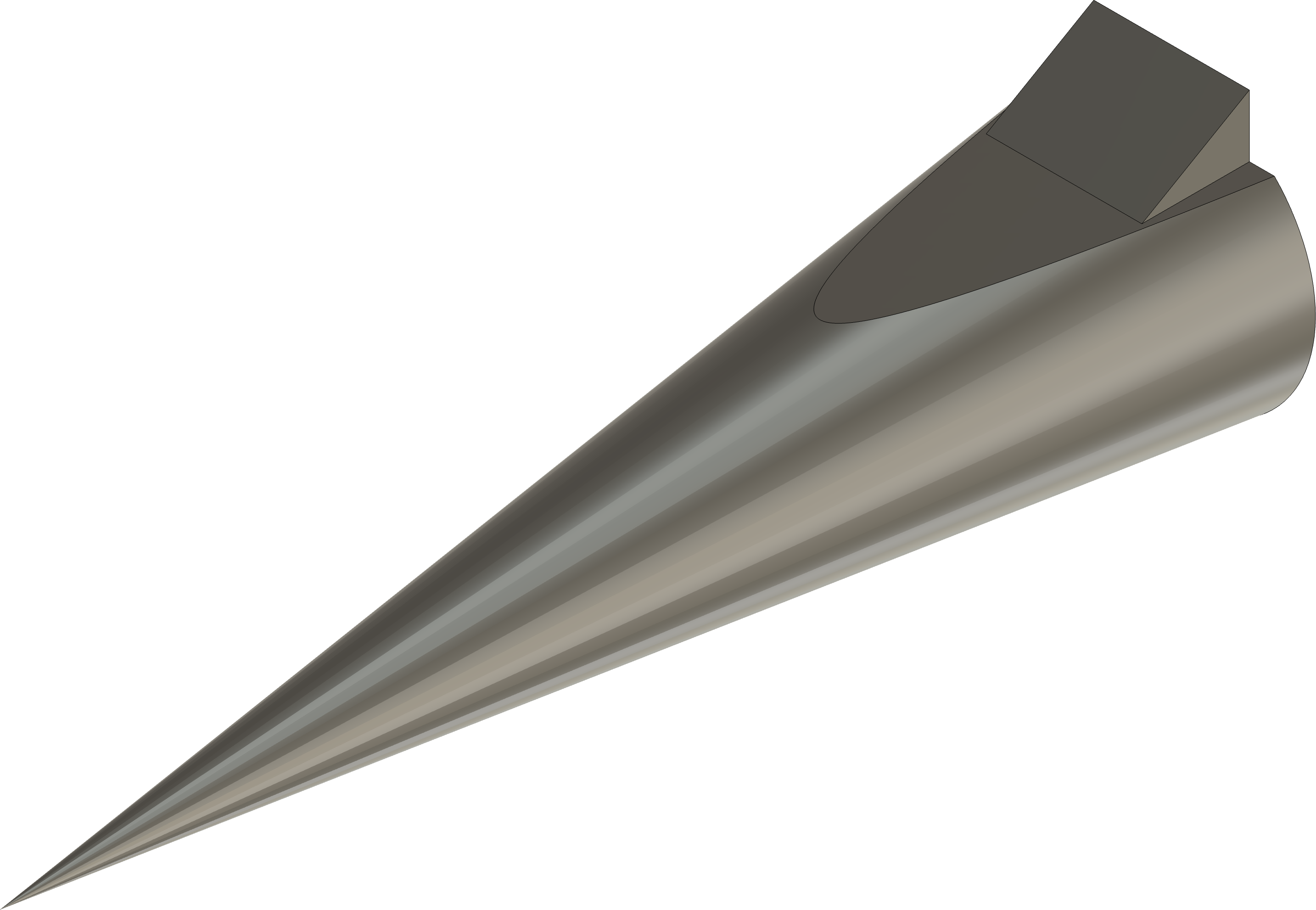};

\end{axis}
\end{tikzpicture}
	\caption{Sliced cone flap geometry.}
	\label{fig:oberkampf}
\end{figure}

\begin{figure}
    \centering
    \begin{tikzpicture}
\begin{groupplot} [
group style={group size = 3 by 2, horizontal sep = 1cm, vertical sep = 1.5cm}]
\nextgroupplot[axis equal image, scaled ticks=false, xticklabels={,,}, yticklabels={,,}, ymin=-0.0508328, ymax=0.0508328, xmin=-0.414, xmax=0.0, width=0.70\textwidth, xlabel=(a)]
\addplot []
graphics [xmin=-0.414,xmax=0.0,ymin=-0.0508328,ymax=0.0508328] { 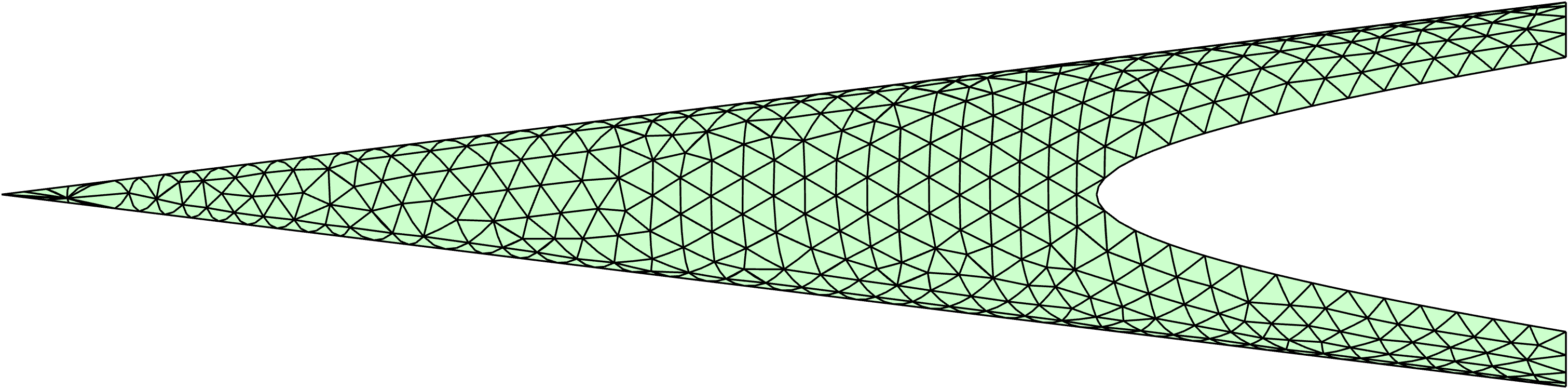};

\nextgroupplot[axis equal image, scaled ticks=false, xticklabels={,,}, yticklabels={,,}, xmin=-0.1242, xmax=0, ymin=-0.0363019, ymax=0.0363019, width=0.30\textwidth, xlabel=(b)]
\addplot []
graphics [xmin=-0.1242,xmax=0,ymin=-0.0363019,ymax=0.0363019] { 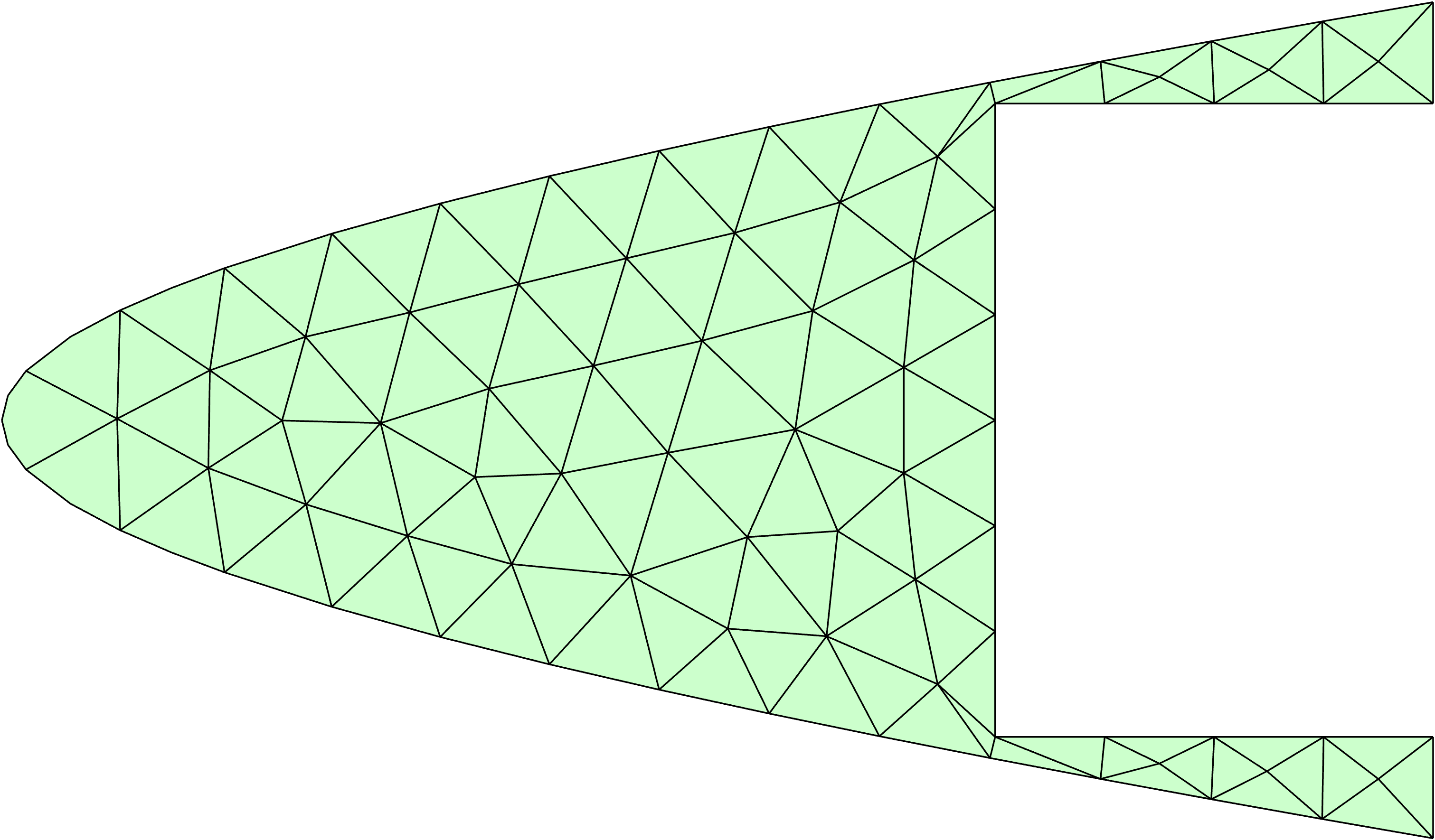};

\nextgroupplot[axis equal image, scaled ticks=false, xticklabels={,,}, yticklabels={,,}, ymin=-0.0275, ymax=0.0275, xmin=-0.038, xmax=0, width=0.24\textwidth, xlabel=(e)]
\addplot []
graphics [xmin=-0.038,xmax=0.0,ymin=-0.0275,ymax=0.0275] { 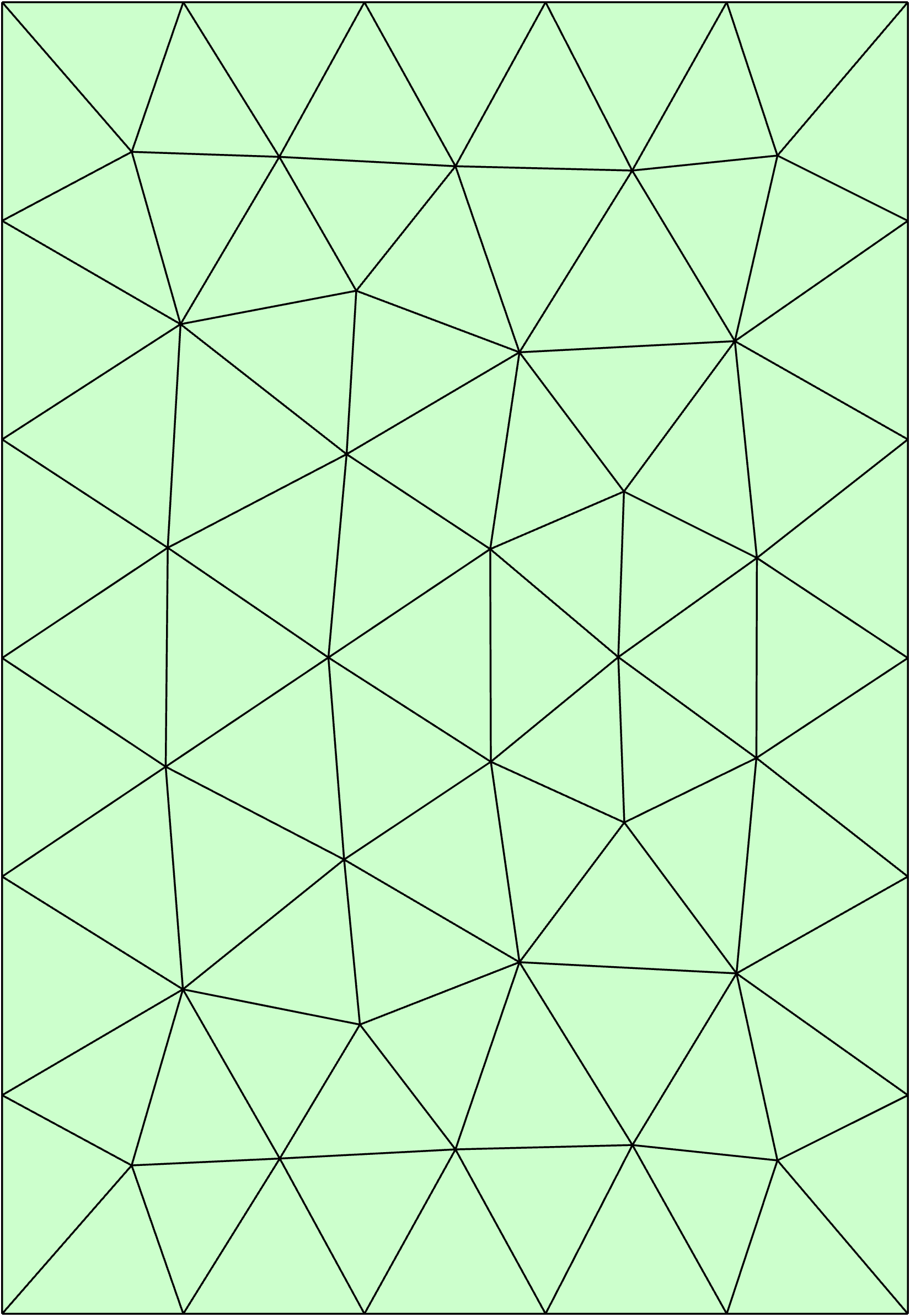};

\nextgroupplot[axis equal image, scaled ticks=false, xticklabels={,,}, yticklabels={,,}, ymin=-0.05, ymax=0.05, xmin=-0.414, xmax=0.0, width=0.70\textwidth, xlabel=(b)]
\addplot []
graphics [xmin=-0.414,xmax=0.0,ymin=-0.0508328,ymax=0.0508328] { 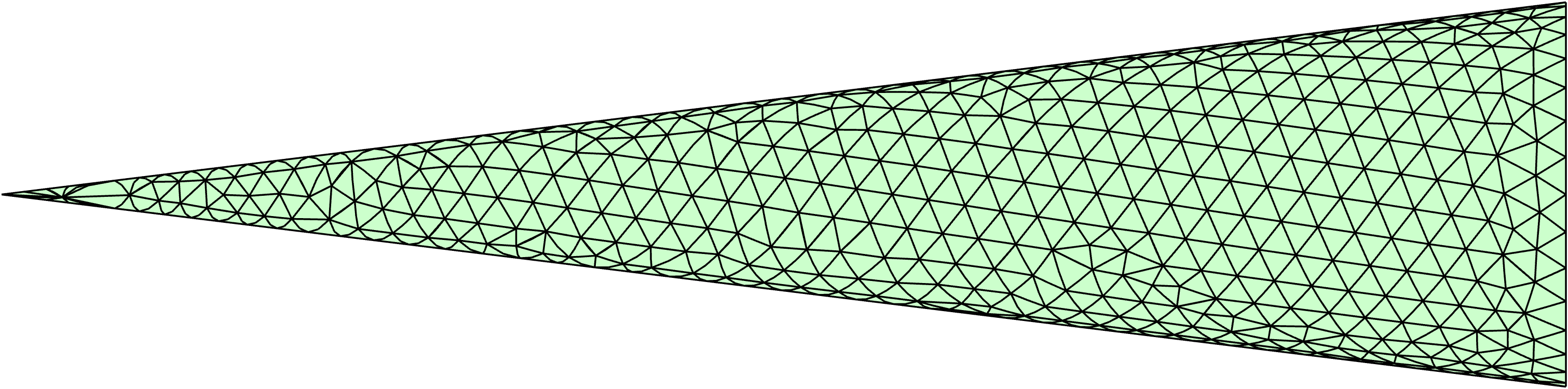};

\nextgroupplot[axis equal image, scaled ticks=false, xticklabels={,,}, yticklabels={,,}, ymin=-0.0508328, ymax=0.0275, xmin=-0.0508328, xmax=0.0508328, width=0.30\textwidth, xlabel=(d)]
\addplot []
graphics [xmin=-0.0508328,xmax=0.0508328,ymin=-0.0508328,ymax=0.0275, includegraphics={angle=90}] { 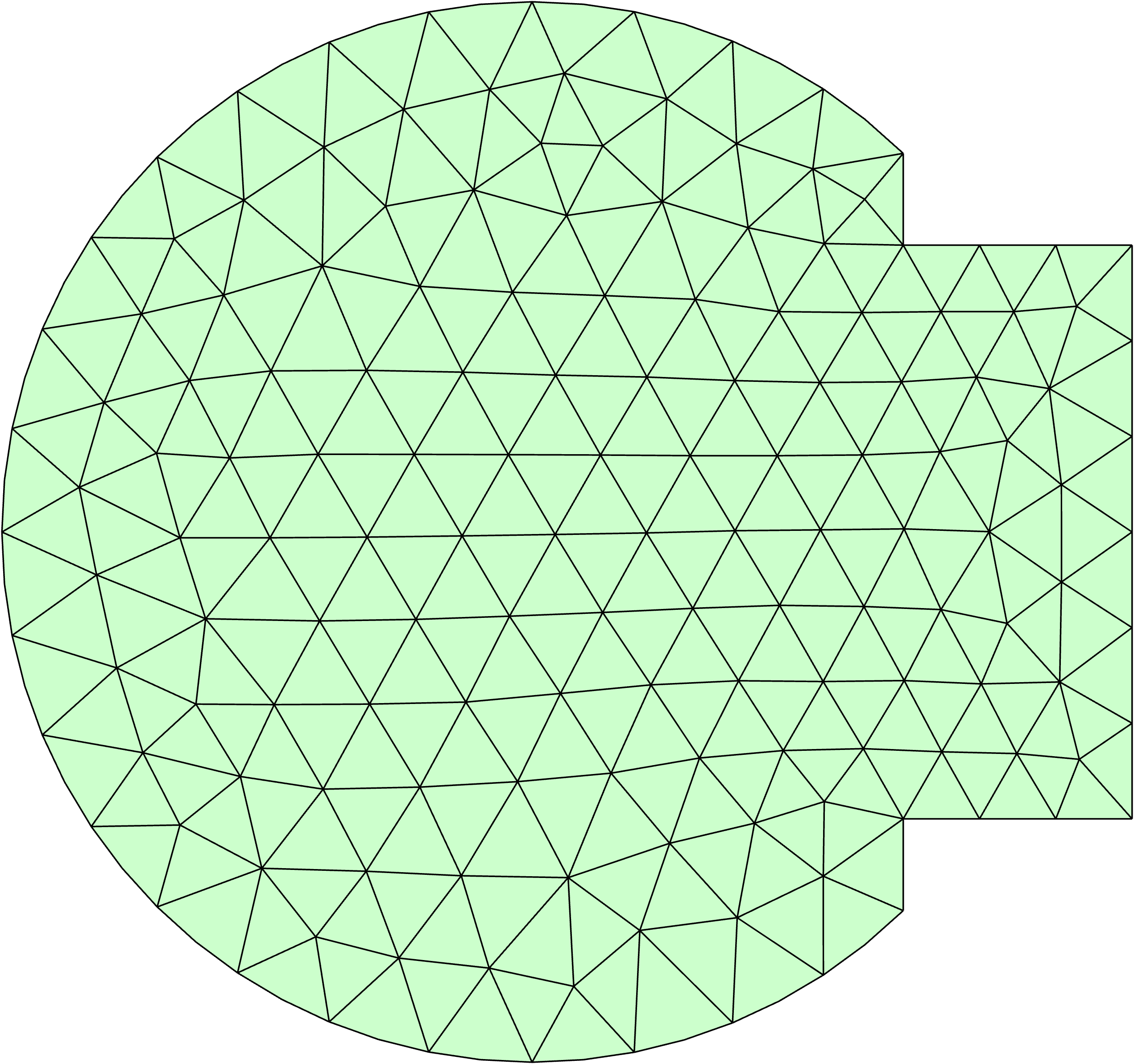};

\nextgroupplot[axis equal image, scaled ticks=false, xticklabels={,,}, yticklabels={,,}, xmin=-0.038, xmax=0.0, ymin=0, ymax=0.0219393, width=0.20\textwidth, xlabel=(f)]
\addplot []
graphics [xmin=-0.038,xmax=0.0,ymin=0.0,ymax=0.0219393, includegraphics={angle=270, origin=c}] { 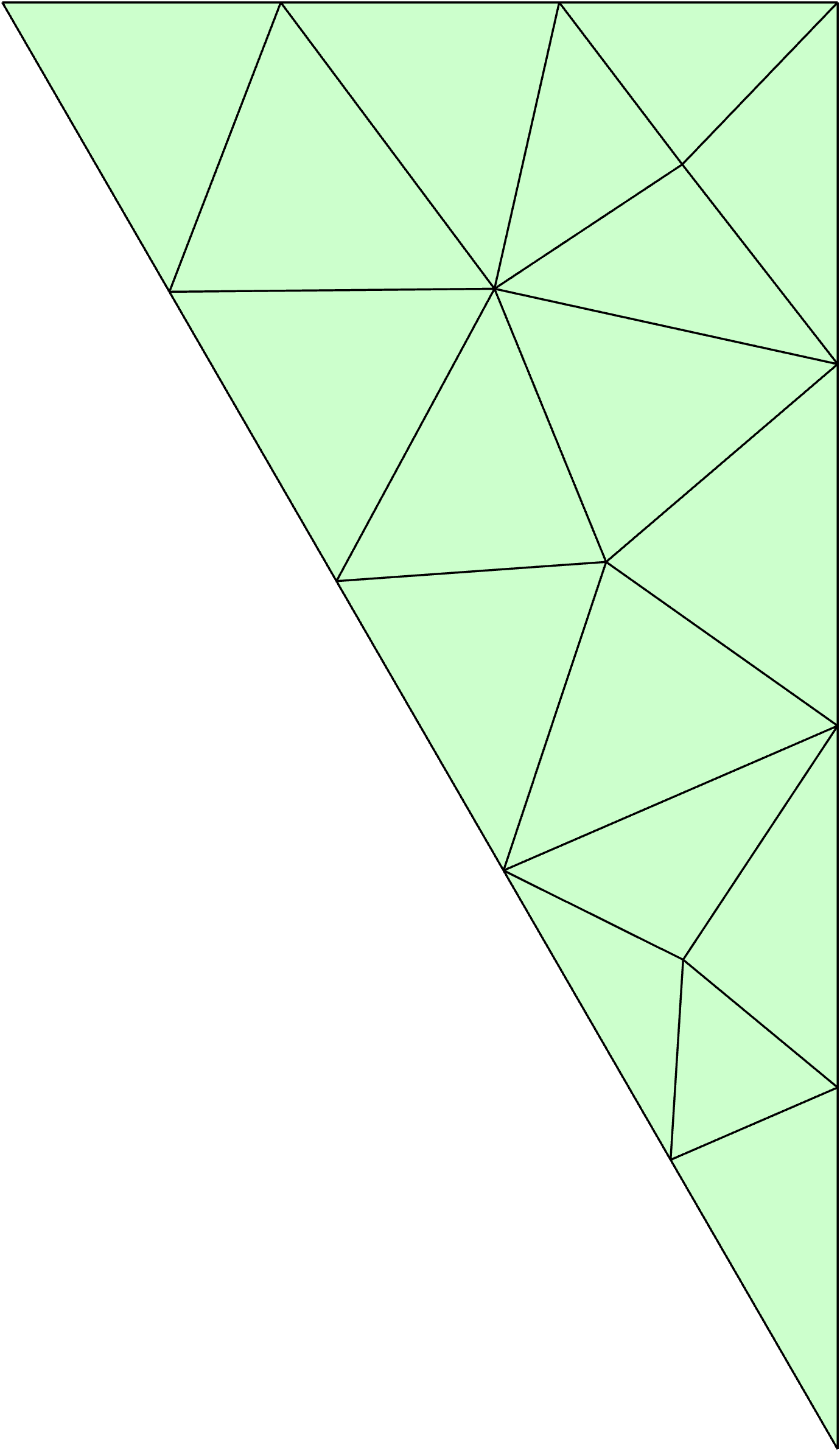};

\end{groupplot}\end{tikzpicture}
	\caption{The seven surfaces of the sliced cone flap geometry: (a) upper cone surface, (b) lower cone surface, (c) slice, (d) back, (e) flap incline, and (f) flap side.}
	\label{fig:oberkampt_surfs}
\end{figure}

We close with a demonstration of mesh-based parametrization for a complex
three-dimensional geometry ($d = 3$), the sliced cone flap \cite{1993_oberkampf}
(Figure~\ref{fig:oberkampf}).
This geometry has seven main surfaces: the upper and lower surface of the cone,
the slice, the back, and the three surfaces of the flap (the incline and two sides)
(Figure~\ref{fig:oberkampt_surfs}). We construct a geometry mesh of all six
surfaces consisting of $2088$ triangular elements of degree $q' = 2$. From these
conforming surface meshes, we also extract the geometry meshes of the intersections.
We construct the parameter domain, $\Rcal_{h',q'}\subset\Rbb^2$ for the upper cone
surface, the slice, and the flap incline using the approach in~\ref{sec:surfproj}
(Figure~\ref{fig:oberkampf_surf_map}). Then,
we define a curve $\Ccal\subset\Rcal_{h',q'}$ (spiral pattern) in the parameter domain
of the three surfaces and apply the mesh-based parametrization to each point
in $\Ccal$, which forms a curve that traverses the corresponding surrogate surfaces in
physical space ($\Rbb^3$) (Figure~\ref{fig:oberkampf_surf_map}). Next, we
construct the parameter domain $\Rcal_{h',q'}\subset\Rbb$ and mesh
$\hat\Rcal_{h',q'}$ of the intersection
of the upper cone and slice surfaces (Figure~\ref{fig:oberkampf_int_map}). Finally, the
mesh-based parametrization of this intersection is used to map five points in
$\Rcal_{h',q'}$ to the intersection of these surrogate surfaces in physical
space ($\Rbb^3$) (Figure~\ref{fig:oberkampf_int_map}).

This example demonstrates that mesh-based parametrization can be used to parametrize
the surfaces and surface intersections of complex geometries. Because it only
requires a conforming mesh of all surfaces, it is an attractive approach to
define the boundary-preserving mappings $\phibold$ for implicit shock tracking.
The fact that each surface and surface intersection are parametrized separately
means that nodes will always remain on their original boundaries. This simplifies
the assignment of boundary conditions to mesh faces and avoids oscillations that
would arise if a face straddles two surfaces with a non-smooth transition between
them (such as between the cone and slice surfaces).

\begin{figure}
   \centering
    \input{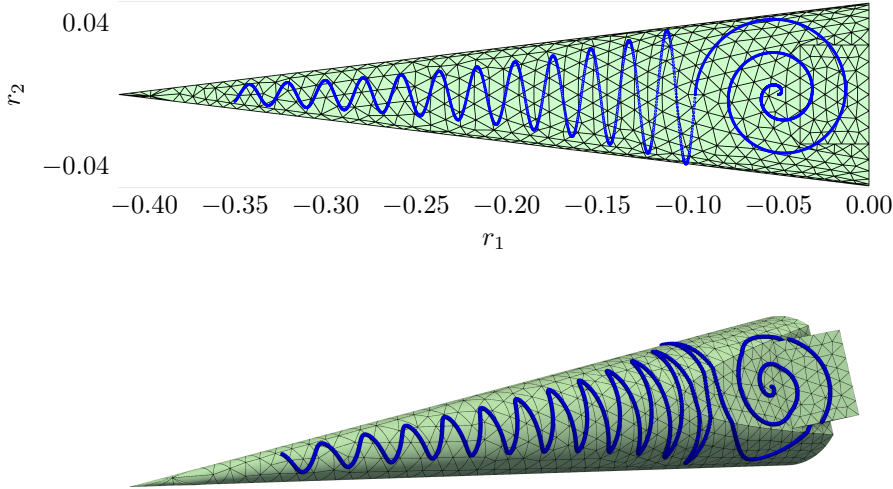}
   \caption{
	   The parameter domain $\Rcal_{h',q'}$ and mesh $\hat\Rcal_{h',q'}$ of the
	   upper cone, slice, and flap incline surfaces (\textit{top}) and
	   the sliced cone flap geometry in physical space (\textit{bottom}).
	   A curve (blue) generated in the parameter domain is guaranteed to
	   lie on the surrogate surfaces when mapped with the mesh-based
	   parametrization.
   }
   \label{fig:oberkampf_surf_map}
\end{figure}
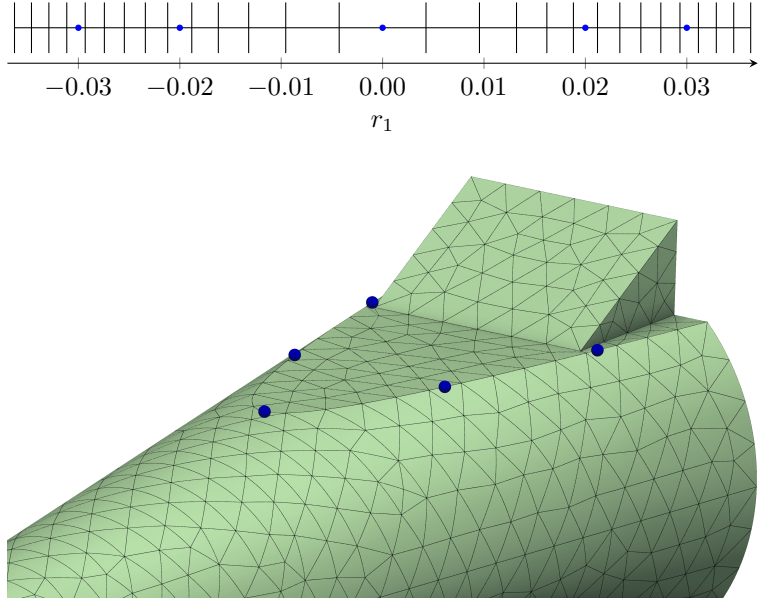
\begin{figure}
   \centering
   \begin{tikzpicture}
\begin{groupplot} [
group style={group size = 1 by 2, horizontal sep = 1.5cm, vertical sep = 1.50cm}]
\nextgroupplot[axis y line=none, axis x line=bottom, axis equal image, enlargelimits=false, legend style={font=\small}, legend cell align=left, legend pos=south east, scaled ticks=false, x tick label style={/pgf/number format/.cd, fixed, fixed zerofill, precision=2, /tikz/.cd}, y tick label style={/pgf/number format/.cd, fixed, fixed zerofill, precision=2, /tikz/.cd}, yticklabels={,,}, xlabel=$r_1$, ymin=-0.0035, ymax=0.0035, xmin=-0.037, xmax=0.037, width=0.7\textwidth]
\addplot [black, forget plot]
coordinates {
(-3.63018862e-02,  0.00000000e+00)
(-3.46276375e-02,  0.00000000e+00)
(-3.29117712e-02,  0.00000000e+00)
(-3.11472341e-02,  0.00000000e+00)
(-2.93265643e-02,  0.00000000e+00)
(-2.74383906e-02,  0.00000000e+00)
(-2.54685053e-02,  0.00000000e+00)
(-2.33965758e-02,  0.00000000e+00)
(-2.11946081e-02,  0.00000000e+00)
(-1.88189778e-02,  0.00000000e+00)
(-1.61998346e-02,  0.00000000e+00)
(-1.32074460e-02,  0.00000000e+00)
(-9.55265735e-03,  0.00000000e+00)
(-4.28078856e-03,  0.00000000e+00)
( 4.28087078e-03,  0.00000000e+00)
( 9.55264049e-03,  0.00000000e+00)
( 1.32074480e-02,  0.00000000e+00)
( 1.61998309e-02,  0.00000000e+00)
( 1.88189787e-02,  0.00000000e+00)
( 2.11945962e-02,  0.00000000e+00)
( 2.33965780e-02,  0.00000000e+00)
( 2.54685023e-02,  0.00000000e+00)
( 2.74383903e-02,  0.00000000e+00)
( 2.93265637e-02,  0.00000000e+00)
( 3.11472316e-02,  0.00000000e+00)
( 3.29117619e-02,  0.00000000e+00)
( 3.46276315e-02,  0.00000000e+00)
( 3.63018862e-02,  0.00000000e+00)};

\addplot [blue, mark=*, mark size=1, mark options={solid, thin}, only marks, forget plot]
coordinates {
( 3.00000000e-02,  0.00000000e+00)
( 2.00000000e-02,  0.00000000e+00)
( 0.00000000e+00,  0.00000000e+00)
(-2.00000000e-02,  0.00000000e+00)
(-3.00000000e-02,  0.00000000e+00)};

\addplot [black, forget plot]
coordinates {
(-3.63018862e-02, -2.50000000e-03)
(-3.63018862e-02,  2.50000000e-03)};

\addplot [black, forget plot]
coordinates {
(-3.46276375e-02, -2.50000000e-03)
(-3.46276375e-02,  2.50000000e-03)};

\addplot [black, forget plot]
coordinates {
(-3.29117712e-02, -2.50000000e-03)
(-3.29117712e-02,  2.50000000e-03)};

\addplot [black, forget plot]
coordinates {
(-3.11472341e-02, -2.50000000e-03)
(-3.11472341e-02,  2.50000000e-03)};

\addplot [black, forget plot]
coordinates {
(-2.93265643e-02, -2.50000000e-03)
(-2.93265643e-02,  2.50000000e-03)};

\addplot [black, forget plot]
coordinates {
(-2.74383906e-02, -2.50000000e-03)
(-2.74383906e-02,  2.50000000e-03)};

\addplot [black, forget plot]
coordinates {
(-2.54685053e-02, -2.50000000e-03)
(-2.54685053e-02,  2.50000000e-03)};

\addplot [black, forget plot]
coordinates {
(-2.33965758e-02, -2.50000000e-03)
(-2.33965758e-02,  2.50000000e-03)};

\addplot [black, forget plot]
coordinates {
(-2.11946081e-02, -2.50000000e-03)
(-2.11946081e-02,  2.50000000e-03)};

\addplot [black, forget plot]
coordinates {
(-1.88189778e-02, -2.50000000e-03)
(-1.88189778e-02,  2.50000000e-03)};

\addplot [black, forget plot]
coordinates {
(-1.61998346e-02, -2.50000000e-03)
(-1.61998346e-02,  2.50000000e-03)};

\addplot [black, forget plot]
coordinates {
(-1.32074460e-02, -2.50000000e-03)
(-1.32074460e-02,  2.50000000e-03)};

\addplot [black, forget plot]
coordinates {
(-9.55265735e-03, -2.50000000e-03)
(-9.55265735e-03,  2.50000000e-03)};

\addplot [black, forget plot]
coordinates {
(-4.28078856e-03, -2.50000000e-03)
(-4.28078856e-03,  2.50000000e-03)};

\addplot [black, forget plot]
coordinates {
( 4.28087078e-03, -2.50000000e-03)
( 4.28087078e-03,  2.50000000e-03)};

\addplot [black, forget plot]
coordinates {
( 9.55264049e-03, -2.50000000e-03)
( 9.55264049e-03,  2.50000000e-03)};

\addplot [black, forget plot]
coordinates {
( 1.32074480e-02, -2.50000000e-03)
( 1.32074480e-02,  2.50000000e-03)};

\addplot [black, forget plot]
coordinates {
( 1.61998309e-02, -2.50000000e-03)
( 1.61998309e-02,  2.50000000e-03)};

\addplot [black, forget plot]
coordinates {
( 1.88189787e-02, -2.50000000e-03)
( 1.88189787e-02,  2.50000000e-03)};

\addplot [black, forget plot]
coordinates {
( 2.11945962e-02, -2.50000000e-03)
( 2.11945962e-02,  2.50000000e-03)};

\addplot [black, forget plot]
coordinates {
( 2.33965780e-02, -2.50000000e-03)
( 2.33965780e-02,  2.50000000e-03)};

\addplot [black, forget plot]
coordinates {
( 2.54685023e-02, -2.50000000e-03)
( 2.54685023e-02,  2.50000000e-03)};

\addplot [black, forget plot]
coordinates {
( 2.74383903e-02, -2.50000000e-03)
( 2.74383903e-02,  2.50000000e-03)};

\addplot [black, forget plot]
coordinates {
( 2.93265637e-02, -2.50000000e-03)
( 2.93265637e-02,  2.50000000e-03)};

\addplot [black, forget plot]
coordinates {
( 3.11472316e-02, -2.50000000e-03)
( 3.11472316e-02,  2.50000000e-03)};

\addplot [black, forget plot]
coordinates {
( 3.29117619e-02, -2.50000000e-03)
( 3.29117619e-02,  2.50000000e-03)};

\addplot [black, forget plot]
coordinates {
( 3.46276315e-02, -2.50000000e-03)
( 3.46276315e-02,  2.50000000e-03)};

\addplot [black, forget plot]
coordinates {
( 3.63018862e-02, -2.50000000e-03)
( 3.63018862e-02,  2.50000000e-03)};

\nextgroupplot[axis equal image, axis y line=none, axis x line=none, axis z line=none, legend style={font=\small}, legend cell align=left, legend pos=south east, scaled ticks=false, xmin=0, xmax=3.84, ymin=0, ymax=2.16, width=0.7\textwidth]
\addplot [width= 0.75\textwidth]
graphics [xmin=0,xmax=3.84,ymin=0,ymax=2.16] { 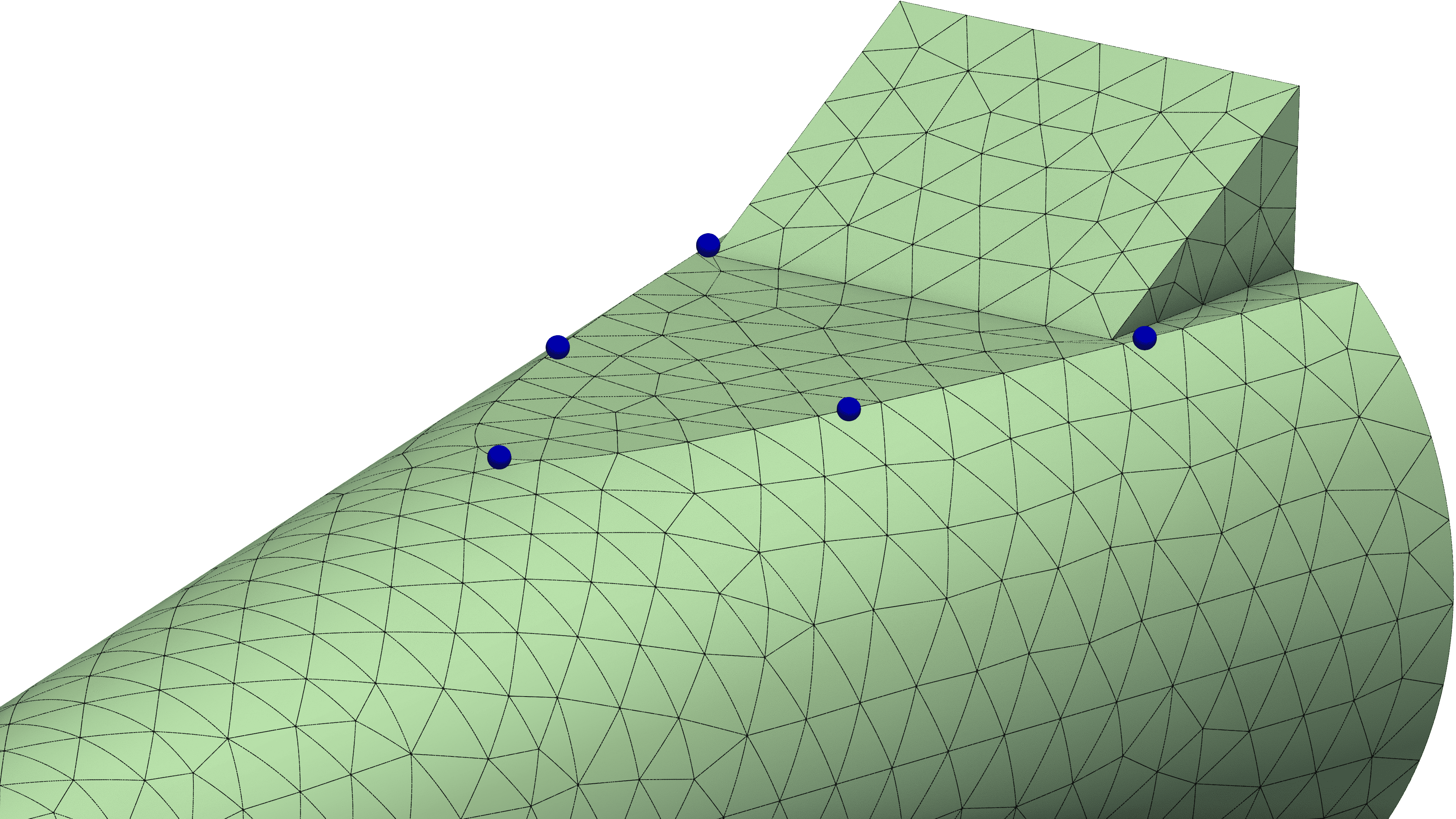};

\end{groupplot}\end{tikzpicture}
   \caption{
	   The parameter domain $\Rcal_{h',q'}$ and mesh $\hat\Rcal_{h',q'}$ of the
           intersection of the upper cone and slice surfaces (\textit{top}) and the
	   sliced cone flap geometry in physical space (\textit{bottom}). Nodes (blue)
	   maps from the one-dimensional parameter domain to the intersection of the
	   upper cone and slice surfaces in physical space.
   }
   \label{fig:oberkampf_int_map}
\end{figure}

\section{Conclusion}
\label{sec:conclude}
A mesh-based parametrization is a mapping from generalized coordinates in the intrinsic
dimension of a geometric object to a high-order approximation of the object itself in the
embedding space, constructed directly from a high-order mesh of the object, i.e., an
analytical expression or CAD representation of the object is not used. We rigorously
formulate mesh-based parametrizations of an arbitrary $d'$-dimensional object embedded
in a $d$-dimensional space using standard tools from high-order finite elements
(Section~\ref{sec:mbp:obj}). Then, we detail an approach to parametrize the nodal
coordinates of a computational mesh such that any node is guaranteed to remain on its
original boundaries by parametrizing each boundary and all boundary intersections using
mesh-based parametrization (Section~\ref{sec:mbp:msh}). Ensuring
nodes remain on their original boundaries guarantees a mesh face will not straddle
multiple boundaries, which would make assigning boundary conditions to such a face
difficult and would lead to poor geometry representation for non-smooth transitions
between boundaries. These boundary-preserving, mesh-based parametrizations are
embedded in an implicit shock tracking framework to allow nodes to freely slide
along their original boundaries in the process of aligning element faces with
non-smooth flow features. It is particularly important to allow boundary nodes
to slide for problems where shocks approach or intersect boundaries
(e.g., transonic flows, shock-boundary layer interactions) because fixing
them would either prohibit shock tracking or highly distort the mesh.
As such, the proposed boundary-preserving, mesh-based parametrizations
extend shock tracking capabilities to a new class of problems. Because mesh-based
parametrizations are defined elementwise, their efficient implementation relies on a
fast, reliable method to locate the element(s) of the parameter domain mesh
$\hat\Rcal_{h',q'}$ in which an arbitrary point $r\in\Rcal_{h',q'}$ lies.
We introduce an algorithm, specialized for high-order meshes, to efficiently
search the elements of a mesh to locate such a point (Section~\ref{sec:mbp:pnteval}).

Two concrete examples of mesh-based parametrizations are provided for analytical
geometries, a Gaussian profile ($d = 2$) and a quarter sphere ($d = 3$), to a
demonstrate the abstract formalism in Section~\ref{sec:mbp:obj} and show the
surrogate objects accurately approximate the true geometry when sufficiently
refined grids (e.g., small $h'$ or large $q'$) are used. Two examples demonstrate
boundary-preserving, mesh-based parametrizations cleanly integrate into the
implicit shock tracking and do not restrict or impede convergence. Finally, we
demonstrate the generality of mesh-based parametrizations using a complex
geometry, the sliced cone flap \cite{1993_oberkampf}. Future research should
investigate the benefits of implicit shock tracking for shock-dominated flows
over increasingly complex and relevant vehicles.


\section*{Acknowledgments}
This work is supported by
AFOSR award numbers FA9550-20-1-0236, FA9550-22-1-0002, FA9550-22-1-0004,
ONR award number N00014-22-1-2299, and
NSF award number CBET-2338843.
The content of this publication does not necessarily reflect the position
or policy of any of these supporters, and no official endorsement should
be inferred.

\appendix
\section{Restriction of embedded objects to their intrinsic dimension via projection}
\label{sec:surfproj}
In this section we detail simple approaches to define the mapping $\Pi_{h',q'}$, and
therefore the parameter space $\Rcal_{h',q'}$, by projecting a surface onto a predefined
hyperplane (\ref{sec:surfproj:surf}) and projecting a curve onto a predefined line
(\ref{sec:surfproj:curve}).

\subsection{Projection of surface onto hyperplane}
\label{sec:surfproj:surf}
Suppose $\Scal$ is a surface, i.e., $d' = d-1$ and define a hyperplane
with unit normal $n\in\Rbb^d$, $\norm{n} = 1$, that passes through $\hat{x}$,
i.e., $\Phi = \left\{x\in\Rbb^d \suchthat (x-\hat{x})\cdot n = 0\right\}$.
The orthogonal projection of any point $x\in\Scal_{h',q'}$ onto $\Phi$
is the solution of $\argmin_{r\in\Rbb^{d-1}} \norm{\hat{x} + A r - x}$, where
$A\in\Rbb^{d\times(d-1)}$ is a matrix whose rows are the vectors of the
nullspace of $n$, i.e., $A = \text{null}(n)^T$, which means $A\cdot n = 0$.
From the analytical solution of this linear least squares problem, we
define the projection-based restriction operator $\Pi_{h',q'}$ as
\begin{equation}
	\Pi_{h',q'} : x \mapsto (A^TA)^{-1}(A^T(x-\hat{x})).
\end{equation}
While simple, this approach has several drawbacks: the need to define the
hyperplane ($\hat{x}$, $n$) and, even for an optimal choice of hyperplane, the
mapping $\Pi_{h',q'}$ will lose injectivity for non-closed surfaces
or surfaces with excessive curvature. Because the framework proposed in this
manuscript handles surfaces and arbitrary intersections of surfaces, a closed
surface or surface with excessive curvature can be subdivided into smaller
open surfaces with less curvature at the expensive of additional surfaces
and intersections that must be parametrized. As such, even this simple
restriction operator works robustly for a wide class of surfaces. 

\subsection{Projection of curve onto line}
\label{sec:surfproj:curve}
Suppose $\Scal$ is a curve, i.e., $d' = 1$ and define a line
with unit tangent $v\in\Rbb^d$, $\norm{v} = 1$, that passes through $\hat{x}$,
i.e., $\Psi = \left\{x\in\Rbb^d \suchthat x = \hat{x} + \alpha v,~\alpha\in\Rbb\right\}$.
The projection of any point $x\in\Scal_{h',q'}$ onto $\Psi$ is the solution of
$\argmin_{\alpha\in\Rbb} \norm{\hat{x} + \alpha v - x}$. From the analytical solution
of this linear least squares problem, we define the projection-based restriction
operator $\Pi_{h',q'}$ as
\begin{equation}
        \Pi_{h',q'} : x \mapsto v \cdot (x-\hat{x}).
\end{equation}
Similar to~\ref{sec:surfproj:surf}, $\Pi_{h',q'}$ may not be injective for
closed curves or curves with high curvature, but this can be circumvented
by further subdividing surfaces until their intersections have been sufficiently
refined that $\Pi_{h',q'}$ is injective for all subdivided curves. 

\bibliographystyle{plain}
\bibliography{master, biblio_intro, biblio, mybib}

\end{document}